\documentclass[twocolumn]{aastex701}
\usepackage{tabularx}

\newcommand{\Mearth}{M_{\oplus}}
\newcommand{\Rearth}{R_{\oplus}}
\newcommand{\Mcore}{M_{\mathrm{core}}}
\newcommand{\fenv}{f_{\mathrm{env}}}
\newcommand{\fc}{f_{\mathrm{core}}}
\newcommand{\unit}[1]{\mathrm{#1}}
\newcommand{\kbb}{k_B/\mathrm{baryon}}
\newcommand{\numcoldplanets}{34}

\newcommand{\planetsolver}{\texttt{PlanetSolver}}
\newcommand{\entropy}{K}

\begin{document}

\title{Do Super-Puffs Defy Core Accretion? Population-Wide Interior Structure Constraints}

\author[orcid=0009-0007-1123-0038]{Nicholas T. Marston}
\affiliation{University of Wisconsin-Madison Department of Astronomy, 475 N. Charter St. Madison, WI 53706, USA}
\email{nmarston@wisc.edu}
\author[orcid=0000-0002-7733-4522]{Juliette Becker} 
\affiliation{University of Wisconsin-Madison Department of Astronomy, 475 N. Charter St. Madison, WI 53706, USA}
\email{juliette.becker@wisc.edu}
\author[orcid=0000-0002-4884-7150]{Alex R. Howe}
\affiliation{The Catholic University of America, 620 Michigan Ave., N.E. Washington, DC 20064}
\affiliation{NASA Goddard Space Flight Center, 8800 Greenbelt Rd, Greenbelt, MD 20771, USA}
\affiliation{Center for Research and Exploration in Space Science and Technology, NASA/GSFC, Greenbelt, MD 20771}
\email{alex.r.howe@nasa.gov}

\begin{abstract}

Sub-Saturn mass planets with extremely low bulk densities $(\rho\lesssim0.3\ \unit{g/cm^3})$, or ``super-puffs'', are one of the most interesting and least understood populations of exoplanets. While many short-period {super-puffs} can be attributed to the effects of high irradiation and star-planet interactions, cold super-puffs appear to challenge the expectations of core accretion theory. We constrain the possible properties of 34 cold super-puffs by computing hydrostatic interior structures using \planetsolver. {We find that 28 planets in our sample can be reproduced by models consistent with core accretion based on their observed masses and radii and adjusting for planet age. We identify HIP~41378~f, Kepler-30~d, Kepler-51~d, Kepler-177~c, TOI-1420~b, and WASP-107~b as planets inconsistent with core accretion theory which necessitate a non-standard explanation (e.g. exo-rings)}. With the exception of TOI-1420~b, core accretion-compatible solutions are possible for these planets if an additional heat source is present. We modify planetary evolution models to determine whether enhanced radiogenic heating or late impacts with sub-planetary mass objects can plausibly inflate sub-Neptunes enough to achieve super-puff densities. We find that the effects of radiogenic heating are insufficient to produce super-puff densities, but that impacts can in many cases produce the necessary inflation for upwards of $1\unit{Gyr}$. We also compile {and present} here an index of all currently known super-puffs.

\end{abstract}


\section{Introduction}

The boundary conditions of the exoplanet census are useful because they reveal not just limits on the scope and outcomes of planet formation, but can also illuminate the relevant physics that might otherwise be occluded by stochasticity in the formation process \citep[e.g.,][]{Batygin2023}. 

One particularly interesting example is the population of super-puff planets, characterized by very low masses relative to their radii, placing them at the extreme low end of bulk density measurements.
While the precise thresholds vary across the literature, super-puffs are typically defined as planets with masses $<30 M_{\oplus}$ and densities $<0.3\ \mathrm{g\,cm^{-3}}$ \citep{Lee2016, Howe_Architectures}. 
Beyond their shared anomalously low densities, this population is diverse and includes planets in a wide range of systems: single- \citep[e.g.,][]{Schanche_2025} and multi-planet \citep[e.g.,][]{Vanderburg2016} systems, planets around both old \citep[e.g.,][]{Cochran_2011_kepler18} and young \citep{Masuda2014} stars, and planets on both short-period, hot \citep[e.g.,][]{Beatty_2017} and long-period, cold orbits \citep[e.g.,][]{Jontof-Hutter_2014_Kepler79}.

The combination of large radii and low masses seen in the super-puff population is broadly consistent with substantial H/He envelopes \citep{Lopez_Fortney_2014_R-M-Relation,Rogers2015}. However, the specific parameter combinations exhibited by individual super-puffs do not present a unifying picture of a homogeneous population.
Part of this heterogeneity could be a consequence of observational limitations: for a planet to be identified as a super-puff, it must have both a measured radius and a measured mass. Consequently, these planets are generally discovered through transit photometry \citep[either by Kepler or TESS;][]{Borucki2010, Guerrero2021} and subsequently have their masses measured either via transit timing variations (TTVs) or follow-up radial velocity (RV) observations.

Because a planet cannot be classified as part of this population without a mass measurement, the planets currently identified as super-puffs are those amenable to such follow-up observations. This typically occurs either because they orbit stars bright enough for precise RV measurements, or because they reside in orbital configurations such as near mean-motion resonances, where TTV amplitudes are enhanced \citep{Hadden_Lithwick_2017} and thus allow a mass solution to be recovered.

As discussed by \citet{Howe_Architectures}, super-puffs appear to frequently reside in multi-planet systems, a correlation that remains true even when considering only planets with radial velocity measurements. At the same time, their host stars also seem more consistent with the population of stars that host giant planets than with those that host sub-Neptunes \citep{Howe2025b}, providing some evidence that their formation pathways may differ substantially from the typical pathways that produce sub-Neptune planets. With that in mind, several questions remain unresolved: how do these planets acquire such low-density envelopes without triggering runaway gas accretion? Does the similarity between their host stars and those of the Jovian population imply that super-puffs are failed Jupiters? Do the observed bulk densities in this population have a singular unifying explanation, or is the population a patchwork of planets with a variety of explanations?

In this paper, rather than treating the super-puff population as a unified group, we examine each planet individually to identify the objects that are most inconsistent with existing theories of planet formation.
In Section \ref{sec:PopulationOverview}, we describe the demographics of the population and discuss existing theories for the observed densities of super-puffs. In Section \ref{sec:InteriorStructureModeling}, we compute possible interior structures for a subset of the population and analyze the results. In Section \ref{sec:ThermalModels}, we examine whether heat-driven inflation caused by radiogenic heating or giant impacts provides a viable explanation for super-puffs. We discuss, summarize, and conclude our results in Sections \ref{sec:Discussion} and \ref{sec:Summary}.

\section{Demographics and Physical Origins of Super-Puff Planets} \label{sec:PopulationOverview}
\subsection{Sample Selection}\label{sec:PopulationOverview-demographics}
The term ``super-puff'' is colloquially used in the literature to describe planets that appear to be sub-Neptunes by mass and that also have very low bulk densities, which appear incompatible with typical mass-radius relations \citep[e.g.,][]{Lopez_Fortney_Miller_2012}. In recent years, the literature has mainly adopted a definitional density cut-off of $\rho<0.3\ \unit{g/cm^3}$ \citep{Piro2020_rings, Liang_2021_Kepler90, Howe_Architectures}, but this cutoff has in the past been considered to be as low as $\rho<0.1\ \unit{g/cm^3}$ \citep{Jontof2019}. The upper mass cut-offs used in the literature are more varied, including $10\,\Mearth$ \citep{Jontof2019}, $15\,\Mearth$ \citep{Piro2020_rings}, $30\,\Mearth$ \citep{Howe_Architectures}, and upwards of $50\,\Mearth$ \citep{Hallatt2021}.
 
\begin{figure}
    \centering
    \includegraphics[width=1\linewidth]{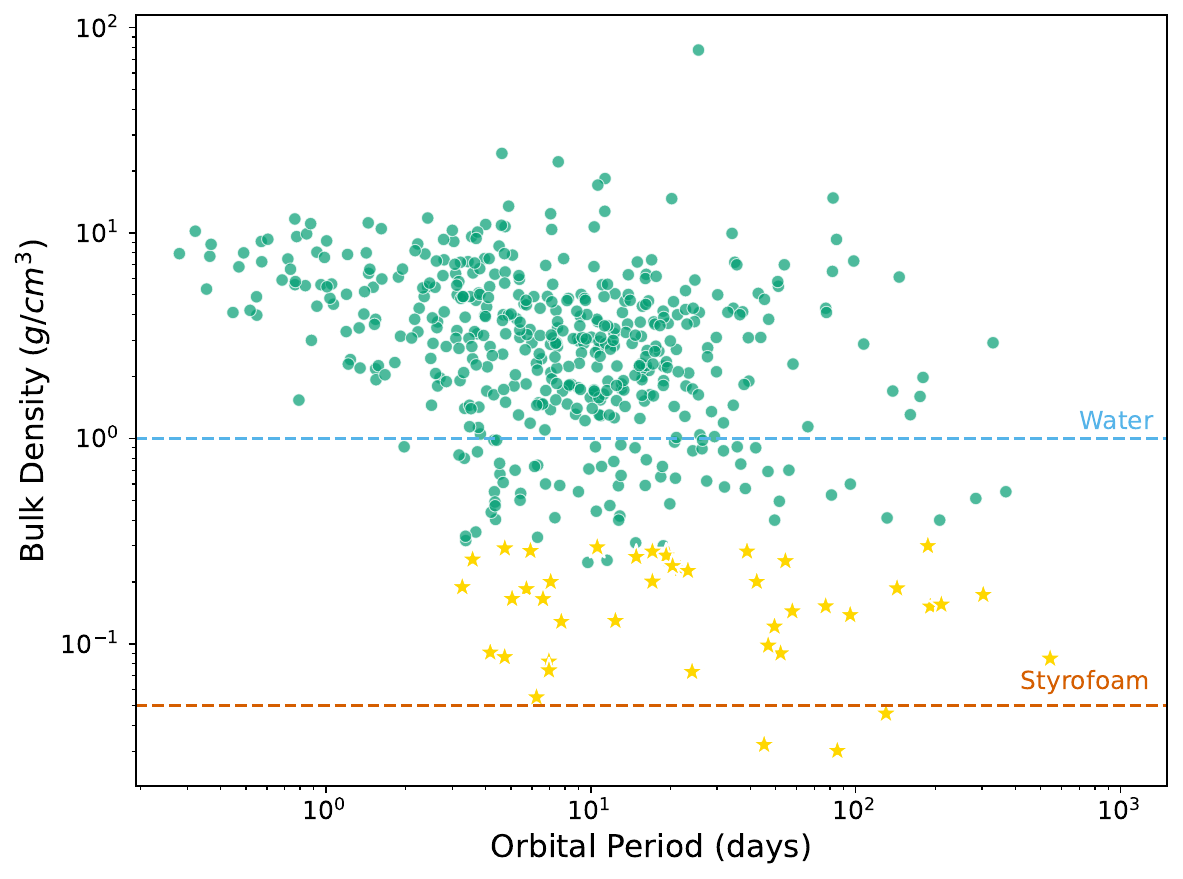}
    \caption{Bulk densities for sub-giant planets with well-constrained reported mass and radius measurements. Planets in our sample are denoted as yellow stars. Data sourced from NASA Exoplanet Archive on 12/23/2025 \citep{Christiansen2025}. \citep[The extreme high density outlier is Kepler-131c, based on the RV solution published by][]{Marcy14}}
    \label{fig:densitydistribution}
\end{figure}

In this work, we consider the possible interior structures for planets that fit into this super-puff population. To construct our sample and ensure we do not miss any interesting low-density planets, we adopted fairly liberal selection criteria of planetary mass $M_p < 60\,\Mearth$ and bulk density $\rho \le 0.30\ \unit{g/cm^{3}}$. We compiled this sample from super-puff planets previously identified in the literature, supplemented by additional candidates drawn from published solutions for confirmed planets on the NASA Exoplanet Archive \citep{Christiansen2025}.
We only include planets in our sample if they have measured masses (via either TTVs or RVs). 
In Figure \ref{fig:densitydistribution}, we show a comparison of our selected population of super-puff planets with the general population of planets for which masses and radii have been measured.

These planets have been observed orbiting host stars on orbits as short as $3.58\unit{d}$ \citep[HATS-8 b,][]{Bayliss_2015_HATS8b} and as long as $542\unit{d}$ \citep[HIP 41378 f,][]{Santerne2019}. 
With the notable exception of Kepler-51 b, c and d, super-puffs typically occur in diverse systems containing multiple planets with a wide range of densities \citep{Howe_Architectures}. Super-puffs have also been cataloged with diverse dynamical properties, including excited inclinations (e.g., \citealt{Espinoza-Retamal_PolarOrbit_2025}\footnote{This work was completed after the analysis of the present paper was complete, and as a result this planet was not included in our sample.}) and several with significant obliquities \citep{Yee_WASP193b2025}. They are also often observed in near-mean-motion resonances \citep[MMRs; e.g.,][]{Hadden_Lithwick_2017}.
For these reasons, super-puffs cannot be considered a homogeneous population with a single explanation. In order to understand the physical origin and structure of these planets, each of them must be studied in the context of its own environment. However, dividing the population into subsets based on shared characteristics can provide insight into this process.

\begin{figure}
    \centering
    \includegraphics[width=1\linewidth]{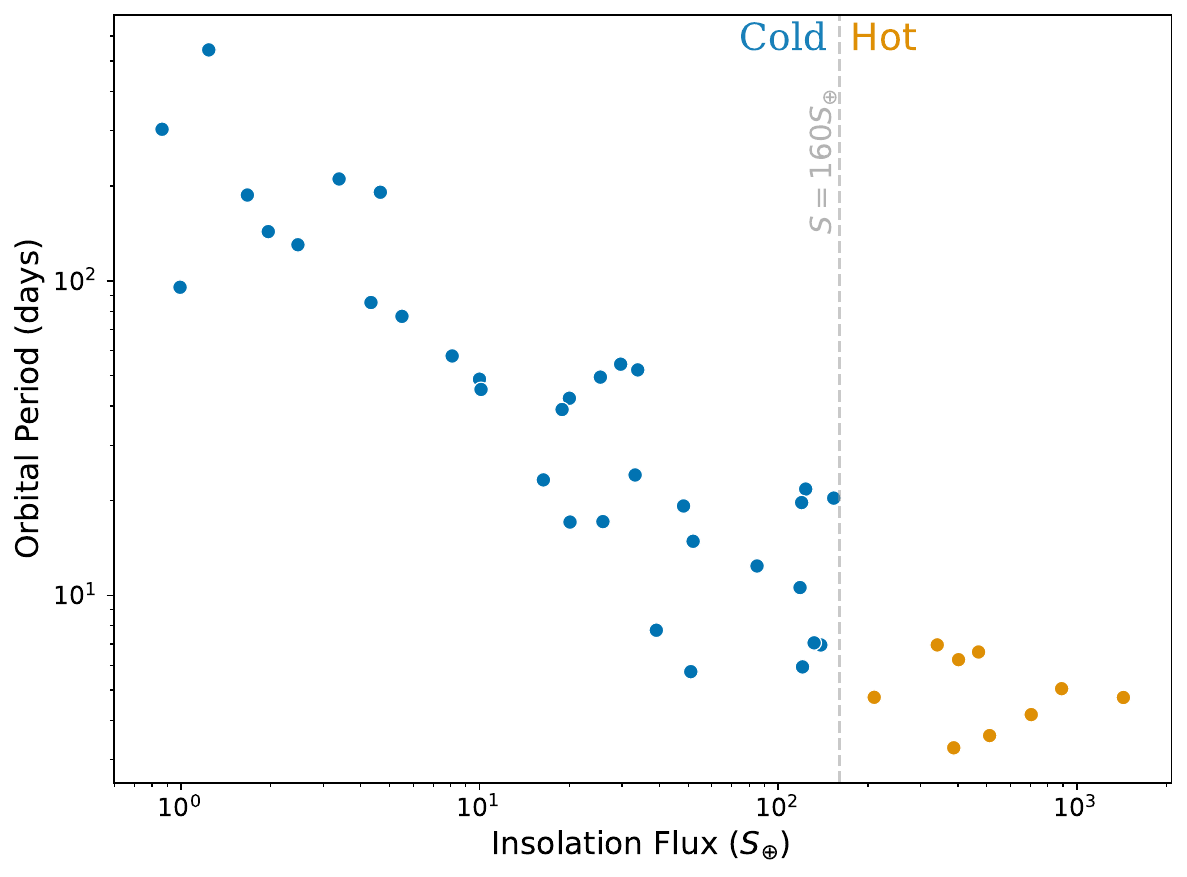}
    \caption{
    The hot-cold distribution of our sample. Shown on the horizontal axis is each planet's insolation flux compared to Earth (units of $S_{\oplus}$). Orbital period (in days) is shown on the vertical axis for comparison. The dashed line denotes the hot-cold boundary used in this work ($160S_{\oplus}$). For the circumbinary Kepler-47 system, we use the time-averaged insolation flux reported in Section 6.4 of \cite{Orosz2019_Kepler47}. } 
    
    \label{fig:irradiationdist}
\end{figure}

A useful distinction is to divide the planet population according to the stellar irradiation they receive. Highly irradiated (``hot'') planets can undergo radius inflation through several mechanisms that are far less effective in colder regions of planetary systems \citep{Demory2011, Thorngren2021}. For the purposes of this analysis, we adopt a semi-arbitrary insolation threshold of \(S = 160\,S_{\oplus}\) to separate hot and cold planets at a natural break in the distribution, and we restrict our study to the cold population. Figure~\ref{fig:irradiationdist} illustrates the distribution of irradiation levels for known super-puffs and marks our adopted boundary between the hot and cold regimes. With this cut applied, the shortest-period planet in our sample is WASP-107 b, with an orbital period of \(P = 5.721\,\unit{d}\) and an insolation flux of \(S = 51\,S_{\oplus}\). 
The planets in Figure \ref{fig:irradiationdist} that are classified as hot planets, which are therefore excluded from further analysis, are listed in Table \ref{tab:hotindex}. The corresponding sample of cold planets is provided in Table \ref{tab:coldindex}.

\begin{deluxetable*}{p{2.5cm}p{2.0cm}p{2.0cm}p{2.0cm}p{0.95cm}p{1.1cm}p{4.0cm}}
\tablecaption{Index of hot super-puff candidates excluded from further analysis based on their high incident stellar irradiation. Parameters listed here are $R_p$: the planet's observed radius in Earth radii, $M_p$: its mass in Earth masses, ${\rho:}$ the planet's bulk density in $\unit{g/cm^3}$, ${S:}$ the incident stellar irradiation in units of the nominal total solar irradiance \citep[solirad or So;][]{Mamajek2025}, and ${P:}$ the orbital period in days. References indicate the source of the adopted planetary parameters.}
\label{tab:hotindex}
\def\arraystretch{2.0}
\tablehead{
\colhead{\shortstack{Planet\\ Name}} & 
\colhead{\shortstack{${R_p}$\\($\Rearth$)}} & 
\colhead{\shortstack{${M_p}$\\($\Mearth$)}} & 
\colhead{\shortstack{${\rho}$\\(cgs)}} & 
\colhead{\shortstack{${S}$\\(So)}} & 
\colhead{\shortstack{${P}$\\(days)}} & 
\colhead{\shortstack{Ref.\\~}}
} 
\startdata
HATS-8 b & $9.79^{+1.38}_{-0.84}$ & $43.86^{+6.00}_{-6.00}$ & $0.26^{+0.07}_{-0.11}$ & 510.82 & 3.584 & \cite{Bayliss_2015_HATS8b} \\
HATS-46 b & $10.12^{+0.70}_{-0.50}$ & $54.98^{+19.70}_{-19.70}$ & $0.29^{+0.11}_{-0.12}$ & 209.77 & 4.742 & \cite{Brahm_Hartman_2018_HATS} \\
HATS-62 b & $11.83^{+0.28}_{-0.28}$ & $<56.90$ & $<$0.19 & 387.40 & 3.277 & \cite{Hartman_2019_HATS} \\
HIP 67522 b & $9.76^{+0.49}_{-0.50}$ & $13.80^{+1.00}_{-1.00}$ & $0.08^{+0.01}_{-0.01}$ & 341.09 & 6.959 & \cite{Thao__2024} \\
KELT-11 b & $15.13^{+1.10}_{-1.10}$ & $54.35^{+4.80}_{-4.80}$ & $0.09^{+0.02}_{-0.02}$ & 1433.00 & 4.736 & \cite{Beatty_2017} \\
TOI-3976 A b & $12.27^{+0.40}_{-0.39}$ & $55.62^{+11.80}_{-11.40}$ & $0.17^{+0.04}_{-0.04}$	 & 469.00 & 6.608 & \cite{Yee_2023_params} \\
WASP-127 b & $14.69^{+0.28}_{-0.33}$ & $52.35^{+6.80}_{-5.47}$ & $0.09^{+0.01}_{-0.01}$ & 703.97 & 4.178 & \cite{Seidel_2020} \\
WASP-193 b & $16.41^{+0.66}_{-0.64}$ & $44.18^{+9.20}_{-9.20}$ & $0.05^{+0.01}_{-0.01}$ & 402.00 & 6.246 & \cite{Barkaoui_2024_WASP193} \\
WASP-195 b & $10.31^{+1.00}_{-1.00}$ & $33.05^{+9.50}_{-9.90}$ & $0.17^{+0.07}_{-0.07}$ & 890.00 & 5.052 & \cite{Schanche_2025} \\
\enddata
\end{deluxetable*}

\subsection{Proposed Explanations}\label{sec:PopulationOverview-explanations}

An intuitive conclusion is that for super-puff planets to have such large radii, they must possess substantial H/He envelopes. 
This suggests that they likely formed in dust-free regions beyond the snow line \citep[where the lower opacity of material accreting onto the envelope allows for more efficient cooling and more accretion;][]{Lee2016} and subsequently migrated into their present, close-in locations. 
This interpretation is supported by the observation that many super-puff planets lie near mean-motion resonances (MMRs), a geometry that is a common signature of disk-driven migration \citep{Batygin2015}. Planets of this size with large enough envelopes to produce super-puff densities are vulnerable to increased rates of atmospheric loss. However, recent work suggests that lower mass super-puffs are able to retain large H/He atmospheres on long timescales \citep{TangFortneyMurrayClay2025_massloss}.

Another possibility is that super-puffs are similar {in composition} to other sub-Neptune-mass planets, but possess additional heat sources that inflate their radii, making them appear anomalously low-density. These heating mechanisms can be an internal process within the planet, or the result of environmental effects and interactions (e.g. planet-star interactions).
One such heat source could be tidal heating: planets on nonzero-eccentricity, short-period orbits may experience significant internal heating due to tidal dissipation \citep{Hut1981}, resulting in an inflated planetary radius at thermal equilibrium \citep{Bodenheimer2001, Ibgui2009}.

Planets with high irradiation levels and partially ionized envelopes can also experience heating and inflation due to Ohmic dissipation \citep{Batygin2010}, where atmospheric winds generate heat in a planet's envelope via electrical resistance. 
This effect could explain the anomalously large radii for some short-period hot Jupiters \citep{Laughlin2011}, mini-Neptunes \citep{Pu2017}, and even some super-puffs such as WASP-107 b \citep{Batygin_2025_WASP107b}.

A further possible explanation for super-puff planet densities is non-hydrostatic atmospheres. In particular, dusty outflows \citep{Gao2020, Ohno2021, Wang2019_dustyoutflows} might increase the apparent transit radius by lifting grains or hazes into (and/or out of) the upper planetary atmosphere. Similarly, extended ring systems may, for some inclination angles, inflate the observed planetary radius \citep{Akinsanmi2020,Piro2020_rings, Ohno2022}. 
Another natural explanation for some degree of radius inflation in young super-puffs is that they have not yet lost their heat of formation \citep{Lopez_Fortney_2014_R-M-Relation}, and as a result retain a high envelope entropy \citep[see, for example, radius evolution tracks in ][]{Lopez_Fortney_Miller_2012, Howe_Burrows_Evolution}. 

While the hypotheses discussed above have been examined relatively extensively in the literature, novel and less studied potential explanations for the low densities of super-puff planets have also been proposed, including the presence of a cloud of material or debris acquired through interactions with comets \citep{Rafizadeh_2025}, recent impacts from smaller bodies that temporarily inflate the planetary radius upon impact \citep{Anderson2012}, and the effects of radiogenic heating. These latter two hypotheses will be revisited in Section \ref{sec:ThermalModels}.

\section{Hydrostatic Interior Structure Modeling}\label{sec:InteriorStructureModeling}
The goal of this section is to examine the cold super-puff population, model the range of possible interior structures for each planet, and identify planets for which a non-standard explanation is necessary. We conclude that such an explanation is needed if all possible interior structures for a planet are inconsistent with the core accretion paradigm, as we assume that planets of these masses should have formed via core accretion. 
The origins of the hot section of our sample (Table \ref{tab:hotindex}) can most likely be attributed to their proximity to their host stars, so we focus our modeling on the more difficult-to-explain cold planets (with parameters given in Table \ref{tab:coldindex}).

In order to determine where a non-standard, non-core-accretion formation explanation for a given planet is necessary, we solve for the parameters that create a hydrostatic planet in agreement with its observed mass and radius.

\begin{figure}
    \centering
    \includegraphics[width=1\linewidth]{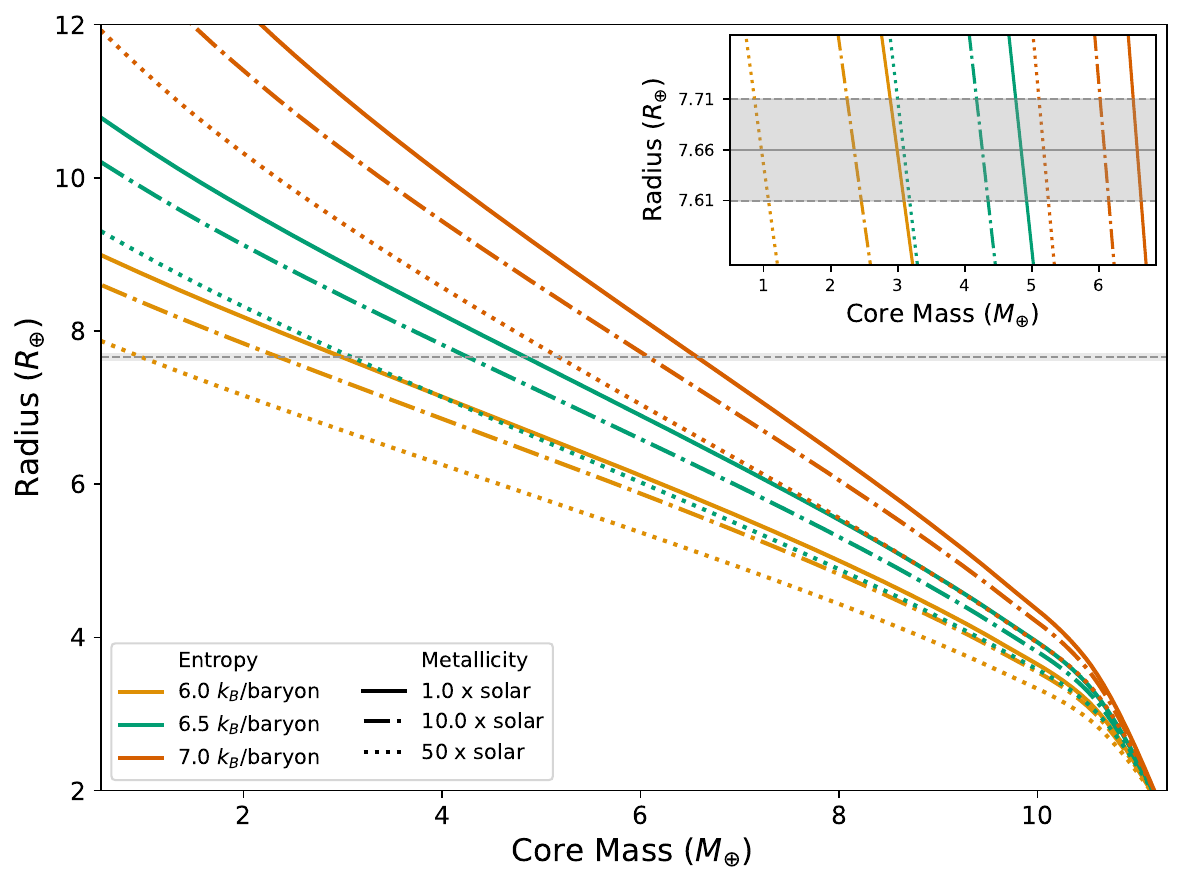}
    \caption{A set of modeled $\Mcore-R_p$ curves for TOI-1338 b, computed for a range of atmospheric specific entropies and metallicities. The inset shows the point where several of the curves achieve TOI-1338 b's measured radius, shown as a solid line with the shaded region representing measurement uncertainty.}
    \label{fig:exampleCMFradius}
\end{figure}

\subsection{Model Description}\label{sec:InteriorStructureModeling-description}

To find the range of plausible interior models for each planet in our sample, we employ a similar method to \citet{Batygin_Stevenson_2013} and \cite{Belkovski_2022} to compute hydrostatic equilibrium (HSE) solutions for a variety of possible planet compositions, then assess which parameter combinations result in a planetary structure consistent with the observationally measured mass and radius. 
We perform these computations using \planetsolver, an open-source implementation of the HSE models described in \cite{Howe2014} and \cite{Howe_Burrows_Evolution}. 

For each planet, we compute a parameter sweep over the core mass fraction ($\fc$), atmospheric entropy ($\entropy$), and atmospheric metallicity ($Z$), varying each parameter and identifying combinations that produce hydrostatic solutions matching the planet's observed mass and radius. 
The equations of HSE are solved using a fourth-order Runge-Kutta method given a combination of equations of state representing the planet's atmospheric and core components. 
We consider four core models, composed of $100\%$ Iron, $100\%$ Ice VII, $100\%$ Perovskite, and a fiducial core model for a ``terrestrial'' planet core-mantle structure. Our fiducial core model, unless otherwise noted, is composed of $67.5\%$ Perovskite and $32.5\%$ Iron \citep{Howe_Burrows_Evolution}. The pure Ice and Iron cores serve as theoretical upper and lower limits on core density, while the Perovskite and terrestrial cores feature more realistic intermediate densities. The atmospheric component of each planet is modeled as a fully convective H$_2$-He envelope. Further details about the implementation of the model, including equations of state and additional assumptions, can be found in \cite{Howe2014}.

For our parameter sweeps, we consider atmospheric metallicities of $0.1,\ 1,\ 10,$ and $100 \times \unit{Solar}$, and atmospheric entropy between $5.5$ and $7.0 \kbb$.  For each combination of $\entropy$ and $Z$, we compute interior structure models with 14 core mass fractions linearly spaced between $0.05$ and $1$, and derive the radius corresponding to these parameters if the model yields a hydrostatically stable solution. From these models, we fit curves to the $\Mcore-R_p$ input-output pairs for each atmospheric configuration, and interpolate intermediate values.  
As an example, $\Mcore-R_p$ curves for TOI-1338 b are shown in Figure \ref{fig:exampleCMFradius} for our fiducial core composition, three values of envelope entropy, and three values of envelope metallicity.
Similar curves were computed for each planet in our sample, using the planet parameters (planet radius, mass, insolation value) given in Table \ref{tab:coldindex}. 
For each planet, we then determine the points at which curves intersect the planet's observed radius, if they exist, as shown in the inset panel of Figure \ref{fig:exampleCMFradius}. Any combination of tested planetary parameters ($\fc,\ K,\ Z$) that reproduces the observed mass and radius is considered a possible internal structure for the planet. The solutions for $\Mcore$ and $\fenv$ for the {terrestrial} core composition are given in Table \ref{tab:coldindex}, and the solutions for all four core compositions are plotted in Figure \ref{fig:syntheticModelResults}.

{Note that our parameter set combines both TTV and RV-derived mass measurements. It has been noted previously that since the TTV and RV methods are sensitive to different planet populations \citep{Steffen2016, Mills2017}, a population with masses derived via both methods can in some cases produce systematic offsets between the two halves of the population. We do not find this to be the case in this analysis, and we present a comparison of TTV and RV-derived masses in the context of our results (Figure \ref{fig:resultdemographics}).}

{Furthermore, we note that the results of our analysis are sensitive to variation of planet mass and radius by $\sigma_M$ and $\sigma_R$. To evaluate the effect of this observational uncertainty on the planetary mass and radius, we run additional sweeps as shown in Figure \ref{fig:syntheticModelResults} for two cases of parameter combinations:  first taking the minimum mass and radius within reported $1\sigma$ uncertainty $(R_p-\sigma_R^-, \ M_p-\sigma_M^-)$, which may maximize $\fc$ in some cases \citep[see e.g.][]{Zeng2019}; then, additionally, taking the case which maximizes bulk density, $(R_p-\sigma_R^-, \ M_p+\sigma_M^+)$. We find that both of these resultant distributions do not differ significantly from the case where measurement error is neglected. Unless otherwise stated, the results presented in this work are derived from the reported best-fit values (neglecting uncertainty), but a note is made when a planet's results differ significantly between parameter combinations. }

In a few cases, the simulations encountered non-physical failure states associated with numerical convergence issues. To mitigate this behavior, an offset was introduced to the core and atmospheric mass fractions on the order of $10^{-5}$, which drastically reduced the issue's occurrence rate. When additional failure states were encountered, the affected structures were re-computed with incrementally increased offsets. At no point did the total applied offset exceed $10^{-4}$, ensuring that the resulting planetary structures were not meaningfully altered by this numerical regularization.

\startlongtable
\begin{deluxetable*}{p{2.5cm}p{1.75cm}p{1.75cm}p{1.75cm}p{0.95cm}p{1.1cm}p{0.5cm}p{0.5cm}p{1.0cm}>{\centering\arraybackslash}p{1.0cm}}
\tablecaption{Index of cold super-puffs included in this work, with their observed parameters and modeled properties. Parameters listed here are ${R_p,\ M_p:}$ the planet's observed radius and mass in Earth units, ${\rho:}$ the planet's bulk density, in $\unit{g/cm^3}$, ${S:}$ insolation flux received by the planet relative to Earth in units of solirad \citep{Mamajek2025}, ${P}:$ orbital period (days), {Quadrant:} The {classification for each planet based on our analysis (described in Section \ref{sec:InteriorStructureModeling}) assuming the best-fit $(M_p, R_p)$ parameter combination}, ${\Mcore:}$ the maximum core mass achieved by our fiducial core-type models for the best-fit mass/radius, ${f_{\mathrm{env}}:}$ the envelope mass corresponding to the maximum core mass (previous column).  Notes. (+): Planet is a circumbinary planet and has variable insolation flux. {(*): Planet is classified in quadrant IV with $1\sigma$ variation $(R_p-\sigma_R^-, \ M_p\pm\sigma_M^{\pm})$.} References: [1] \cite{Vanderburg2016, Becker2019, Santerne2019, Belkovski_2022}, [2] \cite{Petigura_2018_K2-24PARAMS}, [3] \cite{Vissapragada2020}, [4] \cite{Cochran_2011_kepler18}, [5] \cite{Mills_2016}, [6] \cite{Sanchis-Ojeda_2012}, [7] \cite{Hadden_Lithwick_2017}, [8] \cite{Orosz2019_Kepler47}, [9] \cite{Masuda2014},  [10] \cite{JontofHutter2014}, [11] \cite{Ofir_2014}, [12] \cite{Weiss_2024}, [13] \cite{Santerne_2016_SOPHIE_PARAMS, Liang_2021_Kepler90, Shaw2025}, [14] \cite{Wang_Liu_2024}, [15] \cite{Ofir_2025}, [16] \cite{Yoshida_TOI1420b_2023}, [17] \cite{Polanski_2024},  [18] \cite{McKee_Montet_2023},  [19] \cite{Tala_Pinto_2025},  [20] \cite{Trifonov_2023},  [21] \cite{Mantovan_Malavolta_2024},  [22] \cite{Livingston2026},  [23] \cite{Bonomo_2023},  [24] \cite{Galarza_2024_PARAMS_TOI1173},  [25] \cite{Močnik_2017}, [26] \cite{Hellier_2017}  }
\label{tab:coldindex}
\def\arraystretch{2.0}
\tablehead{
\colhead{\shortstack{Planet\\ Name}} & 
\colhead{\shortstack{$R_p$\\($\Rearth$)}} & 
\colhead{\shortstack{$M_p$\\($\Mearth$)}} & 
\colhead{\shortstack{$\rho$\\(cgs)}} & 
\colhead{\shortstack{$S$\\(So)}} & 
\colhead{\shortstack{$P$\\(days)}} & 
\colhead{\shortstack{Quad-\\rant}} & 
\colhead{\shortstack{$\Mcore$\\($M_{\oplus}$)}} & 
\colhead{\shortstack{$f_{\mathrm{env}}$\\~}} & 
\colhead{\shortstack{Ref.\\~}}
} 
\startdata 
HIP 41378 f & $9.20^{+0.10}_{-0.10}$ & $12.00^{+3.00}_{-3.00}$ & $0.08^{+0.02}_{-0.02}$ & 1.24 & 542.08 & IV & 5.14 & $57.18\%$ & [1] \\     
K2-24 c & $7.50^{+0.30}_{-0.30}$ & $15.40^{+1.90}_{-1.80}$ & $0.20^{+0.03}_{-0.03}$ & 20.00 & 42.339 & I & 8.94 & $41.93\%$ & [2] \\         
K2-141 c & $7.00^{+4.60}_{-2.80}$ & $<8.00$ & $<0.13$ & 39.10 & 7.748 & I & 5.42 & $32.24\%$ & [23] \\                        
Kepler-9 b & $8.09^{+0.21}_{-0.21}$ & $25.90^{+9.63}_{-9.63}$ & $0.27^{+0.10}_{-0.10}$ & 48.23 & 19.271 & III & 12.85 & $50.40\%$ & [12] \\  
Kepler-9 c & $8.11^{+0.22}_{-0.22}$ & $27.29^{+14.37}_{-14.37}$ & $0.28^{+0.15}_{-0.15}$ & 18.90 & 38.908 & III & 13.46 & $50.69\%$ & [12] \\
Kepler-18 d & $6.98^{+0.33}_{-0.33}$ & $16.4^{+1.40}_{-1.40}$ & $0.27^{+0.04}_{-0.04}$ & 51.92 & 14.859 & II & 10.4 & $36.57\%$ & [4] \\     
Kepler-30 d & $8.80^{+0.50}_{-0.50}$ & $23.10^{+2.70}_{-2.70}$ & $0.19^{+0.04}_{-0.04}$ & 1.96 & 143.344 & IV & 9.57 & $58.59\%$ & [6] \\    
Kepler-33 d & $4.60^{+1.00}_{-0.90}$ & $4.10^{+1.70}_{-2.00}$ & $0.23^{+0.17}_{-0.19}$ & 123.69 & 21.776 & I & 3.65 & $10.89\%$ & [7] \\     
Kepler-47 d & $7.04^{+0.66}_{-0.49}$ & $19.02^{+23.84}_{-11.67}$ & $0.30^{+0.38}_{-0.20}$ & (+) & 187.366 & II & 11.87 & $37.60\%$ & [8] \\  
Kepler-47 c & $4.65^{+0.09}_{-0.07}$ & $3.17^{+2.18}_{-1.25}$ & $0.17^{+0.12}_{-0.07}$ & (+) & 303.227 & I & 2.85 & $10.03\%$ & [8] \\       
Kepler-51 b & $7.10^{+0.30}_{-0.30}$ & $2.10^{+1.50}_{-0.80}$ & $0.03^{+0.02}_{-0.01}$ & 10.12 & 45.154 & I & 1.74 & $17.08\%$ & [9] \\      
Kepler-51 c & $9.00^{+2.80}_{-1.70}$ & $4.00^{+0.40}_{-0.40}$ & $0.03^{+0.02}_{-0.03}$ & 4.33 & 85.312 & I & 2.57 & $35.72\%$ & [9] \\       
Kepler-51 d & $9.70^{+0.50}_{-0.50}$ & $7.60^{+1.10}_{-1.10}$ & $0.05^{+0.01}_{-0.01}$ & 2.47 & 130.194 & IV & 3.53 & $53.50\%$ & [9] \\     
Kepler-79 d & $7.16^{+0.13}_{-0.16}$ & $6.00^{+2.10}_{-1.60}$ & $0.09^{+0.03}_{-0.02}$ & 33.87 & 52.09 & I & 4.17 & $30.49\%$ & [10] \\      
Kepler-87 c & $6.14^{+0.29}_{-0.29}$ & $6.40^{+0.80}_{-0.80}$ & $0.15^{+0.03}_{-0.03}$ & 4.66 & 191.232 & I & 4.89 & $23.57\%$ & [11] \\     
Kepler-89 e & $6.11^{+0.14}_{-0.14}$ & $10.49^{+0.98}_{-0.95}$ & $0.25^{+0.03}_{-0.03}$ & 29.70 & 54.32 & I & 7.74 & $26.22\%$ & [12, 15] \\ 
Kepler-90 g & $8.10^{+0.80}_{-0.80}$ & $15.00^{+0.90}_{-0.80}$ & $0.16^{+0.05}_{-0.05}$ & 3.39 & 210.603 & I & 7.76 & $48.24\%$ & [13] \\    
Kepler-177 c & $8.73^{+0.36}_{-0.34}$ & $14.70^{+2.70}_{-2.50}$ & $0.12^{+0.03}_{-0.03}$ & 25.40 & 49.409 & IV & 6.65 & $54.73\%$ & [3] \\   
Kepler-223 e & $4.60^{+0.27}_{-0.41}$ & $4.80^{+1.40}_{-1.20}$ & $0.27^{+0.11}_{-0.08}$ & 119.88 & 19.726 & I & 4.24 & $11.58\%$ & [5] \\    
Kepler-359 c & $4.80^{+1.00}_{-0.90}$ & $2.90^{+2.40}_{-1.90}$ & $0.14^{+0.14}_{-0.13}$ & 8.10 & 57.693 & I & 2.6 & $10.24\%$ & [7] \\       
Kepler-359 d & $4.60^{+0.90}_{-0.90}$ & $2.70^{+2.50}_{-1.50}$ & $0.15^{+0.17}_{-0.12}$ & 5.50 & 77.083 & I & 2.46 & $9.04\%$ & [7] \\       
TOI-216.02 & $7.84^{+0.21}_{-0.19}$ & $17.60^{+0.60}_{-0.60}$ & $0.20^{+0.02}_{-0.02}$ & 25.90 & 17.099 & I & 9.44 & $46.36\%$ & [18] \\     
TOI-1173 A b & $9.02^{+0.16}_{-0.15}$ & $28.30^{+4.10}_{-4.00}$ & $0.21^{+0.03}_{-0.03}$ & 132.00 & 7.065 & III* & 10.79 & $61.88\%$ & [24] \\
TOI-1338 b & $7.66^{+0.05}_{-0.05}$ & $11.30^{+2.10}_{-2.10}$ & $0.14^{+0.03}_{-0.03}$ & 0.99 & 95.4 & I & 6.61 & $41.46\%$ & [14] \\        
TOI-1420 b & $11.89^{+0.33}_{-0.33}$ & $25.10^{+3.80}_{-3.80}$ & $0.08^{+0.01}_{-0.01}$ & 139.00 & 6.956 & IV & 1.68 & $93.31\%$ & [16] \\   
TOI-1836 b & $8.28^{+0.20}_{-0.17}$ & $24.70^{+6.10}_{-5.90}$ & $0.24^{+0.06}_{-0.06}$ & 153.52 & 20.381 & III & 11.72 & $52.57\%$ & [17] \\ 
TOI-2328 b & $9.98^{+0.45}_{-0.56}$ & $50.87^{+6.36}_{-6.36}$ & $0.28^{+0.06}_{-0.05}$ & 20.10 & 17.102 & III & 13.36 & $73.74\%$ & [19] \\  
TOI-2525 b & $8.68^{+0.11}_{-0.11}$ & $27.00^{+2.00}_{-2.00}$ & $0.23^{+0.02}_{-0.02}$ & 16.36 & 23.286 & III & 11.44 & $57.65\%$ & [20] \\  
TOI-5398 b & $10.30^{+0.40}_{-0.40}$ & $58.70^{+5.70}_{-5.60}$ & $0.30^{+0.04}_{-0.04}$ & 118.41 & 10.591 & III & 13.29 & $77.36\%$ & [21] \\
V1298 Tau b & $9.95^{+0.37}_{-0.35}$ & $13.10^{+5.30}_{-5.30}$ & $0.07^{+0.03}_{-0.03}$ & 35.00 & 24.14 & IV & 4.5 & $65.65\%$ & [22] \\     
V1298 Tau d & $6.34^{+0.30}_{-0.30}$ & $6.00^{+0.70}_{-0.70}$ & $0.13^{+0.02}_{-0.02}$ & 85.00 & 12.4 & I & 4.53 & $24.49\%$ & [22] \\       
V1298 Tau e & $9.50^{+0.51}_{-0.49}$ & $15.30^{+4.20}_{-4.20}$ & $0.10^{+0.03}_{-0.03}$ & 10.00 & 46.77 & IV & 5.62 & $63.29\%$ & [22] \\    
WASP-107 b & $10.40^{+0.20}_{-0.20}$ & $37.80^{+4.40}_{-4.40}$ & $0.18^{+0.02}_{-0.02}$ & 51.00 & 5.721 & IV & 7.86 & $79.20\%$ & [25] \\    
WASP-139 b & $8.97^{+0.60}_{-0.60}$ & $37.19^{+5.40}_{-5.40}$ & $0.28^{+0.07}_{-0.07}$ & 120.71 & 5.924 & III & 14.35 & $61.41\%$ & [26]  
\enddata
\end{deluxetable*}

\subsection{Model Results}\label{sec:InteriorStructureModeling-results}

\begin{figure*}
    \centering
    \includegraphics[width=1\linewidth]{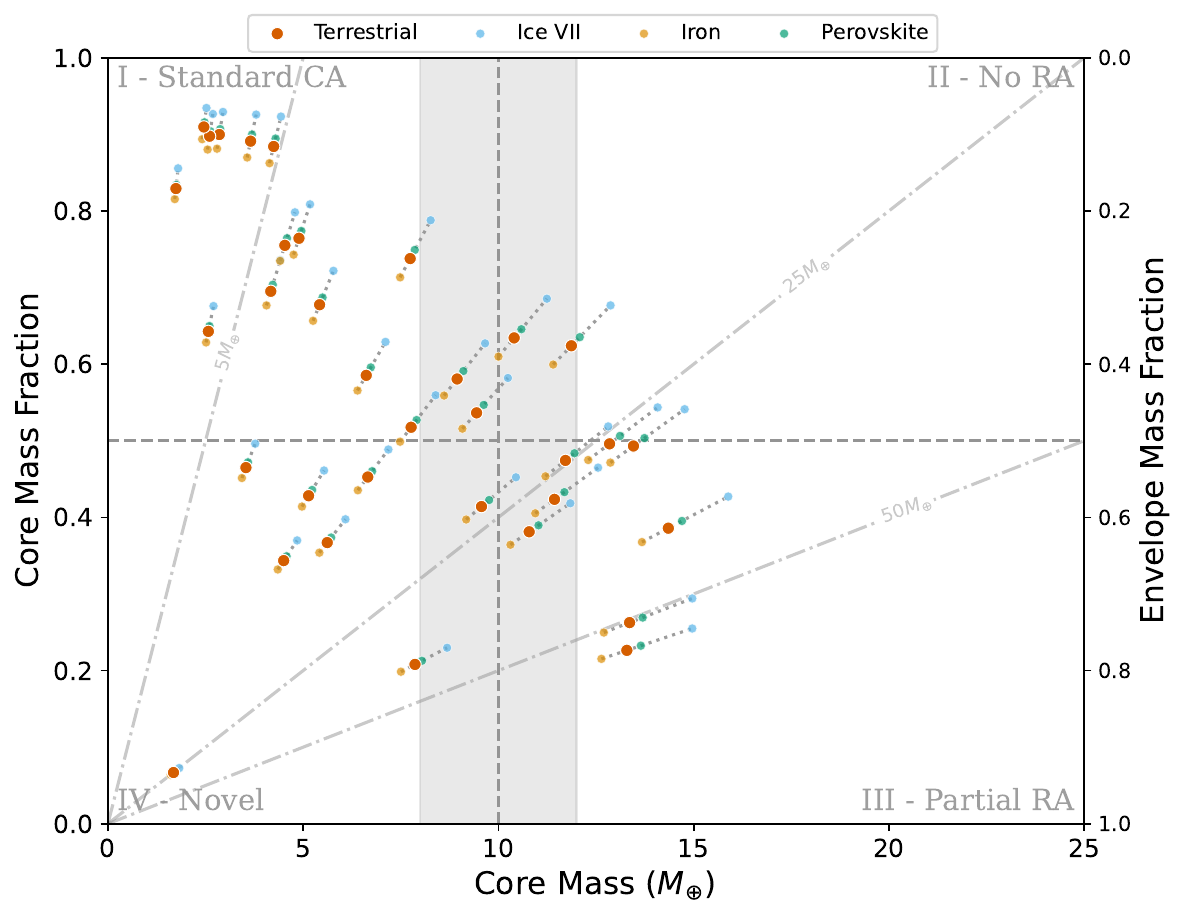} 
    \caption{The interior structure solution that achieved the highest $\fc$ for each planet in our sample with $\entropy=7.0 \kbb$, shown with respect to the Core Mass $\Mcore$ (horizontal axis) and the Core (and Envelope) Mass Fraction $\fc$ ($\fenv$) (vertical axis). Note that we find the $\fc$ for a given solution to be monotonically increasing with $\entropy\in[5.5, 7.0]\kbb$. Results corresponding to each of the core compositions we considered are shown, indicated by color, with dotted lines connecting the set of points corresponding to a particular planet. Estimates of the critical points for runaway accretion (RA) to occur are shown as dashed lines at $\fc = 0.5$ and $\Mcore = 10\,\Mearth$, with shading to indicate a region of uncertainty as the exact cutoff may depend on planet formation location and disk parameters \citep{Piso2015}. Dash-dotted lines of constant total mass are included for reference. The plot is divided by these lines into quadrants corresponding to the inferred physical interpretations: planets consistent with typical core accretion (I - Standard CA), planets which did not undergo runaway accretion despite meeting the requirements (II - No RA), planets which underwent a brief phase of runaway accretion (III - Partial RA), and planets inconsistent with any standard theory (IV - Novel). {Note that the overlapping points in the bottom left of the figure correspond to TOI-1420 b.}}
    \label{fig:syntheticModelResults}
\end{figure*} 

\begin{figure}

    \includegraphics[width=1\linewidth]{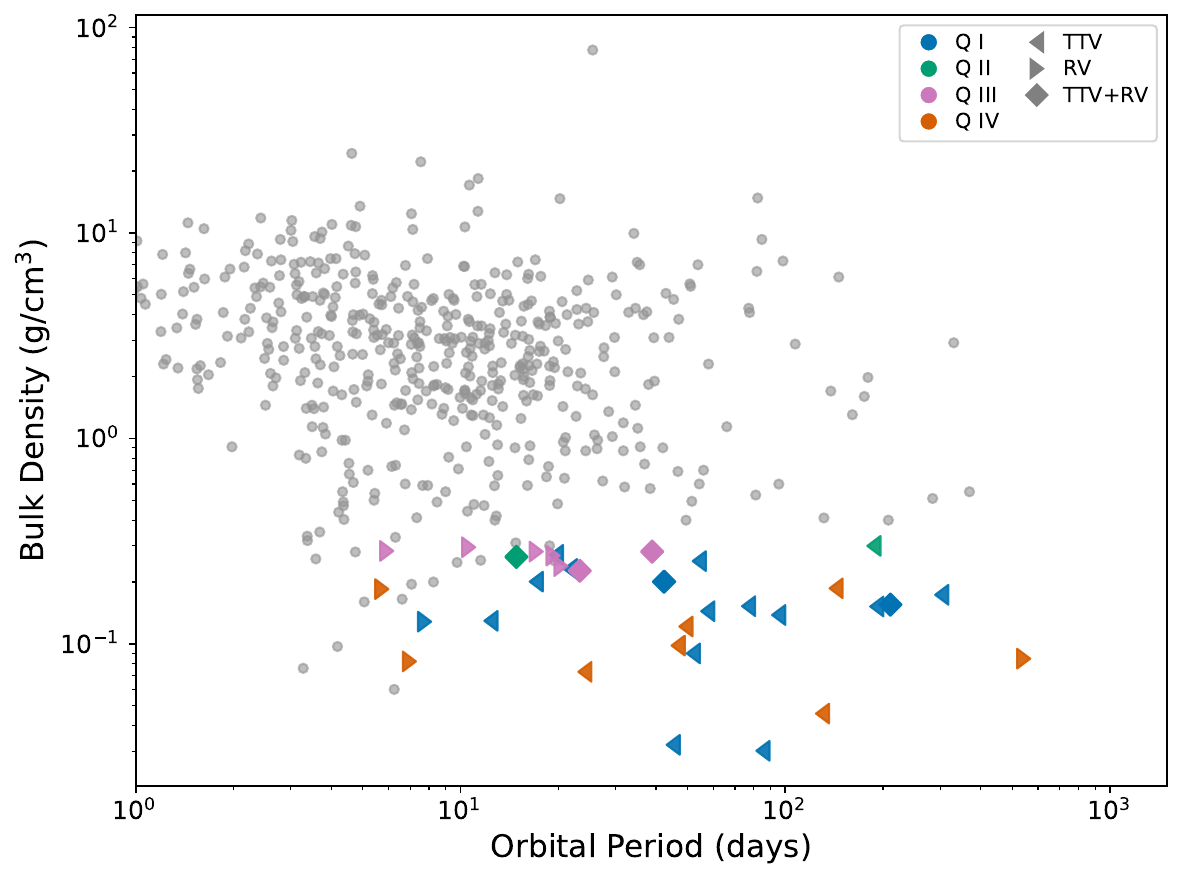} 
    \caption{{The measured orbital period and densities for our modeled cold super-puffs, in the context of the broader exoplanet sample as a whole (grey points). Super-puff points are color coded according to their classification in Section \ref{sec:InteriorStructureModeling-results}. Left-facing triangles, right-facing triangles, and diamonds denote planets whose adopted parameters are consistent with TTV measurements, RV measurements, or both, respectively. The vertical axis shows the maximum density within mass and radius measurement uncertainties, and the horizontal axis shows the planet's orbital period. For comparison, best-fit densities for the remainder of the exoplanet sample are plotted as grey circles. }}
    \label{fig:resultdemographics}
\end{figure}

The computed core mass and core mass fraction corresponding to the hydrostatic solutions for each planet in the sample given in Table \ref{tab:coldindex} are shown in Figure \ref{fig:syntheticModelResults}.
To guide the eye, the parameter space is divided into four quadrants (I, II, III, and IV). Boundaries between quadrants are set at a core mass of 10 $M_{\oplus}$ \citep[the approximate core mass at which a planet may undergo runaway gas accretion;][]{Pollack1996} and a core mass fraction of 50\% \citep[corresponding to the crossover mass where runaway accretion will occur once a planet reaches the aforementioned core mass;][]{Mizuno1980}. 
Each quadrant thus reflects a different degree of consistency with the expectations of core accretion theory, with Quadrants I, II, and III being consistent in ways we will describe in the following subsections, while Quadrant IV is inconsistent without additional non-standard hypotheses. {In Figure \ref{fig:resultdemographics}, we show a density-orbital period distribution which indicates the quadrant classification and mass measurement method for each planet in our sample.}

Since we do not know the true core compositions of the observed super-puff population, each planet shown in Figure~\ref{fig:syntheticModelResults} is represented by four connected points, corresponding to our four tested possible core compositions. 
For some planets, variations in interior composition lead to slight shifts in quadrant placement.
The computed core mass and envelope fraction combinations in Figure~\ref{fig:syntheticModelResults} correspond to the most conservative set of solutions within our parameter space ($K=7.0\kbb$, $Z=0.1\times \unit{Solar}$; that is, {the atmospheric properties} producing the largest modeled radii).

\textit{Quadrant I.} Of the \numcoldplanets\ cold super-puffs in our sample, {16} can be reproduced by models with a terrestrial core composition with $\fc > 50\%$ and $\Mcore < 10\,\Mearth$, {increasing to 22 when using $(R_p-\sigma_R^-, \ M_p-\sigma_M^-)$ and decreasing to 14 when using $(R_p-\sigma_R^-, \ M_p+\sigma_M^+)$.} 

These structures are consistent with expectations for sub-Saturn mass planets (sub-Neptunes, Neptunes, etc.) that did not reach the required critical core mass to begin runaway accretion, which is estimated to be 10 $M_{\oplus}$ \citep{Pollack1996}, but may vary slightly depending on disk parameters \citep{Rafikov2006, Piso2015}.

\textit{Quadrant II}. An additional {2} planets reached the $10\,\Mearth$ threshold for the onset of runaway accretion when assuming a terrestrial core composition. However, these planets were unable to sustain runaway growth and therefore did not become Jupiter-mass objects. Instead, they acquired large gaseous envelopes ($\fenv > 50\%$) while remaining low in total mass. These planets are easily consistent with the expectations of core accretion theory, as they simply failed to complete the runaway accretion process, which could occur for multiple reasons. Runaway accretion depends on the planet's ability to cool and contract, thereby allowing more gas to fall within the hill sphere \citep{Pollack1996, Lee2015cool}. An external heating source can slow the planet's cooling rate such that the accreting gas fills the planet's hill sphere, cutting off accretion. This could feasibly occur due to intense radiation fields in clusters, proximity to the host star, or environments that prevent the disk from cooling and thus limit the radiative cooling of the planet. {With alterations to planet mass and radius within $1\sigma$ uncertainty, as many as six planets meet the criteria for this classification with a terrestrial core composition.}

\textit{Quadrant III}. Planets that did not accrete a significant envelope ($\fenv < 0.5$) despite having a large core over $10\,\Mearth$, which would be large enough to begin runaway gas accretion, are consistent with the expectations of core accretion theory and could be explained by a variety of scenarios that limit the amount of gas available to the accreting planet. Such scenarios could include formation in a gas-poor environment, perhaps due to photoevaporation of the protoplanetary disk by nearby stars \citep{Johnstone1998, Adams2004}, or planet formation that occurs immediately before standard disk dissipation \citep{Adams2021}. There are {eight} planets in our sample that fall into this category when assuming a terrestrial core composition, {becoming as few as four when allowing planet mass and radius to vary by reported $1\sigma$ uncertainty.}

\textit{Quadrant IV.} {Planets with low core masses $\Mcore < 10\,M_{\oplus}$ and high envelope mass fractions of $\fenv > 0.5$ violate the expectations of core accretion theory, as such planets are too small to have accreted their large envelopes normally \citep{Pollack1996}. This suggests non-standard explanations are required for their observed masses and radii. }
With our fiducial mixed core composition, eight planets fall into this category. {These planets are HIP 41378 f, Kepler-30 d, Kepler-51 d, Kepler-177 c, TOI-1420 b, V1298 Tau b, V1298 Tau e, and WASP-107 b. With the exception of Kepler-30 d and Kepler-51 d, these classifications do not vary when considering $(R_p-\sigma_R^-, \ M_p-\sigma_M^-)$ or $R_p-\sigma_R^-, \ M_p+\sigma_M^+$. Kepler-30 d is the only planet with a quadrant IV terrestrial core type solution which features a solution corresponding to a different core type which falls in another quadrant. However, an additional planet, TOI-1173 A b, is added to this category when adopting $(R_p-\sigma_R^-, \ M_p-\sigma_M^-)$ as our input parameters. Furthermore, we note that the terrestrial solution for TOI-1836 b falls just outside this classification in the $(R_p-\sigma_R^-, \ M_p-\sigma_M^-)$ case, with $\fc\approx50.3\%,\ \Mcore\approx9.465\,\Mearth$. }

These planets cannot be reproduced {under our set of assumptions—which include a hydrostatic, two-layer structure—}with physically realistic combinations of core mass and envelope fraction, even when the most favorable atmospheric properties {are taken}. This suggests that their observed mass and radii are consequences not of the standard runaway accretion pathway for gas giants, but by some other mechanism(s). 

In the following subsections, we will provide brief discussions of possible explanations for the structures of each of these planets.

\subsubsection{HIP 41378 f}
Previous work \citep{Belkovski_2022} found that HIP 41378 f has a combination of a high envelope mass fraction and low core mass, and is thus inconsistent with the expectations of core accretion theory. 

We find that the minimum envelope mass fraction required to produce HIP 41378 f's observed mass and radius is $\sim59\%$, while the corresponding core mass for our fiducial core composition is $6.2\,\Mearth$. This means that a novel scenario is still required to explain such a high observed envelope fraction.

It has been proposed that HIP~41378~f hosts an opaque ring system whose obliquity can reproduce the observed transit radius \citep{Akinsanmi2020,Piro2020_rings}. Subsequent observational campaigns have found no evidence inconsistent with this hypothesis \citep{Alam2022}.
While the transit data to date are not yet precise enough to measure HIP 41378 f's obliquity and can provide only constraints \citep{Price2025}, \citet{Lu2025} found that there are dynamically plausible scenarios that would lead a planet like HIP 41378 f to have an oblique exo-ring system that could reproduce the observed transit shape and required planetary obliquity. Future observations with \textit{JWST} could help provide additional evidence for or against this hypothesis, but as it stands, HIP~41378~f is a compelling case in the search for exo-rings, aided by the incompatibility of its inferred core mass and envelope mass fraction with the expectations of core accretion theory.

\subsubsection{Kepler-30 d}
{Kepler-30 d is the outermost confirmed planet in the three-planet system Kepler-30 \citep{Sanchis-Ojeda_2012, Fabrycky2012}, which orbits a young, somewhat magnetically active, Sun-like host star \citep{Freitas2021}. We find that the best-fit core mass for Kepler-30 d is $\Mcore =9.57\,\Mearth$, with an envelope fraction of $\fenv = 58.59\%$. 
The system has unusual mass ratios, with the innermost planet being a typical Neptune-sized exoplanet, the middle planet being a gas giant (Jupiter-mass with an orbital period of around 60 days), and the outer planet, Kepler-30 d, having a bulk density of only 0.19 g/cm$^{3}$. This geometry places the Kepler-30 system into the relatively unpopulated `strongly inverted mass ratio' architecture class defined by \citet{Howe_Architectures}. Previous models suggest that it is possible that this system used to be part of a resonant chain \citep{Panichi2018}.}

{Though $10\,\Mearth$ is commonly used to define the crossover mass (the core mass required to begin the runaway accretion process), planets further out in the protoplanetary disk may have lower crossover masses \citep{Piso2015}. As such, Kepler-30 d's core mass of $9.57\,\Mearth$ is large enough that it may have begun to accrete large amounts of gas if located at an appropriate orbital radius, and the process could have stalled or terminated before the planet was able to reach a Saturn or Jovian-like mass, just like the planet structures in Quadrant II of Figure \ref{fig:syntheticModelResults}. When adopting the mass and radius which maximized Kepler-30 d's bulk density within $1\sigma$ uncertainty, we find a much higher core mass of $12.14\,\Mearth$ ($52.95\%$).}

\subsubsection{Kepler-51 d}
The three confirmed super-puffs in the Kepler-51 system are among the most well-studied planets in our sample. Several hypotheses have been proposed to explain the low density of these planets. 
The set of proposed hypotheses for Kepler-51 d, the coldest in the system, include high-altitude hazes \citep{Gao2020_deflating}, dusty outflows \citep{Wang2019_dustyoutflows}, and a ring system \citep{Piro2020_rings}. A JWST NIRSpec-PRISM transmission spectrum \citep{Libby-Roberts-K51d-JWST} of Kepler-51 d's atmosphere found that it likely has a low-metallicity atmosphere including a large high-altitude haze layer, but that the hazes alone cannot account for the observed radius, and an abnormally large H/He envelope is still required. The transmission spectra also allowed constraints on the properties of a ring system around Kepler-51 d which could account for the observed radius, and it was found that the lifetime of such a system would be $0.1\ \mathrm{Myr}$, requiring a recent ring-forming event, suggesting that this explanation is unlikely \citep[but not impossible;][]{Libby-Roberts-K51d-JWST}.

We find that the two innermost planets (Kepler-51 b and c) do not require a non-standard explanation {in any parameter set}, {but that terrestrial solutions for Kepler-51 d are classified as novel in the best-fit case}. The minimum atmospheric mass fraction achieved with our terrestrial core type {in the best-fit case} was $\fenv\approx53.50\%$, with a core mass of $3.53\,\Mearth$. {We note that solutions with typical atmospheric properties were found for some input parameters, particularly when using $(R_p-\sigma_R^-, \ M_p-\sigma_M^-)$, which yielded $\fenv=46.65$\% and $\Mcore=3.47\,\Mearth$. Additionally, our maximum bulk density results are consistent with \cite{Gao2020_deflating} in suggesting that} a sub $50\%$ envelope mass structure may reproduce a mass and radius within Kepler-51 d's measurement uncertainty with a core composed of primarily ice, however we leave further investigation into this possibility to future work.

\subsubsection{Kepler-177 c}
\label{sec:Kepler177}
We find that the minimum envelope mass fraction consistent with best-fit measurements for Kepler-177 c is $54.73\%$, corresponding to a core mass of $6.65\,\Mearth$. It is possible that Kepler-177 c underwent a brief phase of runaway accretion; however, Kepler-177 c's lower core mass renders this scenario less probable \citep[but potentially still possible for a planet forming at larger distances, $\sim5$ AU or beyond, assuming that accretion does not heat the planetary envelope too much;][]{Piso2015}. Therefore, Kepler-177 c may not require a non-standard explanation, just long-scale disk migration.

\cite{Piro2020_rings} investigated planetary ring systems as an explanation for several super-puffs, considering ring material, orientation, and stability. 
They determined that Kepler-177 c is an attractive candidate to be explained by rings and a promising prospect for observational confirmation.  \cite{Gao2020_deflating} investigated radius inflation due to photochemical hazes, and it was determined that a non-standard explanation is not required if the planet has an intrinsic temperature $T_{\mathrm{int}}\approx75 \unit{K}$, which they find to be reasonable for Kepler-177 c given the planet's age ($4.37_{-2.55}^{+3.63}\unit{Gyr}$). 
However, we note that both \cite{Piro2020_rings} and \cite{Gao2020_deflating} use older parameter sets which had a lower measured planetary radius \citep[$R_p = 7.1^{+3.71}_{-0.72}\,\Rearth$, which came from the TTV fit of][]{Xie2014} for Kepler-177 c, as compared to our adopted value \citep[$R_p = 8.73^{+0.36}_{-0.34} \,\Rearth$, which comes from diffuser-assisted photometry from][]{Vissapragada2020}. 
This newer, larger radius makes the existing explanations less feasible.

\subsubsection{TOI-1173 A b}
\label{sec:TOI-1173}
TOI-1173 consists of the wide binary TOI-1173 A and B and the super-puff planet TOI-1173 A b \citep{Galarza_2024_PARAMS_TOI1173}. The stellar chemical abundances of TOI-1173 A are enhanced in refractory elements, which is rare for a planet as similar in mass to the Sun as TOI-1173 A is (0.91 $M_{\odot}$). This can be taken as evidence that at least one planet on an orbit interior to TOI-1173 A b was engulfed by the host star \citep{Galarza_2024_engulfment_TOI1173}. 
The dynamics of the system could suggest that TOI-1173 A b was placed on its current short period orbit via the von Zeipel-Lidov-Kozai mechanisms and subsequent tidal circularization, which may have resulted in the inward migration and engulfment of interior lower mass planets and makes TOI-1173 A a planet with a uniquely active and recent dynamical history. 

{When using best-fit solutions from the literature for the measured planetary mass and radius, we find that the observed parameters of TOI-1173 A b can be reproduced by models with $\fenv\approx61.88\%$ and $\Mcore\approx10.79\,\Mearth$, consistent with the core accretion picture. When selecting mass and radius values that maximize $\fc$ for this planet $(R_p-\sigma_R^-, \ M_p-\sigma_M^-)$, we find a solution with $\fenv\approx59.62\%$ and $\Mcore\approx9.81\,\Mearth$, which would place this planet in Quadrant IV. As such, our classification scheme finds that TOI-1173 A b is barely consistent with the expectations of core accretion theory; however, its low density has been repeatedly noted in the literature.}

\citet{Galarza_2024_PARAMS_TOI1173} found that the radius of TOI-1173 A b cannot be explained by rings because the planet is expected to be tidally locked, and \citet{Yee_Vissapragada_2025_popcorn} found that tidal heating is also not a viable explanation because TOI-1173 A b's orbit is not measurably eccentric. Further observations could help evaluate the feasibility of atmospheric escape or high altitude hazes as inflation mechanisms.

{Similar to Kepler-30 d, TOI-1173 A b's core mass is large enough that it may have begun to rapidly accrete large amounts of gas, but was unable to sustain this process for the duration required to form a Jovian-like planet. If this is the case, the binary companion TOI-1173 B could have played a role by affecting the protoplanetary disk around TOI-1173 A.}

\subsubsection{TOI-1420 b}
TOI-1420\,b has a mass of $25.1 \pm 3.8\,\Mearth$, a radius of $11.89 \pm 0.33\,\Rearth$, and a very low mean density of $\sim 0.10\,\mathrm{g\,cm^{-3}}$ \citep{Yoshida_TOI1420b_2023}. 
{Visible in Fig. \ref{fig:syntheticModelResults} as the set of overlapping points in the bottom left corner of the plot,} we find that all hydrostatic solutions for TOI-1420 b possess extremely large atmospheric mass fractions, with the fiducial core solution (given in Table \ref{tab:coldindex}) having $\fenv=93.31\%,\ \Mcore=1.68\,\Mearth$. Even alternate core compositions do not result in a remotely feasible composition according to the expectations of core accretion theory. 
There are no known mechanisms by which a core of this size can accumulate and retain such an envelope, and thus a novel explanation for this planet is required. 
A follow-up observation determined TOI-1420 b likely has an outflowing atmosphere which could account for some of the observed inflation \citep{Vissapragada_2024_TOI1420B}, but further observations and analysis are necessary to determine the significance of this effect. This planet exhibits the largest physical inconsistencies and is the least well explained by literature hypotheses within our sample.

\subsubsection{V1298 Tau b \& e}
V1298 Tau is a very young system \citep[$23\pm4\ \mathrm{Myr}$,][]{David_2019_V1298TauDiscovery} that hosts four known planets, three of which are low density {\citep{Karalis_2025, Livingston2026}}. 
V1298~Tau~b has a mass of $\sim13.1\,M_{\oplus}$ and a radius of $10\,R_{\oplus}$, with an inferred core mass of $\sim4.5\,M_{\oplus}$ and an inferred envelope mass fraction of $65.65\%$; V1298~Tau~d has a mass of $6\,\Mearth$ and a radius of $6.3\,R_{\oplus}$, corresponding to a core mass of $\sim4.53\,M_{\oplus}$ and an envelope fraction of $24.49\%$; and V1298~Tau~e has a mass of $15.3\,M_{\oplus}$ and a radius of $9.5\,R_{\oplus}$, with a core mass of $\sim 5.62\,M_{\oplus}$ and an envelope fraction of $63.29\%$. Of these three planets, two of them (V1298 Tau b \& e) have core masses and envelope fractions inconsistent with the expectations of core accretion theory for mature systems. However, the young inferred ages of the planets likely explain their low observed densities; due to the age of the system, V1298 Tau b and e are likely in the early stages of evolution, and still retain much of their initial entropy from formation.

Computing an additional set of models with $\entropy>7.0\kbb$ for V1298 Tau b \& e, we find that when allowing up to $\entropy=7.5\kbb$, both planets can be reproduced with reasonable core and atmospheric masses. Meanwhile, planet formation models frequently assume envelope entropy significantly higher than $\entropy=7.5\kbb$. Cold start planet formation models \citep[applicable for planets forming via core accretion but where the heat from accretion shocks has radiated away;][]{Mordasini2012_part1} typically begin with envelope entropy around 7-9 $\kbb$, while hot start models \citep[applicable for accreting giant planets formed via disk instability or for planets that do not cool effectively;][]{Marley2007} start closer to 11 $\kbb$.
We conclude that the additional entropy required to produce these solutions with $\fc>0.5$ is consistent with the age of the system, and no further explanation is required. 

\subsubsection{WASP-107 b}
We find that WASP-107 b's observed mass and radius suggest an anomalous envelope mass fraction of $\fenv=79.20\%$, with $\Mcore=7.86\,\Mearth$, consistent with previous work that has identified WASP-107 b as requiring a non-standard explanation \citep{Piaulet2021}. 

The dynamics of the WASP-107 system have been the subject of multiple works that propose tidal heating as the source of WASP-107 b's atmospheric inflation \citep{Sethi_Millholland_2025_WASP107b,Millholland_Petigura_Batygin_2020_wASP107b}. However, recent studies have cast doubt on the tidal hypothesis, noting that the tidal quality factor required for sufficient inflation is improbable, that timescales for circularization are short \citep{Batygin_2025_WASP107b}, and that the eccentricity of WASP-107 b's orbit is not sufficient to tidally inflate the planet significantly \citep{Yee_Vissapragada_2025_popcorn}. 

Motivated by JWST observations suggesting a high atmospheric metallicity \citep{Sing2024, Welbanks2024}, \cite{Batygin_2025_WASP107b} proposed that it is not tidal interactions, but \textit{electromagnetic} star-planet interactions (Ohmic dissipation) that heat the planet's atmosphere, thus driving inflation. 

While Ohmic dissipation is generally expected to be most effective in hot Jupiters, WASP-107 b offers a promising Neptune-sized case in which this mechanism may also operate.

\section{Thermal Evolution Modeling}
\label{sec:ThermalModels}
In the previous section, we solved the equation of HSE for the present-day measurements 

(given in Table \ref{tab:coldindex}) for each planet in our sample, using favorable draws from the observational posteriors, to test whether each known super-puff is consistent with the expectations of core accretion theory. Some planets were inconsistent with the theory, requiring additional hypotheses. 
In this section, we will consider two of these hypotheses that have not been generally considered in the literature to explain super-puff densities: radiogenic heating and recent impact events. These models involve thermal evolution of planet states rather than static HSE models, so we utilize the mode of \planetsolver\ that allows for the self-consistent evolution of planetary structure over time, including both mass loss (via photo-evaporation, stellar wind ablation, and core-powered mechanisms) and standard cooling as the planet ages.

Unlike in the previous section, where we used the measured planet parameters, we here use a control and examine whether these two alternative hypotheses could inflate the sub-Neptune to super-puff densities. 
Control planet models were defined with characteristics intended to be representative of the sub-Neptunes that comprise a large part of our sample. We choose the initial conditions, defined as the state immediately after the planet has finished the initial accretion phase, such that the control model retains the properties of a typical sub-Neptune throughout its lifetime. Our control models are constructed based on the initial conditions and method of \cite{Howe_Burrows_Evolution}. We set the planet's total initial mass to be $8\,\Mearth$, orbiting a sun-like star at a distance of $0.3$AU. We choose an atmospheric metallicity $Z=30\times \unit{Solar}$, though we note that varying this figure has a relatively insignificant impact on the model planet's structure \citep{Howe_Burrows_Evolution}. We use our fiducial core, composed of $32.5\%$ Iron and $67.5\%$ Perovskite, and compute models for $10\%$, $20\%$ and $30\%$ envelope mass. We fit for the starting entropy that results in an initial radius of $10\,\Rearth$, which we adopt as the radius of the planet immediately following accretion. All models were evolved for at least $10\unit{Gyr}$.  When unperturbed, each control model cools and contracts according to \cite{Lopez_Fortney_Miller_2012}, with the $10\%$, $20\%$, $30\%$ $\fenv$ models respectively contracting to $3.44, 3.72,$ and $4.11\,\Rearth$ by $2\ \unit{Gyr}$.

These control models represent the standard expected evolution of a sub-Neptune planet, and do not result in super-puff densities for mature planets. 

Since our control planet model is defined by its initial parameters, envelope mass loss is relatively modest, as expected for standard sub-Neptunes. In nearly all cases, atmospheric loss amounts to less than $2.5\%$ by an age of $4\ \unit{Gyr}$, and does not have a significant effect on the planet's total mass or radius. 
In each of the following sections, we construct additional models that test our two novel hypotheses (radiogenic heating and recent impact events), and compare these to the control models to determine if either hypothesis could take a normal sub-Neptune and inflate it to become a super-puff.

\subsection{Radiogenic heating}\label{sec:Radiogenics}
\subsubsection{Radiogenic Model description}\label{sec:RadiogenicModel}
Internal radioactive decay accounts for a significant fraction of a planet's internal heat budget, the primary contributing isotopes throughout a planet's evolution being $^{40}$K, $^{232}$Th, $^{235}$U, and $^{238}$U \citep{Barnes2019}. To simulate the effects of this mechanism, the majority of planetary evolution models use estimations of Earth's radionuclide abundances, which may not be representative of planets forming in different environments or circumstances \citep{Fatuzzo2015, Nimmo2020, Tanglin2025}. 

We aim to determine whether increased heating of a planet's interior due to higher radionuclide abundances can inflate a planet's atmosphere sufficiently to become a super-puff. To this end, we modify the evolution model described by \cite{Howe_Burrows_Evolution} to allow varying radionuclide abundances. We model the evolution of planetary interior structures from formation to $10\ \unit{Gyr}$ considering radionuclide abundances up to $100$ times that of Earth, in order to produce an order-of-magnitude estimation of the effects on atmospheric entropy and radius.

The planet’s thermal evolution is described by equating its energy budget to the change in entropy $\entropy$ as 
\begin{equation}
    \int_{\Mcore}^{M_p}dm\frac{Td\entropy}{dt}=-L_{\mathrm{int}}+L_{\mathrm{radio}}-c_V\Mcore\frac{dT_{\mathrm{core}}}{dt}
\end{equation}

The core radiogenic heating contribution for element $i$, with heating power $\varepsilon_i$ and mean lifetime $\tau_i$, is defined as:
\begin{equation}
    L_{\mathrm{radio},i} = \Mcore\times \varepsilon_i\exp{(-t/\tau_i)}
\end{equation}
To simulate the effects of increased presence of radioactive isotopes, a constant multiplier $\alpha$ is applied uniformly to the term corresponding to each nuclide:
\begin{equation}
    L_{\mathrm{radio}} = \alpha\times\sum_{i}{L_{\mathrm{radio,i}}}\ ,
\end{equation}

which sufficiently approximates the evolution of a planet with a radionuclide abundance of $\alpha\times A_{\oplus}$ for the purpose of this experiment.
\subsubsection{Radiogenic model results}\label{sec:RadiogenicResults}
\begin{figure*}
    \centering
    \includegraphics[width=1\linewidth]{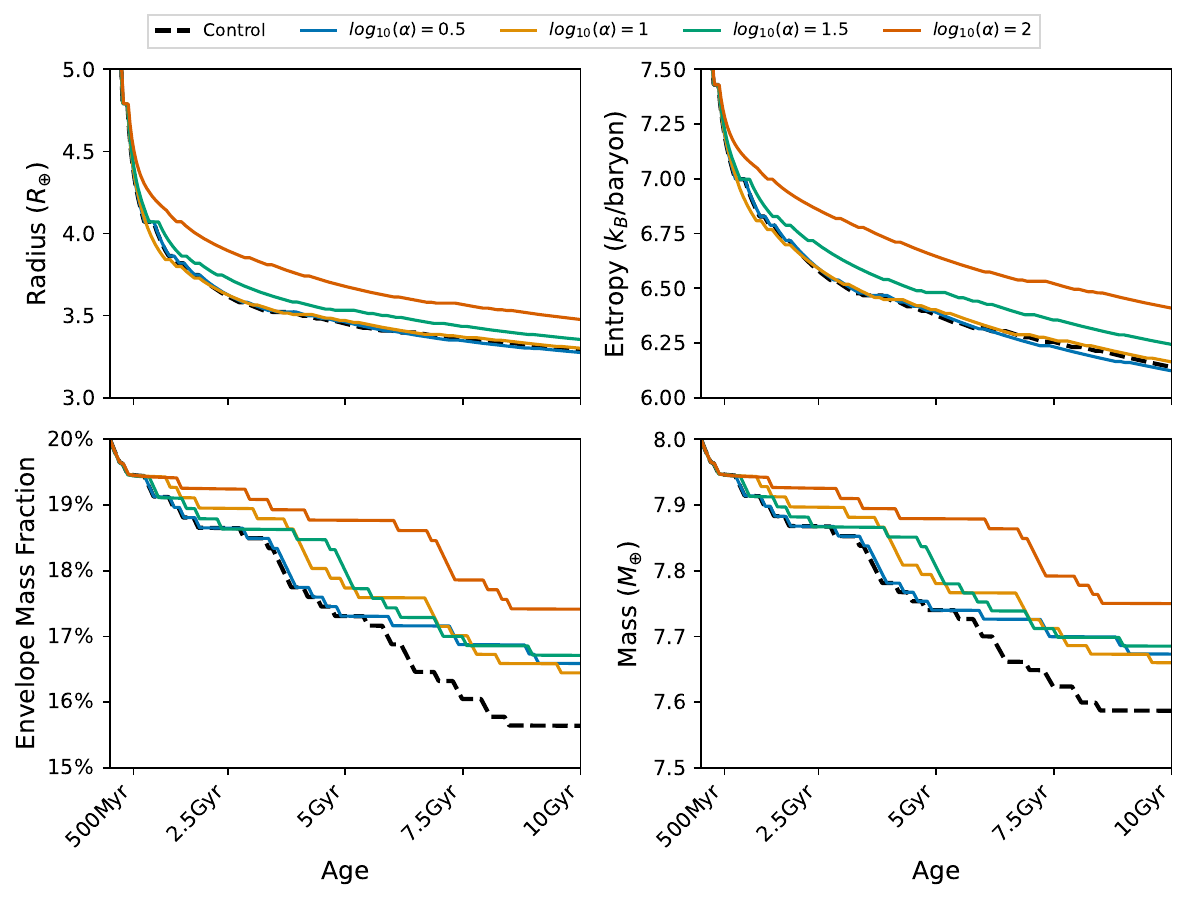}
    \caption{Examples of the evolution of Radius (top left), Entropy (top right), Envelope Mass Fraction $\fenv$ (bottom left), and Mass (bottom right) for $20\%$ envelope mass planet models with enhanced radionuclide abundances $A=\alpha \times A_{\oplus}$. Evolution tracks corresponding to $\log_{10}\alpha=0.5,\ 1.0,\ 1.5,$ and $2$ are shown. We derive the effects of increased radiogenic heating by comparing each track to the $20\%$ $\fenv$ control planet (dashed lines above). Note that the uneven steps are a consequence of the code's use of entropy, rather than time, as the iterating variable.}
    \label{fig:radiogenictracks}
\end{figure*}

\begin{figure}
    \centering
    \includegraphics[width=1\linewidth]{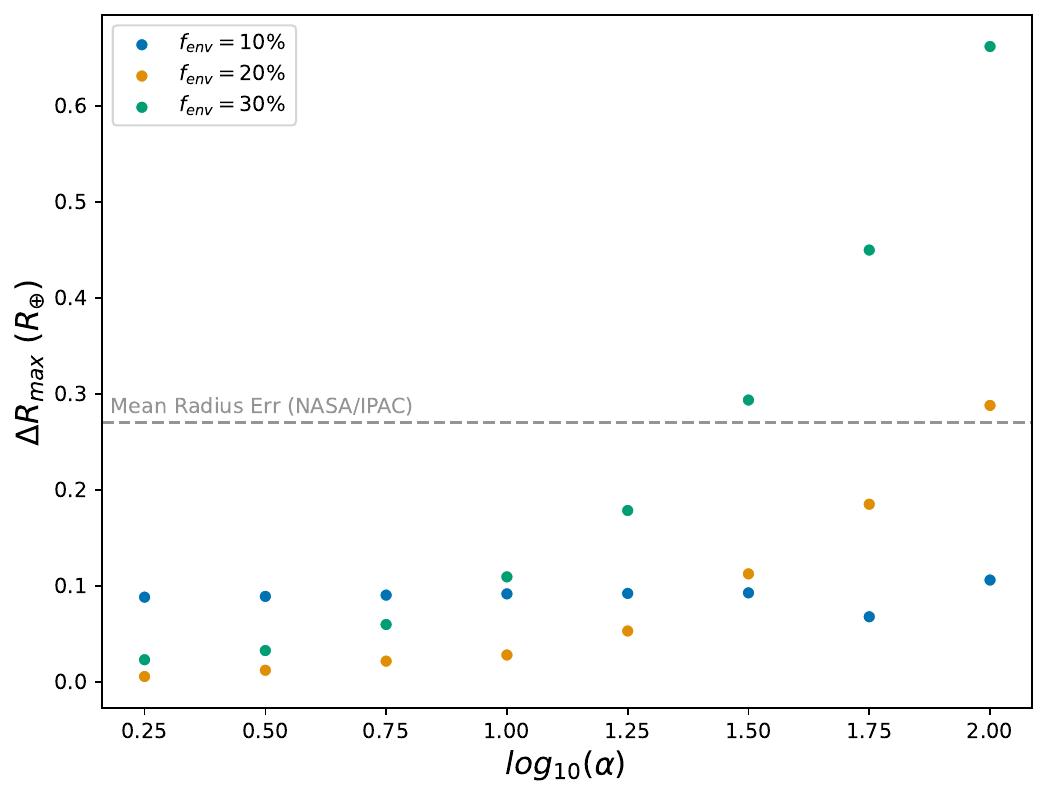}
    \caption{The maximum radius inflation, $\Delta R = R(t)-R_{\mathrm{control}}(t)$, achieved with $10\%,\ 20\%,\ 30\%$ envelope mass fraction models for a range of enhanced radiogenic abundances, where model abundance $A=\alpha \times A_{\oplus}$ for each of $^{40}$K, $^{232}$Th, $^{235}$U, and $^{238}$U. For context, the $\Delta R$ needed to reach $0.3\ \mathrm{g/cm^3}$ and become a super-puff is $>1\,\Rearth$ for all models. These results suggest that radiogenic heating is not sufficient to produce planets with super-puff densities. }
    \label{fig:radiogenicresults}
\end{figure}

Evolutionary tracks are calculated by decrementing atmospheric entropy at intervals determined by the evolving planet's properties, subject to step size controls, and thus timestamps are not consistent across simulations. To enable comparison between models, we interpolate between time steps to define planet properties at $10^4$ time steps between formation and $10\ \unit{Gyr}$. Examples of $R_p$, $K$, $\fenv$, and $M_p$ evolution tracks for models with varied levels of radiogenic heating are shown in Fig. \ref{fig:radiogenictracks}. We quantify the effects of increased radionuclide abundances by comparing enhanced models with corresponding controls at each time step. Figure \ref{fig:radiogenicresults} shows the maximum inflation achieved by each model, calculated as the largest difference between enhanced and control radii at the same age. We find that the additional heating from radiogenic enhancement was not sufficient to explain the entropy increases noted in Section \ref{sec:ImpactModelResults}, and furthermore that differences do not become observable for abundances less than $\sim30\times A_{\oplus}$. Our models achieve observable levels of inflation in the $\fenv=30\%$ case for $A \gtrsim30\times A_{\oplus}$, and $A \gtrsim100\times A_{\oplus}$ for the $\fenv=20\%$ model. No $10\%$ envelope mass models achieved observable inflation. Based on these results, we conclude that radiogenic heating is unlikely to produce sufficient atmospheric inflation to achieve super-puff densities.

\subsection{Giant Impacts}\label{sec:ImpactModels}
\subsubsection{Impact model description}\label{sec:ImpactModelDesc}
To study the magnitude and duration of the effects of giant impacts on sub-Neptune atmospheres, we again utilize a modified version of the thermal evolution model of \cite{Howe_Burrows_Evolution}. Impact events were modeled based on \cite{Biersteker_Schlichting_2019}, with $v_{\mathrm{imp}}\sim v_{\mathrm{esc}}$, and the impact energy given by:
\[E_{\mathrm{imp}}=\eta M_{\mathrm{imp}}\frac{GM_p}{R_{c}}\]
Where $R_{c}$ is the radius of the core, $M_p$ is the planet's mass, $M_{\mathrm{imp}}$ is the mass of the impacting object, and $\eta=0.5$ is the fraction of the impact energy available for heating. Considering the atmosphere to be isothermal with $T=T_{\mathrm{eq}}$, the change in atmospheric entropy is related to impact energy as $\Delta \entropy = E_{\mathrm{imp}}/{T_{\mathrm{eq}}}$. Impacts are simulated as an instantaneous injection of entropy into the control planet’s atmosphere at a given time $t_{\mathrm{imp}}$, using the assumption that the impact's energy is transferred directly to the planet's atmosphere. 

\begin{figure*}
    \centering
    \includegraphics[width=1\linewidth]{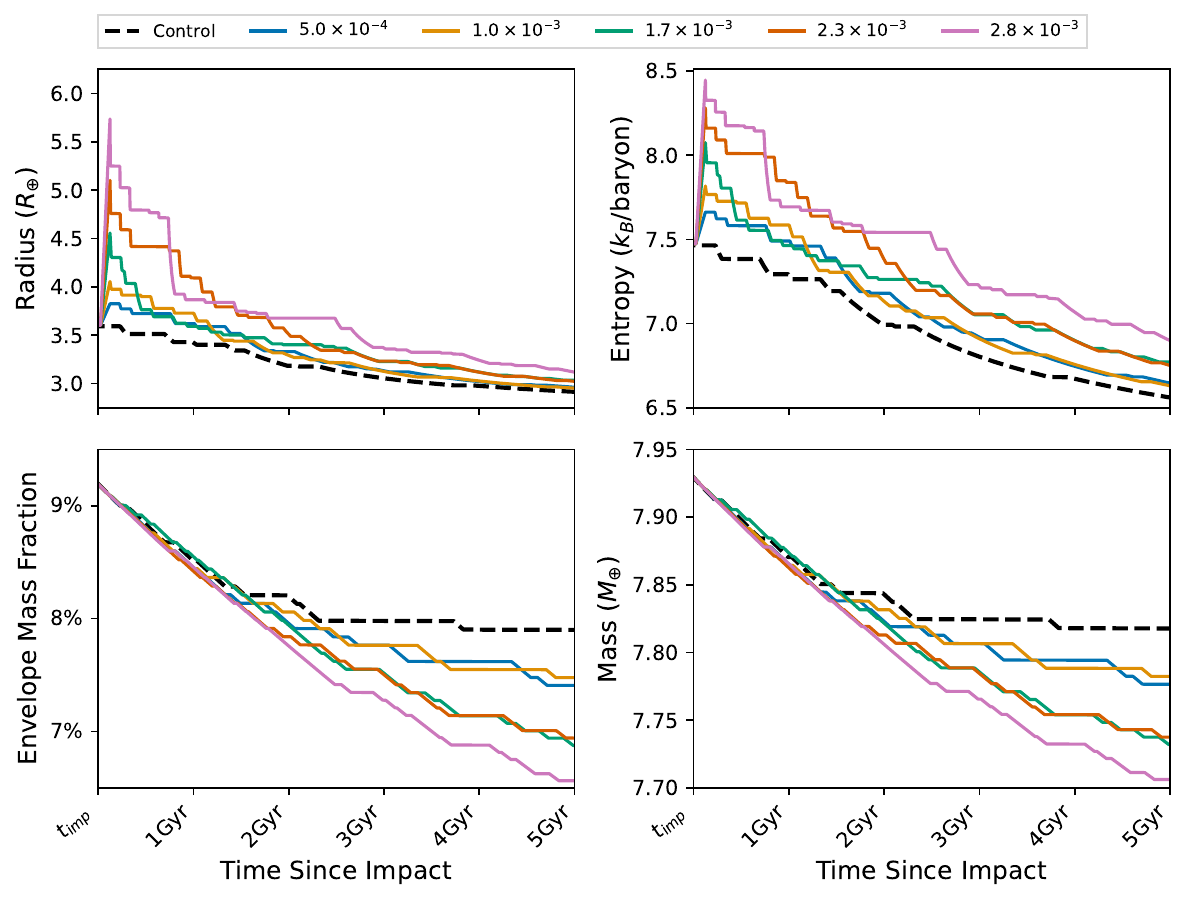}
    \caption{Examples of the evolution of Radius (top left), Entropy (top right), Envelope Mass Fraction $\fenv$ (bottom left), and Mass (bottom right) after an impact event for our $10\%$ initial envelope mass model. Impactor masses of $M_{\mathrm{imp}}/\Mcore=5\times10^{-4},$ and $1.0,\ 1.7,\ 2.3,\ 2.8\ \times 10^{-3}$ are shown. Note that the uneven steps are a consequence of the code's use of entropy, rather than time, as the iterating variable.}
    \label{fig:imapacttracks}
\end{figure*}

 The planet models evolve and cool normally after impact, and parameters were recorded over time for comparison with the corresponding control model. Our complete parameter sweep includes, with corresponding control models, $\fenv\,\mathord{=}\,[0.1,\ 0.2,\ 0.3]$, $t_{\mathrm{imp}}\,\mathord{=}\,[250\unit{Myr},\ 1\unit{Gyr},\ 3\unit{Gyr}]$, and $M_{\mathrm{imp}}/\Mcore \in [10^{-4}, 10^{-2}]$. We consider only impact energies well below the threshold for catastrophic mass loss due to shocks or boil-off \citep{Biersteker_Schlichting_2019}, which corresponds to the range $M_{\mathrm{imp}}/\Mcore \lesssim 0.01$. Additionally, {the thermal evolution prescriptions used in this work are not intended} for planets with $R_p\gtrsim10\,\Rearth$ following the initial contraction phase, and numerical integration becomes unreliable when impacts cause models to significantly exceed this limit. For this reason, simulated impacts which result in radii higher than $10\,\Rearth$ were discarded. An example plot comparing the evolution of radius, atmospheric entropy, and density of control and impact over time is shown in Figure \ref{fig:imapacttracks}. Note that we additionally considered impacts occurring at later times between $1$ and $3\unit{Gyr}$, and observed no significant difference in the magnitude or longevity of the effects.

\subsubsection{Impact model results}\label{sec:ImpactModelResults}

\begin{figure*}
    \centering
    \includegraphics[width=1\linewidth]{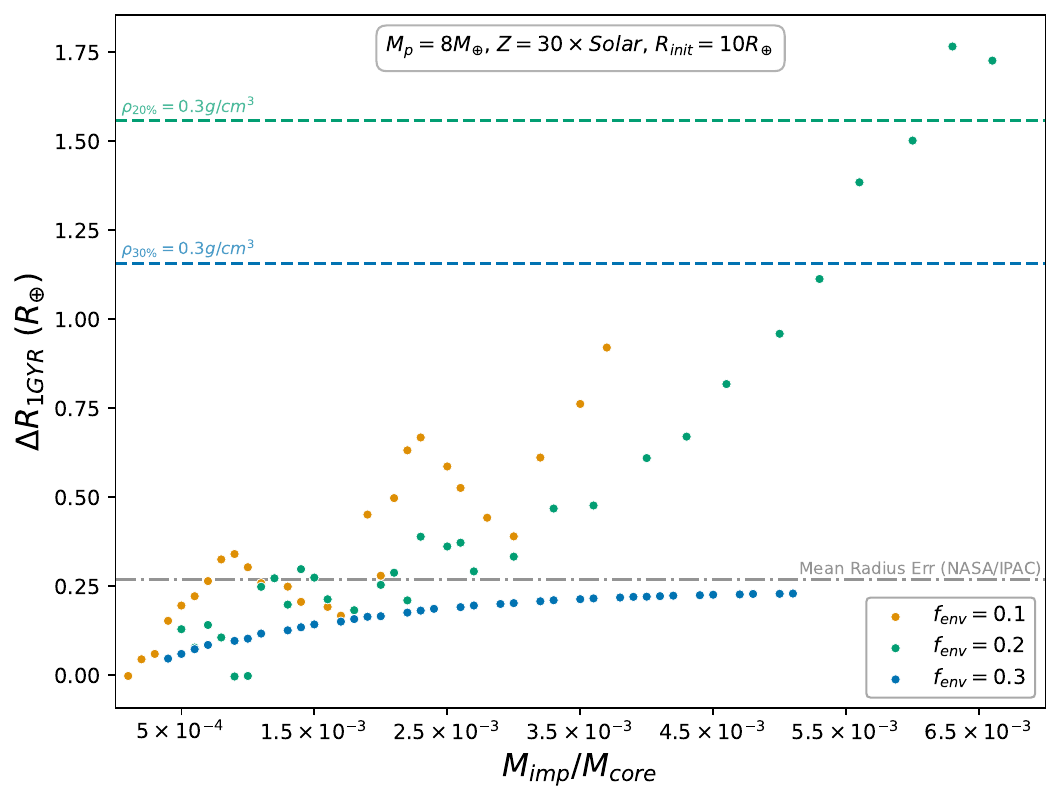}
    \caption{The maximum radius inflation, $\Delta R = R(t)-R_{\mathrm{control}}(t)$, achieved with the $10\%,\ 20\%,\ 30\%$ envelope mass fraction models described above with respect to the ratio of impact mass to core mass. The two dashed lines represent the approximate radius inflation required to lower the planet's bulk density below $0.3g/cm^3$ for $20\%$ (top) and $30\%$ (middle) envelope mass models. The dot-dashed line denotes the point at which inflation exceeds the mean upper measurement error listed on NASA/IPAC Exoplanet Archive for planetary radii.} 
    \label{fig:imapactresults}
\end{figure*}

The difference in radii between impacted and control models after allowing the planet to cool for $1\ \unit{Gyr}$ post-impact is shown in Figure \ref{fig:imapactresults}. Our results suggest that giant impact events can drive significant and observable atmospheric inflation in sub-Neptunes similar to our control models. Specifically, we find that impact events with impactor to core mass ratios $M_{\mathrm{imp}}/\Mcore < 3\times10^{-3} $ can lead to observably inflated radii persisting on timescales $\tau_{\mathrm{obs}} \gtrsim 1\ \unit{Gyr}$ for both the $20\%$ and $30\%$ models. Furthermore, the resulting inflation in our $20\%$ envelope mass model achieves super-puff densities ($\lesssim0.3\ \unit{g/cm^3}$) for $M_{\mathrm{imp}}/\Mcore \gtrsim6\times10^{-3}$. 

We observe that inflation appears in most cases to increase with impact energy, up to the point where our methods are no longer applicable. We consider only small impactors below the threshold to cause substantial mass loss or alterations to the planet's core, and it is probable that inflation continues to increase with larger impactor masses up to the point that mass loss becomes dominant. Even neglecting this conjecture, our results indicate that the additional heat added to a planet's atmosphere is capable of producing super-puffs that remain as such for upwards of $1\ \unit{Gyr}$. We emphasize that these models serve only as a proof of concept, but that the results warrant further investigation into the viability of giant impacts as a super-puff formation channel.

\section{Discussion}\label{sec:Discussion}

In this paper, we have completed two complementary lines of analysis. In the first (Section \ref{sec:InteriorStructureModeling}), we computed hydrostatic solutions for the population of known super-puffs to assess whether super-puff interior structures are consistent with being built via core accretion. In the second (Section \ref{sec:ThermalModels}), we assessed in general terms two possible explanations for super-puff densities that involve the thermal evolution of hydrostatic models: recent impacts from smaller bodies and radiogenic heating. 

{Our interior model analysis (Section \ref{sec:InteriorStructureModeling}) aims to place constraints on the core and atmospheric mass fraction for each super-puff}. Core mass values that are too small (less than roughly $10\,\Mearth$) with envelope fractions that are too large (over $50\%$) show tension with the expectations of core accretion theory, suggesting that alternative explanations for their observed masses and radii are needed. 
In our models and results, we consider a range of core compositions including pure iron and pure ice, which provide {theoretical upper and lower limits on core density}. As neither of these core types are likely to occur in nature, our results presented in Table \ref{tab:coldindex} used a fiducial core composition ($67.5\%$ perovskite and $32.5\%$ iron), which corresponds to an intermediate density. 
Future work constraining the core composition of sub-Saturn planets will test the validity of our assumption, and may warrant a more complex or multi-layer interior model.

Our thermal evolution analysis (Section \ref{sec:ThermalModels}) examines how planetary densities may change over time over two types of thermal evolution: impacts, which impart energy to a planet's envelope at one time, and the energy is then radiated away; and radiogenic heating, in which enhanced levels of radioactive elements contribute an increased internal heat flux throughout the lifetime of the planet. We find that while recent impacts can explain increased planetary radii, radiogenic heating must be at unphysically high levels to cause any appreciable decrease in planetary density. Future work aiming to explain super-puff planet densities should consider the effect of impacts, but can likely disregard the effect of radiogenic heating. 

We additionally stress that our thermal evolution models are only intended to determine whether giant impacts or high radioactivity are capable of inflating typical sub-Neptunes to super-puff densities. These models do not constitute a robust investigation of these mechanisms, nor do they test whether these mechanisms can explain the particular planets in the cold super-puff sample given in Table \ref{tab:coldindex}. Rather, they provide only an order-of-magnitude estimation of atmospheric inflation.

\subsection{Robustness of the Quadrant Classification}
\label{sec:robustness}
Our quadrant-based classification (Figure~\ref{fig:syntheticModelResults}; Table~\ref{tab:coldindex}) is intended to be conservative: we aim to identify planets that are inconsistent with core accretion expectations based on hydrostatic interior structure models, rather than to precisely infer each planet’s true core and envelope properties. 

As a result, a planet being sorted into the quadrants that are consistent with core accretion theory (I, II, III) does not necessarily mean that those planets are truly consistent with core accretion, just that there is a consistent solution within the present observational errors. 

There are several sources of uncertainty that can shift a planet’s inferred $(\Mcore, f_{\mathrm{env}})$ and therefore its quadrant assignment: its measured mass and radius, its true core composition, the thermodynamic state and composition of its envelope, and the boundary conditions used in the structure calculation.

Fundamentally, these first two parameters are the primary drivers of planet classification: a super-puff can only be identified if both its mass and radius are measured, and its classification depends entirely on these quantities. Consequently, observational uncertainties and systematic errors can lead to misclassifications, either causing true super-puffs to be missed or resulting in planets that are not intrinsically low-density being classified as super-puffs.

This is an issue that certainly plagues the super-puff sample. Many systems in our sample exhibit significantly different mass and radius measurements depending on the observational technique or data set used. For example, in this work, we used the most modern solution from \citet{Petigura_2018_K2-24PARAMS} for K2-24 c, which reports a planetary radius of $R_p = 7.50^{+0.30}_{-0.30}\,\Rearth$ and a mass of $M_p = 15.40^{+1.90}_{-1.80}\,\Mearth$. 
{This solution yields an interior structure in Quadrant I, consistent with formation via core accretion. Other solutions previous to the 2018 solution gave significantly different physical parameters, ranging from a radius of $5.59\,\Rearth$ reported as a candidate solution by \citet{Vanderburg2016a} to a radius of $9.8\,\Rearth \pm1.2\,\Rearth$ reported by \citet{Crossfield2016}. For this planet, the former solution yields a standard interior structure consistent with core accretion (Quadrant II), while the latter would correspond to a super-puff inconsistent with core accretion (Quadrant IV). }
Similarly, as discussed in Section \ref{sec:Kepler177}, Kepler-177 c, one of the Quadrant IV planets apparently inconsistent with core accretion, has multiple literature solutions with different radii. Population-wide systematic errors \citep{Han2025} may also systematically distort the inferred $(\Mcore, f_{\mathrm{env}})$ distribution across the sample.

Of the other physical parameters that can visibly affect quadrant placement (core composition and envelope entropy), we do not know the true core compositions, and as shown in Figure \ref{fig:syntheticModelResults}, they can make a relatively modest difference in the inferred core mass and envelope fraction. Entropy has a much stronger effect on the inferred interior structure, and the solutions shown in Figure~\ref{fig:syntheticModelResults} adopt the most favorable entropy values that are still consistent with the expected range for mature planets.
{Finally, the relationship between the observed planet parameters (mass and radius) and our inferred interior properties (core mass and envelope fraction) can be nonlinear. For most of our Quadrant IV planets, the $(R_p-\sigma_R^-, \ M_p-\sigma_M^-)$ parameter combination maximizes $\fc$, likely because of lower planetary surface gravity which allows for a higher atmospheric scale height for a given envelope mass. Similarly, the $(R_p-\sigma_R^-, \ M_p+\sigma_M^+)$ parameter combination maximizes both bulk density and total core mass. However, TOI-1420 b, our planet most inconsistent with core accretion, exhibits a more massive core in the $(R_p-\sigma_R^-, \ M_p-\sigma_M^-)$ case, unlike every other Quadrant IV planet that shows less massive cores in that case. }

\subsection{The Likelihood and Frequency of Impact Events}
In Section~\ref{sec:ImpactModels}, we assess how a recent impact from a low-mass object can deposit heat and entropy into a planet’s envelope, thereby increasing its apparent radius. As shown in Figure \ref{fig:imapacttracks}, such an event produces a transient and time-limited episode of radius inflation, during which a planet may temporarily exhibit super-puff–like radii and bulk densities.

While this mechanism likely contributes to the observed radius distribution of planets overall \citep{Chance2022}, a clear limitation of this explanation for super-puff densities is the expected frequency and likelihood of such impact events as a function of stellar age. 
{For a sub-Neptune to be inflated to a super-puff density, the impact event must be both recent (likely within no more than 1 Gyr, since the inflation lasts for 50 Myr - 1 Gyr in our models) and from a sufficiently massive impactor (Figure \ref{fig:imapactresults}). If we assume $\sim 34/6000$ planets have known enhanced densities that could be explained by recent impact events, this requires an impact rate of at least $\approx10^{-11}$/yr per planet. While this rate is modest, sustaining it over stellar ages of several Gyr (typical ages in our sample) requires the existence of long-lived reservoirs of Moon-to-Mars-sized objects. However, impact events are most common in young planetary systems}, when substantial debris remains from the planet formation process \citep{Raymond2009}, likely within the first $\sim10$--$100\ \mathrm{Myr}$ of a system’s lifetime. As planetary systems age, the reservoir of material capable of impacting Neptune-sized planets is progressively depleted, and giant impacts become significantly less common, especially after 40 Myr of age \citet{Quintana2016}. 
Because the majority of known super-puff planets orbit relatively old stars, impact-driven inflation cannot be a ubiquitous explanation for super-puff densities, although it may play an important role for a subset of systems.

\subsection{Caveats and Future Work}
In this work, we consider which super-puffs are the most inconsistent with the core accretion paradigm and identify several targets, enumerated in Section \ref{sec:InteriorStructureModeling-results}. As illustrated in that section, our interior structure models have a strong dependence on the planet's atmospheric entropy. 
Because atmospheric entropy is not an easily observable quantity and such measurements have not been made for the planets in our sample, our results are dependent on the validity of our assumptions regarding each planet's entropy. The entropies considered in our structure models span the expected range for planets older than $\sim1\ \unit{Gyr}$ \citep{Lopez_Fortney_2014_R-M-Relation}, which reflects the majority of our sample. Young planets, on the other hand, may retain a significant amount of initial heat and their radii will be proportionally larger for the same core mass and envelope fraction. For example, the planets in the V1298 Tau system appear inconsistent with the core accretion paradigm when modeled with entropy values typical of planets older than $1\ \unit{Gyr}$, but are well described by core accretion when allowing entropy values expected for the system's young age \citep[$23\pm4\ \mathrm{Myr}$,][]{David_2019_V1298TauDiscovery}. 

The methods used here to model the interior structures and evolution of planets provide a means of studying planetary interiors while minimizing the assumptions made, as a large range of possible core composition and envelope entropy values were considered. 
However, this approach comes with some limitations. Our interior structure models make use of a simplified model which considers only a convective atmosphere, and neglects regions where heat is transferred by non-convective processes. This approach is common; however, recent work suggests that the inclusion of non-convective atmospheric components may alter the expected radii of Neptunes and sub-Neptunes \citep{Eberlein_Helled_2025, Tang_2025_structure, TangFortneyMurrayClay2025_massloss}. 

Furthermore, the method of \cite{Howe_Burrows_Evolution} used here to compute planetary evolution is intended to model sub-Neptune/Neptune-sized planets, and becomes unreliable when radii exceed $10\,\Rearth$. This limitation prevents consideration of inflation events or mechanisms which cause planets to exceed this size. To consider more energetic impactors than were considered in our work in Section \ref{sec:ThermalModels} would require an alternative model or improvements on the current \planetsolver\ model.

Many planets in the super-puff population have masses derived from TTVs, which preferentially yield precise mass constraints in near-resonant systems. This introduces two related issues. First, TTV-inferred masses can differ between analyses depending on dynamical assumptions (e.g., treatment of eccentricities, priors, and the number of interacting planets modeled). Second, the population of ``well-measured'' super-puffs that have mass measurements made only with TTVs (which will include much of the Kepler sample) will itself be shaped by detectability, {potentially biasing the observed planet population towards or against particular system dynamical histories}. As a result, for systems where multiple independent mass determinations exist that are not consistent (e.g., different TTV solutions, or TTV vs.\ RV constraints), quadrant assignments should be considered provisional until the mass scale is reconciled.

To be securely classified as inconsistent with core accretion theory (which can be taken as evidence that some interesting alternative hypothesis is at play), a planet must be robustly and unambiguously inconsistent across measurements. In our sample, a good example of this is TOI-1420 b, which is not even close to consistent with the core accretion picture in any published measurement, while something more like Kepler-177 c that resides close to the boundary and has multiple solutions with significantly different measured radii values could be misclassified based on measurement errors.

%
%
%
%
%
%
%
%
%
%
\section{Summary \& Conclusions}\label{sec:Summary}

This work examines the likely interior structures of super-puff planets using two complementary approaches. First, we modeled the assumed current hydrostatic solutions for {super-puffs} in the observational sample to assess which planets are consistent with the expectations of core accretion theory and which are not, and therefore require nonstandard explanations for their interior structures. Next, we explored two potential novel explanations for super-puffs: impact-driven inflation and radiogenic heating.

Our results show that the majority of known {super-puffs} do not require novel formation mechanisms, as their {inferred physical properties} are consistent with the expectations of core accretion theory. {Of the 34 known super-puffs in our sample, we identify six planets (HIP 41378 f, Kepler-30 d, Kepler-51 d, Kepler-177 c, TOI-1420 b, and WASP-107 b) as inconsistent with these expectations. The inferred properties of V1298 Tau b and V1298 Tau e resemble those of the above six novel planets, but this can most likely be attributed to the age of the system.} These planets present interesting opportunities for further observational follow-up and theoretical modeling efforts, and such efforts may eventually confirm a novel explanation such as circumplanetary rings. {An additional planet, TOI-1173~A~b, has solutions inconsistent with core accretion theory when non-best-fit combinations of radius and mass are assumed.}

Of the two additional explanations we considered (impact-driven inflation and radiogenic heating), impact-driven inflation could explain some small fraction of the super-puffs, while radiogenic heating is unlikely to be sufficient to drive observed super-puff densities. 
Our results overall support the idea that super-puff planets are a heterogeneous population built by a diversity of physical mechanisms.

\begin{acknowledgments}
We thank Fred Adams for substantially useful suggestions that shaped the direction of the manuscript. 
    We thank Dan Fabrycky, Zifan Lin, and Adam Distler for useful conversations. 

This research has made use of the NASA Exoplanet Archive, which is operated by the California Institute of Technology, under contract with the National Aeronautics and Space Administration under the Exoplanet Exploration Program. This research has made use of the Astrophysics Data System, funded by NASA under Cooperative Agreement 80NSSC21M0056.
\end{acknowledgments}

\facilities{Exoplanet Archive}

\software{Astropy \citep{Astropy_v5}, 
astroquery \citep{astroquery},
numpy \citep{numpy},
SciPy \citep{scipy}, 
seaborn \citep{seaborn}, 
matplotlib \citep{Hunter:Matplotlib}, 
IPython \citep{PER-GRA:IPython}, 
Jupyter \citep{Kluyver:2016aa}, 
pandas \citep{mckinney-proc-scipy-2010},
\planetsolver \citep{Howe_Burrows_Evolution}
}




\bibliography{PASPsample701}{}

@ARTICLE{Guerrero2021,
       author = {{Guerrero}, Natalia M. and {Seager}, S. and {Huang}, Chelsea X. and {Vanderburg}, Andrew and {Garcia Soto}, Aylin and {Mireles}, Ismael and {Hesse}, Katharine and {Fong}, William and {Glidden}, Ana and {Shporer}, Avi and {Latham}, David W. and {Collins}, Karen A. and {Quinn}, Samuel N. and {Burt}, Jennifer and {Dragomir}, Diana and {Crossfield}, Ian and {Vanderspek}, Roland and {Fausnaugh}, Michael and {Burke}, Christopher J. and {Ricker}, George and {Daylan}, Tansu and {Essack}, Zahra and {G{\"u}nther}, Maximilian N. and {Osborn}, Hugh P. and {Pepper}, Joshua and {Rowden}, Pamela and {Sha}, Lizhou and {Villanueva}, Jr., Steven and {Yahalomi}, Daniel A. and {Yu}, Liang and {Ballard}, Sarah and {Batalha}, Natalie M. and {Berardo}, David and {Chontos}, Ashley and {Dittmann}, Jason A. and {Esquerdo}, Gilbert A. and {Mikal-Evans}, Thomas and {Jayaraman}, Rahul and {Krishnamurthy}, Akshata and {Louie}, Dana R. and {Mehrle}, Nicholas and {Niraula}, Prajwal and {Rackham}, Benjamin V. and {Rodriguez}, Joseph E. and {Rowden}, Stephen J.~L. and {Sousa-Silva}, Clara and {Watanabe}, David and {Wong}, Ian and {Zhan}, Zhuchang and {Zivanovic}, Goran and {Christiansen}, Jessie L. and {Ciardi}, David R. and {Swain}, Melanie A. and {Lund}, Michael B. and {Mullally}, Susan E. and {Fleming}, Scott W. and {Rodriguez}, David R. and {Boyd}, Patricia T. and {Quintana}, Elisa V. and {Barclay}, Thomas and {Col{\'o}n}, Knicole D. and {Rinehart}, S.~A. and {Schlieder}, Joshua E. and {Clampin}, Mark and {Jenkins}, Jon M. and {Twicken}, Joseph D. and {Caldwell}, Douglas A. and {Coughlin}, Jeffrey L. and {Henze}, Chris and {Lissauer}, Jack J. and {Morris}, Robert L. and {Rose}, Mark E. and {Smith}, Jeffrey C. and {Tenenbaum}, Peter and {Ting}, Eric B. and {Wohler}, Bill and {Bakos}, G. {\'A}. and {Bean}, Jacob L. and {Berta-Thompson}, Zachory K. and {Bieryla}, Allyson and {Bouma}, Luke G. and {Buchhave}, Lars A. and {Butler}, Nathaniel and {Charbonneau}, David and {Doty}, John P. and {Ge}, Jian and {Holman}, Matthew J. and {Howard}, Andrew W. and {Kaltenegger}, Lisa and {Kane}, Stephen R. and {Kjeldsen}, Hans and {Kreidberg}, Laura and {Lin}, Douglas N.~C. and {Minsky}, Charlotte and {Narita}, Norio and {Paegert}, Martin and {P{\'a}l}, Andr{\'a}s and {Palle}, Enric and {Sasselov}, Dimitar D. and {Spencer}, Alton and {Sozzetti}, Alessandro and {Stassun}, Keivan G. and {Torres}, Guillermo and {Udry}, Stephane and {Winn}, Joshua N.},
        title = "{The TESS Objects of Interest Catalog from the TESS Prime Mission}",
      journal = {\apjs},
         year = 2021,
        month = jun,
       volume = {254},
       number = {2},
          eid = {39},
        pages = {39},
          doi = {10.3847/1538-4365/abefe1},
archivePrefix = {arXiv},
       eprint = {2103.12538},
 primaryClass = {astro-ph.EP},
       adsurl = {https://ui.adsabs.harvard.edu/abs/2021ApJS..254...39G}
}

@ARTICLE{Borucki2010,
       author = {{Borucki}, William J. and {Koch}, David and {Basri}, Gibor and {Batalha}, Natalie and {Brown}, Timothy and {Caldwell}, Douglas and {Caldwell}, John and {Christensen-Dalsgaard}, J{\o}rgen and {Cochran}, William D. and {DeVore}, Edna and {Dunham}, Edward W. and {Dupree}, Andrea K. and {Gautier}, Thomas N. and {Geary}, John C. and {Gilliland}, Ronald and {Gould}, Alan and {Howell}, Steve B. and {Jenkins}, Jon M. and {Kondo}, Yoji and {Latham}, David W. and {Marcy}, Geoffrey W. and {Meibom}, S{\o}ren and {Kjeldsen}, Hans and {Lissauer}, Jack J. and {Monet}, David G. and {Morrison}, David and {Sasselov}, Dimitar and {Tarter}, Jill and {Boss}, Alan and {Brownlee}, Don and {Owen}, Toby and {Buzasi}, Derek and {Charbonneau}, David and {Doyle}, Laurance and {Fortney}, Jonathan and {Ford}, Eric B. and {Holman}, Matthew J. and {Seager}, Sara and {Steffen}, Jason H. and {Welsh}, William F. and {Rowe}, Jason and {Anderson}, Howard and {Buchhave}, Lars and {Ciardi}, David and {Walkowicz}, Lucianne and {Sherry}, William and {Horch}, Elliott and {Isaacson}, Howard and {Everett}, Mark E. and {Fischer}, Debra and {Torres}, Guillermo and {Johnson}, John Asher and {Endl}, Michael and {MacQueen}, Phillip and {Bryson}, Stephen T. and {Dotson}, Jessie and {Haas}, Michael and {Kolodziejczak}, Jeffrey and {Van Cleve}, Jeffrey and {Chandrasekaran}, Hema and {Twicken}, Joseph D. and {Quintana}, Elisa V. and {Clarke}, Bruce D. and {Allen}, Christopher and {Li}, Jie and {Wu}, Haley and {Tenenbaum}, Peter and {Verner}, Ekaterina and {Bruhweiler}, Frederick and {Barnes}, Jason and {Prsa}, Andrej},
        title = "{Kepler Planet-Detection Mission: Introduction and First Results}",
      journal = {Science},
         year = 2010,
        month = feb,
       volume = {327},
       number = {5968},
        pages = {977},
          doi = {10.1126/science.1185402},
       adsurl = {https://ui.adsabs.harvard.edu/abs/2010Sci...327..977B}
}

@ARTICLE{Rogers2015,
       author = {{Rogers}, Leslie A.},
        title = "{Most 1.6 Earth-radius Planets are Not Rocky}",
      journal = {\apj},
         year = 2015,
        month = mar,
       volume = {801},
       number = {1},
          eid = {41},
        pages = {41},
          doi = {10.1088/0004-637X/801/1/41},
archivePrefix = {arXiv},
       eprint = {1407.4457},
 primaryClass = {astro-ph.EP},
       adsurl = {https://ui.adsabs.harvard.edu/abs/2015ApJ...801...41R}
}

@article{Nimmo2020,
	title = {Radiogenic Heating and Its Influence on Rocky Planet Dynamos and Habitability},
	volume = {903},
	issn = {2041-8205},
	url = {https://dx.doi.org/10.3847/2041-8213/abc251},
	doi = {10.3847/2041-8213/abc251},
	pages = {L37},
	number = {2},
	journaltitle = {The Astrophysical Journal Letters},
	shortjournal = {{ApJL}},
	author = {Nimmo, Francis and Primack, Joel and Faber, S. M. and Ramirez-Ruiz, Enrico and Safarzadeh, Mohammadtaher},
	urldate = {2024-02-24},
	date = {2020-11},
        year = 2020,
	langid = {english},
	note = {Publisher: The American Astronomical Society},
}

@ARTICLE{Fatuzzo2015,
       author = {{Fatuzzo}, Marco and {Adams}, Fred C.},
        title = "{Distributions of Long-lived Radioactive Nuclei Provided by Star-forming Environments}",
      journal = {\apj},
         year = 2015,
        month = nov,
       volume = {813},
       number = {1},
          eid = {55},
        pages = {55},
          doi = {10.1088/0004-637X/813/1/55},
archivePrefix = {arXiv},
       eprint = {1510.05158},
 primaryClass = {astro-ph.EP},
       adsurl = {https://ui.adsabs.harvard.edu/abs/2015ApJ...813...55F}
}

@ARTICLE{Quintana2016,
       author = {{Quintana}, Elisa V. and {Barclay}, Thomas and {Borucki}, William J. and {Rowe}, Jason F. and {Chambers}, John E.},
        title = "{The Frequency of Giant Impacts on Earth-like Worlds}",
      journal = {\apj},
         year = 2016,
        month = apr,
       volume = {821},
       number = {2},
          eid = {126},
        pages = {126},
          doi = {10.3847/0004-637X/821/2/126},
archivePrefix = {arXiv},
       eprint = {1511.03663},
 primaryClass = {astro-ph.EP},
       adsurl = {https://ui.adsabs.harvard.edu/abs/2016ApJ...821..126Q}
}

@ARTICLE{Raymond2009,
       author = {{Raymond}, Sean N. and {O'Brien}, David P. and {Morbidelli}, Alessandro and {Kaib}, Nathan A.},
        title = "{Building the terrestrial planets: Constrained accretion in the inner Solar System}",
      journal = {\icarus},
         year = 2009,
        month = oct,
       volume = {203},
       number = {2},
        pages = {644-662},
          doi = {10.1016/j.icarus.2009.05.016},
archivePrefix = {arXiv},
       eprint = {0905.3750},
 primaryClass = {astro-ph.EP},
       adsurl = {https://ui.adsabs.harvard.edu/abs/2009Icar..203..644R}
}

@ARTICLE{Marley2007,
       author = {{Marley}, Mark S. and {Fortney}, Jonathan J. and {Hubickyj}, Olenka and {Bodenheimer}, Peter and {Lissauer}, Jack J.},
        title = "{On the Luminosity of Young Jupiters}",
      journal = {\apj},
         year = 2007,
        month = jan,
       volume = {655},
       number = {1},
        pages = {541-549},
          doi = {10.1086/509759},
archivePrefix = {arXiv},
       eprint = {astro-ph/0609739},
 primaryClass = {astro-ph},
       adsurl = {https://ui.adsabs.harvard.edu/abs/2007ApJ...655..541M}
}

@ARTICLE{Hut1981,
       author = {{Hut}, P.},
        title = "{Tidal evolution in close binary systems.}",
      journal = {\aap},
         year = 1981,
        month = jun,
       volume = {99},
        pages = {126-140},
       adsurl = {https://ui.adsabs.harvard.edu/abs/1981A&A....99..126H}
}

@ARTICLE{Becker2019,
       author = {{Becker}, Juliette C. and {Vanderburg}, Andrew and {Rodriguez}, Joseph E. and {Omohundro}, Mark and {Adams}, Fred C. and {Stassun}, Keivan G. and {Yao}, Xinyu and {Hartman}, Joel and {Pepper}, Joshua and {Bakos}, Gaspar and {Barentsen}, Geert and {Beatty}, Thomas G. and {Bhatti}, Waqas and {Chontos}, Ashley and {Collier Cameron}, Andrew and {Hellier}, Coel and {Huber}, Daniel and {James}, David and {Kuhn}, Rudolf B. and {Lund}, Michael B. and {Pollacco}, Don and {Siverd}, Robert J. and {Stevens}, Daniel J. and {Cardoso}, Jos{\'e} Vin{\'\i}cius de Miranda and {West}, Richard},
        title = "{A Discrete Set of Possible Transit Ephemerides for Two Long-period Gas Giants Orbiting HIP 41378}",
      journal = {\aj},
         year = 2019,
        month = jan,
       volume = {157},
       number = {1},
          eid = {19},
        pages = {19},
          doi = {10.3847/1538-3881/aaf0a2},
archivePrefix = {arXiv},
       eprint = {1809.10688},
 primaryClass = {astro-ph.EP},
       adsurl = {https://ui.adsabs.harvard.edu/abs/2019AJ....157...19B}
}

@ARTICLE{Vanderburg2016,
       author = {{Vanderburg}, Andrew and {Becker}, Juliette C. and {Kristiansen}, Martti H. and {Bieryla}, Allyson and {Duev}, Dmitry A. and {Jensen-Clem}, Rebecca and {Morton}, Timothy D. and {Latham}, David W. and {Adams}, Fred C. and {Baranec}, Christoph and {Berlind}, Perry and {Calkins}, Michael L. and {Esquerdo}, Gilbert A. and {Kulkarni}, Shrinivas and {Law}, Nicholas M. and {Riddle}, Reed and {Salama}, Ma{\"\i}ssa and {Schmitt}, Allan R.},
        title = "{Five Planets Transiting a Ninth Magnitude Star}",
      journal = {\apjl},
         year = 2016,
        month = aug,
       volume = {827},
       number = {1},
          eid = {L10},
        pages = {L10},
          doi = {10.3847/2041-8205/827/1/L10},
archivePrefix = {arXiv},
       eprint = {1606.08441},
 primaryClass = {astro-ph.EP},
       adsurl = {https://ui.adsabs.harvard.edu/abs/2016ApJ...827L..10V}
}

@ARTICLE{Han2025,
       author = {{Han}, Te and {Robertson}, Paul and {Brandt}, Timothy D. and {Kanodia}, Shubham and {Ca{\~n}as}, Caleb and {Shporer}, Avi and {Ricker}, George and {Beard}, Corey},
        title = "{Hundreds of TESS Exoplanets Might Be Larger than We Thought}",
      journal = {\apjl},
         year = 2025,
        month = jul,
       volume = {988},
       number = {1},
          eid = {L4},
        pages = {L4},
          doi = {10.3847/2041-8213/ade794},
archivePrefix = {arXiv},
       eprint = {2506.19985},
 primaryClass = {astro-ph.EP},
       adsurl = {https://ui.adsabs.harvard.edu/abs/2025ApJ...988L...4H}
}

@ARTICLE{Xie2014,
       author = {{Xie}, Ji-Wei},
        title = "{Transit Timing Variation of Near-resonance Planetary Pairs. II. Confirmation of 30 Planets in 15 Multiple-planet Systems}",
      journal = {\apjs},
         year = 2014,
        month = feb,
       volume = {210},
       number = {2},
          eid = {25},
        pages = {25},
          doi = {10.1088/0067-0049/210/2/25},
archivePrefix = {arXiv},
       eprint = {1309.2329},
 primaryClass = {astro-ph.EP},
       adsurl = {https://ui.adsabs.harvard.edu/abs/2014ApJS..210...25X}
}

@ARTICLE{Lee2015cool,
       author = {{Lee}, Eve J. and {Chiang}, Eugene},
        title = "{To Cool is to Accrete: Analytic Scalings for Nebular Accretion of Planetary Atmospheres}",
      journal = {\apj},
         year = 2015,
        month = sep,
       volume = {811},
       number = {1},
          eid = {41},
        pages = {41},
          doi = {10.1088/0004-637X/811/1/41},
archivePrefix = {arXiv},
       eprint = {1508.05096},
 primaryClass = {astro-ph.EP},
       adsurl = {https://ui.adsabs.harvard.edu/abs/2015ApJ...811...41L}
}

@ARTICLE{Mizuno1980,
       author = {{Mizuno}, H.},
        title = "{Formation of the Giant Planets}",
      journal = {Progress of Theoretical Physics},
         year = 1980,
        month = aug,
       volume = {64},
       number = {2},
        pages = {544-557},
          doi = {10.1143/PTP.64.544},
       adsurl = {https://ui.adsabs.harvard.edu/abs/1980PThPh..64..544M}
}

@ARTICLE{Johnstone1998,
       author = {{Johnstone}, Doug and {Hollenbach}, David and {Bally}, John},
        title = "{Photoevaporation of Disks and Clumps by Nearby Massive Stars: Application to Disk Destruction in the Orion Nebula}",
      journal = {\apj},
         year = 1998,
        month = may,
       volume = {499},
       number = {2},
        pages = {758-776},
          doi = {10.1086/305658},
       adsurl = {https://ui.adsabs.harvard.edu/abs/1998ApJ...499..758J}
}

@ARTICLE{Adams2004,
       author = {{Adams}, Fred C. and {Hollenbach}, David and {Laughlin}, Gregory and {Gorti}, Uma},
        title = "{Photoevaporation of Circumstellar Disks Due to External Far-Ultraviolet Radiation in Stellar Aggregates}",
      journal = {\apj},
         year = 2004,
        month = aug,
       volume = {611},
       number = {1},
        pages = {360-379},
          doi = {10.1086/421989},
archivePrefix = {arXiv},
       eprint = {astro-ph/0404383},
 primaryClass = {astro-ph},
       adsurl = {https://ui.adsabs.harvard.edu/abs/2004ApJ...611..360A}
}

@ARTICLE{Piso2015,
       author = {{Piso}, Ana-Maria A. and {Youdin}, Andrew N. and {Murray-Clay}, Ruth A.},
        title = "{Minimum Core Masses for Giant Planet Formation with Realistic Equations of State and Opacities}",
      journal = {\apj},
         year = 2015,
        month = feb,
       volume = {800},
       number = {2},
          eid = {82},
        pages = {82},
          doi = {10.1088/0004-637X/800/2/82},
archivePrefix = {arXiv},
       eprint = {1412.5185},
 primaryClass = {astro-ph.EP},
       adsurl = {https://ui.adsabs.harvard.edu/abs/2015ApJ...800...82P}
}

@ARTICLE{Rafikov2006,
       author = {{Rafikov}, Roman R.},
        title = "{Atmospheres of Protoplanetary Cores: Critical Mass for Nucleated Instability}",
      journal = {\apj},
         year = 2006,
        month = sep,
       volume = {648},
       number = {1},
        pages = {666-682},
          doi = {10.1086/505695},
archivePrefix = {arXiv},
       eprint = {astro-ph/0405507},
 primaryClass = {astro-ph},
       adsurl = {https://ui.adsabs.harvard.edu/abs/2006ApJ...648..666R}
}

@ARTICLE{Price2025,
       author = {{Price}, Ellen M. and {Becker}, Juliette and {de Beurs}, Zo{\"e} L. and {Rogers}, Leslie A. and {Vanderburg}, Andrew},
        title = "{A Long Spin Period for a Sub-Neptune-mass Exoplanet}",
      journal = {\apjl},
         year = 2025,
        month = mar,
       volume = {981},
       number = {1},
          eid = {L7},
        pages = {L7},
          doi = {10.3847/2041-8213/adb42b},
archivePrefix = {arXiv},
       eprint = {2410.05408},
 primaryClass = {astro-ph.EP},
       adsurl = {https://ui.adsabs.harvard.edu/abs/2025ApJ...981L...7P}
}

@ARTICLE{Batygin2010,
       author = {{Batygin}, Konstantin and {Stevenson}, David J.},
        title = "{Inflating Hot Jupiters with Ohmic Dissipation}",
      journal = {\apjl},
         year = 2010,
        month = may,
       volume = {714},
       number = {2},
        pages = {L238-L243},
          doi = {10.1088/2041-8205/714/2/L238},
archivePrefix = {arXiv},
       eprint = {1002.3650},
 primaryClass = {astro-ph.EP},
       adsurl = {https://ui.adsabs.harvard.edu/abs/2010ApJ...714L.238B}
}

@ARTICLE{Laughlin2011,
       author = {{Laughlin}, Gregory and {Crismani}, Matteo and {Adams}, Fred C.},
        title = "{On the Anomalous Radii of the Transiting Extrasolar Planets}",
      journal = {\apjl},
         year = 2011,
        month = mar,
       volume = {729},
       number = {1},
          eid = {L7},
        pages = {L7},
          doi = {10.1088/2041-8205/729/1/L7},
archivePrefix = {arXiv},
       eprint = {1101.5827},
 primaryClass = {astro-ph.EP},
       adsurl = {https://ui.adsabs.harvard.edu/abs/2011ApJ...729L...7L}
}

@ARTICLE{Ibgui2009,
       author = {{Ibgui}, Laurent and {Burrows}, Adam},
        title = "{Coupled Evolution with Tides of the Radius and Orbit of Transiting Giant Planets: General Results}",
      journal = {\apj},
         year = 2009,
        month = aug,
       volume = {700},
       number = {2},
        pages = {1921-1932},
          doi = {10.1088/0004-637X/700/2/1921},
archivePrefix = {arXiv},
       eprint = {0902.3998},
 primaryClass = {astro-ph.EP},
       adsurl = {https://ui.adsabs.harvard.edu/abs/2009ApJ...700.1921I}
}

@ARTICLE{Bodenheimer2001,
       author = {{Bodenheimer}, Peter and {Lin}, D.~N.~C. and {Mardling}, R.~A.},
        title = "{On the Tidal Inflation of Short-Period Extrasolar Planets}",
      journal = {\apj},
         year = 2001,
        month = feb,
       volume = {548},
       number = {1},
        pages = {466-472},
          doi = {10.1086/318667},
       adsurl = {https://ui.adsabs.harvard.edu/abs/2001ApJ...548..466B}
}

@ARTICLE{Batygin2015,
       author = {{Batygin}, Konstantin},
        title = "{Capture of planets into mean-motion resonances and the origins of extrasolar orbital architectures}",
      journal = {\mnras},
         year = 2015,
        month = aug,
       volume = {451},
       number = {3},
        pages = {2589-2609},
          doi = {10.1093/mnras/stv1063},
archivePrefix = {arXiv},
       eprint = {1505.01778},
 primaryClass = {astro-ph.EP},
       adsurl = {https://ui.adsabs.harvard.edu/abs/2015MNRAS.451.2589B}
}

@ARTICLE{Gao2020_deflating,
       author = {{Gao}, Peter and {Zhang}, Xi},
        title = "{Deflating Super-puffs: Impact of Photochemical Hazes on the Observed Mass-Radius Relationship of Low-mass Planets}",
      journal = {\apj},
         year = 2020,
        month = feb,
       volume = {890},
       number = {2},
          eid = {93},
        pages = {93},
          doi = {10.3847/1538-4357/ab6a9b},
archivePrefix = {arXiv},
       eprint = {2001.00055},
 primaryClass = {astro-ph.EP},
       adsurl = {https://ui.adsabs.harvard.edu/abs/2020ApJ...890...93G}
}

@ARTICLE{Lu2025,
       author = {{Lu}, Tiger and {Li}, Gongjie and {Cassese}, Ben and {Lin}, D.~N.~C.},
        title = "{The Dynamical History of HIP-41378 f{\textemdash}Oblique Exorings Masquerading as a Puffy Planet}",
      journal = {\apj},
         year = 2025,
        month = feb,
       volume = {980},
       number = {1},
          eid = {39},
        pages = {39},
          doi = {10.3847/1538-4357/ada4b2},
archivePrefix = {arXiv},
       eprint = {2410.00641},
 primaryClass = {astro-ph.EP},
       adsurl = {https://ui.adsabs.harvard.edu/abs/2025ApJ...980...39L}
}

@ARTICLE{Piro2020_rings,
       author = {{Piro}, Anthony L. and {Vissapragada}, Shreyas},
        title = "{Exploring Whether Super-puffs can be Explained as Ringed Exoplanets}",
      journal = {\aj},
         year = 2020,
        month = apr,
       volume = {159},
       number = {4},
          eid = {131},
        pages = {131},
          doi = {10.3847/1538-3881/ab7192},
archivePrefix = {arXiv},
       eprint = {1911.09673},
 primaryClass = {astro-ph.EP},
       adsurl = {https://ui.adsabs.harvard.edu/abs/2020AJ....159..131P}
}

@ARTICLE{Batygin2023,
       author = {{Batygin}, Konstantin and {Adams}, Fred C. and {Becker}, Juliette},
        title = "{The Origin of Universality in the Inner Edges of Planetary Systems}",
      journal = {\apjl},
         year = 2023,
        month = jul,
       volume = {951},
       number = {1},
          eid = {L19},
        pages = {L19},
          doi = {10.3847/2041-8213/acdb5d},
archivePrefix = {arXiv},
       eprint = {2306.08822},
 primaryClass = {astro-ph.EP},
       adsurl = {https://ui.adsabs.harvard.edu/abs/2023ApJ...951L..19B}
}

@article{Batygin_2025_WASP107b,
       author = {{Batygin}, Konstantin},
        title = "{From Tides to Currents: Unraveling the Mechanism that Powers WASP-107b's Internal Heat Flux}",
      journal = {\apj},
         year = 2025,
        month = may,
       volume = {985},
       number = {1},
          eid = {87},
        pages = {87},
          doi = {10.3847/1538-4357/adccc4},
archivePrefix = {arXiv},
       eprint = {2505.01581},
 primaryClass = {astro-ph.EP},
       adsurl = {https://ui.adsabs.harvard.edu/abs/2025ApJ...985...87B}
}

@article{Batygin_Stevenson_2013, title={Mass-Radius Relationships for Very Low Mass Gaseous Planets}, volume={769}, ISSN={2041-8205, 2041-8213}, DOI={10.1088/2041-8205/769/1/L9}, abstractNote={Recently, the Kepler spacecraft has detected a sizable aggregate of objects, characterized by giant-planet-like radii and modest levels of stellar irradiation. With the exception of a handful of objects, the physical nature, and specifically the average densities, of these bodies remain unknown. Here, we propose that the detected giant planet radii may partially belong to planets somewhat less massive than Uranus and Neptune. Accordingly, in this work, we seek to identify a physically sound upper limit to planetary radii at low masses and moderate equilibrium temperatures. As a guiding example, we analyze the interior structure of the Neptune-mass planet Kepler-30d and show that it is acutely deficient in heavy elements, especially compared with its solar system counterparts. Subsequently, we perform numerical simulations of planetary thermal evolution and in agreement with previous studies, show that generally, 10 - 20 Earth mass, multi-billion year old planets, composed of high density cores and extended H/He envelopes can have radii that firmly reside in the giant planet range. We subject our results to stability criteria based on extreme ultraviolet radiation, as well as Roche-lobe overflow driven mass-loss and construct mass-radius relationships for the considered objects. We conclude by discussing observational avenues that may be used to confirm or repudiate the existence of putative low mass, gas-dominated planets.}, note={arXiv:1304.5157 [astro-ph]}, number={1}, journal={The Astrophysical Journal}, author={Batygin, Konstantin and Stevenson, David J.}, year={2013}, month=may, pages={L9} }

@article{Belkovski_2022, title={A Multi-Planet System’s Sole Super-Puff: Exploring Allowable Physical Parameters for the Cold Super-Puff HIP 41378 f}, volume={163}, ISSN={0004-6256, 1538-3881}, DOI={10.3847/1538-3881/ac6353}, abstractNote={The census of known exoplanets exhibits a variety of physical parameters, including densities that are measured to span the range from less dense than styrofoam to more dense than iron. These densities represent a large diversity of interior structures. Despite this staggering diversity, recent analyses have shown that the densities of planets that orbit a common star exhibit remarkable uniformity. A fascinating exception to this is the system HIP 41378 (also known as K2-93), which contains a super-puff planet, HIP 41378 f, as well as several planets with more typical bulk densities. The range of densities in this system begs the question of what physical processes are responsible for the disparate planetary structures in this system. In this paper, we consider how the densities of the planets in the HIP 41378 system would have changed over time as the host star evolved and the planets’ atmospheres were subsequently affected by the evolving insolation level. We also present a range of allowable core masses for HIP 41378 f based on the measured planet parameters, and comment on the feasibility of the proposed existence of planetary rings around HIP 41378 f as an explanation for its current low density.}, note={arXiv:2203.17180 [astro-ph]}, number={6}, journal={The Astronomical Journal}, author={Belkovski, Michelle and Becker, Juliette and Howe, Alex and Malsky, Isaac and Batygin, Konstantin}, year={2022}, month=jun, pages={277} }

@article{Biersteker_Schlichting_2019, title={Atmospheric mass-loss due to giant impacts: the importance of the thermal component for hydrogen-helium envelopes}, volume={485}, ISSN={0035-8711}, DOI={10.1093/mnras/stz738}, note={ADS Bibcode: 2019MNRAS.485.4454B}, journal={Monthly Notices of the Royal Astronomical Society}, author={Biersteker, John B. and Schlichting, Hilke E.}, year={2019}, month=may, pages={4454–4463} }

@article{Howe_Architectures, title={Architecture Classification for Extrasolar Planetary Systems}, volume={169}, ISSN={0004-6256, 1538-3881}, DOI={10.3847/1538-3881/adabdb},  note={arXiv:2501.08191 [astro-ph]}, number={3}, journal={The Astronomical Journal}, author={Howe, Alex R. and Becker, Juliette C. and Stark, Christopher C. and Adams, Fred C.}, year={2025}, month=mar, pages={149} }

@article{Howe_Burrows_Evolution, title={Evolutionary Models of Super-Earths and Mini-Neptunes Incorporating Cooling and Mass Loss}, volume={808}, ISSN={0004-637X}, DOI={10.1088/0004-637X/808/2/150}, abstractNote={We construct models of the structural evolution of super-Earth- and mini-Neptune-type exoplanets with H2-He envelopes, incorporating radiative cooling and XUV-driven mass loss. We conduct a parameter study of these models, focusing on initial mass, radius, and envelope mass fractions, as well as orbital distance, metallicity, and the specific prescription for mass loss. From these calculations, we investigate how the observed masses and radii of exoplanets today relate to the distribution of their initial conditions. Orbital distance and the initial envelope mass fraction are the most important factors determining planetary evolution, particularly radius evolution. Initial mass also becomes important below a “turnoff mass,” which varies with orbital distance, with mass-radius curves being approximately flat for higher masses. Initial radius is the least important parameter we study, with very little difference between the hot start and cold start limits after an age of 100 Myr. Model sets with no mass loss fail to produce results consistent with observations, but a plausible range of mass-loss scenarios is allowed. In addition, we present scenarios for the formation of the Kepler-11 planets. Our best fit to observations of Kepler-11b and Kepler-11c involves formation beyond the snow line, after which they moved inward, circularized, and underwent a reduced degree of mass loss.}, note={ADS Bibcode: 2015ApJ...808..150H}, journal={The Astrophysical Journal}, publisher={IOP}, author={Howe, Alex R. and Burrows, Adam}, year={2015}, month=aug, pages={150} }

@article{Lopez_Fortney_2014_R-M-Relation, title={Understanding the Mass-Radius Relation for Sub-neptunes: Radius as a Proxy for Composition}, volume={792}, ISSN={0004-637X}, DOI={10.1088/0004-637X/792/1/1}, abstractNote={Transiting planet surveys like Kepler have provided a wealth of information on the distribution of planetary radii, particularly for the new populations of super-Earth- and sub-Neptune-sized planets. In order to aid in the physical interpretation of these radii, we compute model radii for low-mass rocky planets with hydrogen-helium envelopes. We provide model radii for planets 1-20 M ⊕, with envelope fractions 0.01%-20%, levels of irradiation 0.1-1000 times Earth’s, and ages from 100 Myr to 10 Gyr. In addition we provide simple analytic fits that summarize how radius depends on each of these parameters. Most importantly, we show that at fixed H/He envelope fraction, radii show little dependence on mass for planets with more than ~1% of their mass in their envelope. Consequently, planetary radius is to a first order a proxy for planetary composition, i.e., H/He envelope fraction, for Neptune- and sub-Neptune-sized planets. We recast the observed mass-radius relationship as a mass-composition relationship and discuss it in light of traditional core accretion theory. We discuss the transition from rocky super-Earths to sub-Neptune planets with large volatile envelopes. We suggest ~1.75 R ⊕ as a physically motivated dividing line between these two populations of planets. Finally, we discuss these results in light of the observed radius occurrence distribution found by Kepler.}, note={ADS Bibcode: 2014ApJ...792....1L}, journal={The Astrophysical Journal}, publisher={IOP}, author={Lopez, Eric D. and Fortney, Jonathan J.}, year={2014}, month=sep, pages={1} }

@article{Lopez_Fortney_Miller_2012, title={How Thermal Evolution and Mass-loss Sculpt Populations of Super-Earths and Sub-Neptunes: Application to the Kepler-11 System and Beyond}, volume={761}, ISSN={0004-637X}, DOI={10.1088/0004-637X/761/1/59}, abstractNote={We use models of thermal evolution and extreme ultraviolet (XUV) driven mass loss to explore the composition and history of low-mass, low-density transiting planets. We investigate the Kepler-11 system in detail and provide estimates of both the current and past planetary compositions. We find that an H/He envelope on Kepler-11b is highly vulnerable to mass loss. By comparing to formation models, we show that in situ formation of the system is extremely difficult. Instead we propose that it is a water-rich system of sub-Neptunes that migrated from beyond the snow line. For the broader population of observed planets, we show that there is a threshold in bulk planet density and incident flux above which no low-mass transiting planets have been observed. We suggest that this threshold is due to the instability of H/He envelopes to XUV-driven mass loss. Importantly, we find that this mass-loss threshold is well reproduced by our thermal evolution/contraction models that incorporate a standard mass-loss prescription. Treating the planets’ contraction history is essential because the planets have significantly larger radii during the early era of high XUV fluxes. Over time low-mass planets with H/He envelopes can be transformed into water-dominated worlds with steam envelopes or rocky super-Earths. Finally, we use this threshold to provide likely minimum masses and radial-velocity amplitudes for the general population of Kepler candidates. Likewise, we use this threshold to provide constraints on the maximum radii of low-mass planets found by radial-velocity surveys.}, note={ADS Bibcode: 2012ApJ...761...59L}, journal={The Astrophysical Journal}, publisher={IOP}, author={Lopez, Eric D. and Fortney, Jonathan J. and Miller, Neil}, year={2012}, month=dec, pages={59} }

@ARTICLE{Mordasini2012_part1,
       author = {{Mordasini}, C. and {Alibert}, Y. and {Klahr}, H. and {Henning}, T.},
        title = "{Characterization of exoplanets from their formation. I. Models of combined planet formation and evolution}",
      journal = {\aap},
         year = 2012,
        month = nov,
       volume = {547},
          eid = {A111},
        pages = {A111},
          doi = {10.1051/0004-6361/201118457},
archivePrefix = {arXiv},
       eprint = {1206.6103},
 primaryClass = {astro-ph.EP},
       adsurl = {https://ui.adsabs.harvard.edu/abs/2012A&A...547A.111M}
}

@article{Rafizadeh_2025, title={Planet–comet interaction as another mechanism for formation of super-puff exoplanets}, volume={539}, rights={https://creativecommons.org/licenses/by/4.0/}, ISSN={0035-8711, 1365-2966}, DOI={10.1093/mnras/staf581}, number={4}, journal={Monthly Notices of the Royal Astronomical Society}, author={Rafizadeh, Hamid A}, year={2025}, month=may, pages={3518–3533}, language={en} }

@article{Vissapragada_2024_TOI1420B, title={Helium in the Extended Atmosphere of the Warm Superpuff TOI-1420b}, volume={167}, ISSN={0004-6256}, DOI={10.3847/1538-3881/ad3241},  note={ADS Bibcode: 2024AJ....167..199V}, journal={The Astronomical Journal}, publisher={IOP}, author={Vissapragada, Shreyas and Greklek-McKeon, Michael and Linssen, Dion and MacLeod, Morgan and Thorngren, Daniel P. and Gao, Peter and Knutson, Heather A. and Latham, David W. and López-Morales, Mercedes and Oklopčić, Antonija and Pérez González, Jorge and Saidel, Morgan and Tumborang, Abigail and Yoshida, Stephanie}, year={2024}, month=may, pages={199} }

@article{Yee_WASP193b2025, title={The Super-puff WASP-193 b is on a Well-aligned Orbit*}, volume={169}, ISSN={0004-6256, 1538-3881}, DOI={10.3847/1538-3881/adba5f},  number={4}, journal={The Astronomical Journal}, author={Yee, Samuel W. and Stefánsson, Gudmundur and Thorngren, Daniel and Monson, Andy and Hartman, Joel D. and Charbonneau, David B. and Teske, Johanna K. and Butler, R. Paul and Crane, Jeffrey D. and Osip, David and Shectman, Stephen A.}, year={2025}, month=apr, pages={225} }

@article{Espinoza-Retamal_PolarOrbit_2025,
       author = {{Espinoza-Retamal}, Juan I. and {Brahm}, Rafael and {Petrovich}, Cristobal and {Jord{\'a}n}, Andr{\'e}s and {Henning}, Thomas and {Trifonov}, Trifon and {Winn}, Joshua N. and {Rea}, Erika and {G{\"u}nther}, Maximilian N. and {Agabi}, Abdelkrim and {Bendjoya}, Philippe and {Bhaskar}, Hareesh and {Bouchy}, Fran{\c{c}}ois and {Catelan}, M{\'a}rcio and {Charalambous}, Carolina and {Deloupy}, Vincent and {Dransfield}, George and {Eberhardt}, Jan and {Espinoza}, N{\'e}stor and {Freckelton}, Alix V. and {Guillot}, Tristan and {Hobson}, Melissa J. and {Jones}, Mat{\'\i}as I. and {Lendl}, Monika and {Mekarnia}, Djamel and {Mu{\~n}oz}, Diego J. and {D. Nielsen}, Louise and {Rojas}, Felipe I. and {Schmider}, Fran{\c{c}}ois-Xavier and {Sedaghati}, Elyar and {Stef{\'a}nsson}, Gu{\dj}mundur and {Striegel}, Stephanie and {Suarez}, Olga and {Tala Pinto}, Marcelo and {Timmermans}, Mathilde and {Triaud}, Amaury H.~M.~J. and {Udry}, St{\'e}phane and {Ulmer-Moll}, Sol{\`e}ne and {Ziegler}, Carl},
        title = "{A Cold and Superpuffy Planet on a Prograde Orbit}",
      journal = {\apjl},
         year = 2026,
        month = jan,
       volume = {996},
       number = {1},
          eid = {L13},
        pages = {L13},
          doi = {10.3847/2041-8213/ae2bfa},
archivePrefix = {arXiv},
       eprint = {2510.00102},
 primaryClass = {astro-ph.EP},
       adsurl = {https://ui.adsabs.harvard.edu/abs/2026ApJ...996L..13E}
}

@article{Orosz2019_Kepler47, title={Discovery of a Third Transiting Planet in the Kepler-47 Circumbinary System}, volume={157}, ISSN={0004-6256, 1538-3881}, DOI={10.3847/1538-3881/ab0ca0}, number={5}, journal={The Astronomical Journal}, author={Orosz, Jerome A. and Welsh, William F. and Haghighipour, Nader and Quarles, Billy and Short, Donald R. and Mills, Sean M. and Satyal, Suman and Torres, Guillermo and Agol, Eric and Fabrycky, Daniel C. and Jontof-Hutter, Daniel and Windmiller, Gur and Müller, Tobias W. A. and Hinse, Tobias C. and Cochran, William D. and Endl, Michael and Ford, Eric B. and Mazeh, Tsevi and Lissauer, Jack J.}, year={2019}, month=may, pages={174} }

@ARTICLE{Welbanks2024,
       author = {{Welbanks}, Luis and {Bell}, Taylor J. and {Beatty}, Thomas G. and {Line}, Michael R. and {Ohno}, Kazumasa and {Fortney}, Jonathan J. and {Schlawin}, Everett and {Greene}, Thomas P. and {Rauscher}, Emily and {McGill}, Peter and {Murphy}, Matthew and {Parmentier}, Vivien and {Tang}, Yao and {Edelman}, Isaac and {Mukherjee}, Sagnick and {Wiser}, Lindsey S. and {Lagage}, Pierre-Olivier and {Dyrek}, Achr{\`e}ne and {Arnold}, Kenneth E.},
        title = "{A high internal heat flux and large core in a warm Neptune exoplanet}",
      journal = {\nat},
         year = 2024,
        month = jun,
       volume = {630},
       number = {8018},
        pages = {836-840},
          doi = {10.1038/s41586-024-07514-w},
archivePrefix = {arXiv},
       eprint = {2405.11018},
 primaryClass = {astro-ph.EP},
       adsurl = {https://ui.adsabs.harvard.edu/abs/2024Natur.630..836W}
}

@ARTICLE{Sing2024,
       author = {{Sing}, David K. and {Rustamkulov}, Zafar and {Thorngren}, Daniel P. and {Barstow}, Joanna K. and {Tremblin}, Pascal and {Alves de Oliveira}, Catarina and {Beck}, Tracy L. and {Birkmann}, Stephan M. and {Challener}, Ryan C. and {Crouzet}, Nicolas and {Espinoza}, N{\'e}stor and {Ferruit}, Pierre and {Giardino}, Giovanna and {Gressier}, Am{\'e}lie and {Lee}, Elspeth K.~H. and {Lewis}, Nikole K. and {Maiolino}, Roberto and {Manjavacas}, Elena and {Rauscher}, Bernard J. and {Sirianni}, Marco and {Valenti}, Jeff A.},
        title = "{A warm Neptune's methane reveals core mass and vigorous atmospheric mixing}",
      journal = {\nat},
         year = 2024,
        month = jun,
       volume = {630},
       number = {8018},
        pages = {831-835},
          doi = {10.1038/s41586-024-07395-z},
archivePrefix = {arXiv},
       eprint = {2405.11027},
 primaryClass = {astro-ph.EP},
       adsurl = {https://ui.adsabs.harvard.edu/abs/2024Natur.630..831S}
}

@ARTICLE{Piaulet2021,
       author = {{Piaulet}, Caroline and {Benneke}, Bj{\"o}rn and {Rubenzahl}, Ryan A. and {Howard}, Andrew W. and {Lee}, Eve J. and {Thorngren}, Daniel and {Angus}, Ruth and {Peterson}, Merrin and {Schlieder}, Joshua E. and {Werner}, Michael and {Kreidberg}, Laura and {Jaouni}, Tareq and {Crossfield}, Ian J.~M. and {Ciardi}, David R. and {Petigura}, Erik A. and {Livingston}, John and {Dressing}, Courtney D. and {Fulton}, Benjamin J. and {Beichman}, Charles and {Christiansen}, Jessie L. and {Gorjian}, Varoujan and {Hardegree-Ullman}, Kevin K. and {Krick}, Jessica and {Sinukoff}, Evan},
        title = "{WASP-107b's Density Is Even Lower: A Case Study for the Physics of Planetary Gas Envelope Accretion and Orbital Migration}",
      journal = {\aj},
         year = 2021,
        month = feb,
       volume = {161},
       number = {2},
          eid = {70},
        pages = {70},
          doi = {10.3847/1538-3881/abcd3c},
archivePrefix = {arXiv},
       eprint = {2011.13444},
 primaryClass = {astro-ph.EP},
       adsurl = {https://ui.adsabs.harvard.edu/abs/2021AJ....161...70P}
}

@unpublished{Tanglin2025,
  author       = {Tanglin, Nathaniel and Becker, Juliette},
  title        = {Simulated Effects of Radioactive Elements on Magnetic Evolution and Atmospheric Mass Loss of Earth-like Planets},
  year         = {2025},
  note         = {submitted to AAS journals}
}

@ARTICLE{Jontof2019,
       author = {{Jontof-Hutter}, Daniel},
        title = "{The Compositional Diversity of Low-Mass Exoplanets}",
      journal = {Annual Review of Earth and Planetary Sciences},
         year = 2019,
        month = may,
       volume = {47},
        pages = {141-171},
          doi = {10.1146/annurev-earth-053018-060352},
archivePrefix = {arXiv},
       eprint = {1911.04598},
 primaryClass = {astro-ph.EP},
       adsurl = {https://ui.adsabs.harvard.edu/abs/2019AREPS..47..141J}
}

@ARTICLE{Crossfield2016,
       author = {{Crossfield}, Ian J.~M. and {Ciardi}, David R. and {Petigura}, Erik A. and {Sinukoff}, Evan and {Schlieder}, Joshua E. and {Howard}, Andrew W. and {Beichman}, Charles A. and {Isaacson}, Howard and {Dressing}, Courtney D. and {Christiansen}, Jessie L. and {Fulton}, Benjamin J. and {L{\'e}pine}, S{\'e}bastien and {Weiss}, Lauren and {Hirsch}, Lea and {Livingston}, John and {Baranec}, Christoph and {Law}, Nicholas M. and {Riddle}, Reed and {Ziegler}, Carl and {Howell}, Steve B. and {Horch}, Elliott and {Everett}, Mark and {Teske}, Johanna and {Martinez}, Arturo O. and {Obermeier}, Christian and {Benneke}, Bj{\"o}rn and {Scott}, Nic and {Deacon}, Niall and {Aller}, Kimberly M. and {Hansen}, Brad M.~S. and {Mancini}, Luigi and {Ciceri}, Simona and {Brahm}, Rafael and {Jord{\'a}n}, Andr{\'e}s and {Knutson}, Heather A. and {Henning}, Thomas and {Bonnefoy}, Micha{\"e}l and {Liu}, Michael C. and {Crepp}, Justin R. and {Lothringer}, Joshua and {Hinz}, Phil and {Bailey}, Vanessa and {Skemer}, Andrew and {Defrere}, Denis},
        title = "{197 Candidates and 104 Validated Planets in K2{\textquoteright}s First Five Fields}",
      journal = {\apjs},
         year = 2016,
        month = sep,
       volume = {226},
       number = {1},
          eid = {7},
        pages = {7},
          doi = {10.3847/0067-0049/226/1/7},
archivePrefix = {arXiv},
       eprint = {1607.05263},
 primaryClass = {astro-ph.EP},
       adsurl = {https://ui.adsabs.harvard.edu/abs/2016ApJS..226....7C}
}

@ARTICLE{Vanderburg2016a,
       author = {{Vanderburg}, Andrew and {Latham}, David W. and {Buchhave}, Lars A. and {Bieryla}, Allyson and {Berlind}, Perry and {Calkins}, Michael L. and {Esquerdo}, Gilbert A. and {Welsh}, Sophie and {Johnson}, John Asher},
        title = "{Planetary Candidates from the First Year of the K2 Mission}",
      journal = {\apjs},
         year = 2016,
        month = jan,
       volume = {222},
       number = {1},
          eid = {14},
        pages = {14},
          doi = {10.3847/0067-0049/222/1/14},
archivePrefix = {arXiv},
       eprint = {1511.07820},
 primaryClass = {astro-ph.EP},
       adsurl = {https://ui.adsabs.harvard.edu/abs/2016ApJS..222...14V}
}

@ARTICLE{Howe2025b,
       author = {{Howe}, Alex R. and {Becker}, Juliette C. and {Adams}, Fred C.},
        title = "{Architectures of Planetary Systems. II. Trends with Host Star Mass and Metallicity}",
      journal = {\aj},
         year = 2026,
        month = mar,
       volume = {171},
       number = {3},
          eid = {148},
        pages = {148},
          doi = {10.3847/1538-3881/ae3aa6},
archivePrefix = {arXiv},
       eprint = {2602.03657},
 primaryClass = {astro-ph.EP},
       adsurl = {https://ui.adsabs.harvard.edu/abs/2026AJ....171..148H}
}

@ARTICLE{Mamajek2025,
       author = {{Mamajek}, Eric E. and {Wright}, Jason T. and {Tuchow}, Noah W. and {Young}, Patrick A. and {Kenworthy}, Matthew A. and {Gilbert}, Emily A.},
        title = "{The Solirad (So) as a Convenient Unit for Quoting Astronomical Irradiances for Planetary Insolations and Exoplanetary Instellations}",
      journal = {\pasp},
         year = 2026,
        month = feb,
       volume = {138},
       number = {2},
          eid = {023001},
        pages = {023001},
          doi = {10.1088/1538-3873/ae37da},
archivePrefix = {arXiv},
       eprint = {2512.20126},
 primaryClass = {astro-ph.EP},
       adsurl = {https://ui.adsabs.harvard.edu/abs/2026PASP..138b3001M}
}

@ARTICLE{Demory2011,
       author = {{Demory}, Brice-Olivier and {Seager}, Sara},
        title = "{Lack of Inflated Radii for Kepler Giant Planet Candidates Receiving Modest Stellar Irradiation}",
      journal = {\apjs},
         year = 2011,
        month = nov,
       volume = {197},
       number = {1},
          eid = {12},
        pages = {12},
          doi = {10.1088/0067-0049/197/1/12},
archivePrefix = {arXiv},
       eprint = {1110.6180},
 primaryClass = {astro-ph.EP},
       adsurl = {https://ui.adsabs.harvard.edu/abs/2011ApJS..197...12D}
}

@ARTICLE{Thorngren2021,
       author = {{Thorngren}, Daniel P. and {Fortney}, Jonathan J. and {Lopez}, Eric D. and {Berger}, Travis A. and {Huber}, Daniel},
        title = "{Slow Cooling and Fast Reinflation for Hot Jupiters}",
      journal = {\apjl},
         year = 2021,
        month = mar,
       volume = {909},
       number = {1},
          eid = {L16},
        pages = {L16},
          doi = {10.3847/2041-8213/abe86d},
archivePrefix = {arXiv},
       eprint = {2101.05285},
 primaryClass = {astro-ph.EP},
       adsurl = {https://ui.adsabs.harvard.edu/abs/2021ApJ...909L..16T}
}

@ARTICLE{Christiansen2025,
       author = {{Christiansen}, Jessie L. and {McElroy}, Douglas L. and {Harbut}, Marcy and {Ciardi}, David R. and {Crane}, Megan and {Good}, John and {Hardegree-Ullman}, Kevin K. and {Kesseli}, Aurora Y. and {Lund}, Michael B. and {Lynn}, Meca and {Muthiar}, Ananda and {Nilsson}, Ricky and {Oluyide}, Toba and {Papin}, Michael and {Rivera}, Amalia and {Swain}, Melanie and {Susemiehl}, Nicholas D. and {Tam}, Raymond and {van Eyken}, Julian and {Beichman}, Charles},
        title = "{The NASA Exoplanet Archive and Exoplanet Follow-up Observing Program: Data, Tools, and Usage}",
      journal = {\psj},
         year = 2025,
        month = aug,
       volume = {6},
       number = {8},
          eid = {186},
        pages = {186},
          doi = {10.3847/PSJ/ade3c2},
archivePrefix = {arXiv},
       eprint = {2506.03299},
 primaryClass = {astro-ph.EP},
       adsurl = {https://ui.adsabs.harvard.edu/abs/2025PSJ.....6..186C}
}

@article{Cochran_2011_kepler18, title={Kepler-18b, c, and d: A System of Three Planets Confirmed by Transit Timing Variations, Light Curve Validation, Warm-Spitzer Photometry, and Radial Velocity Measurements}, volume={197}, ISSN={0067-0049}, DOI={10.1088/0067-0049/197/1/7}, note={ADS Bibcode: 2011ApJS..197....7C}, journal={The Astrophysical Journal Supplement Series}, publisher={IOP}, author={Cochran, William D. and Fabrycky, Daniel C. and Torres, Guillermo and Fressin, François and Désert, Jean-Michel and Ragozzine, Darin and Sasselov, Dimitar and Fortney, Jonathan J. and Rowe, Jason F. and Brugamyer, Erik J. and Bryson, Stephen T. and Carter, Joshua A. and Ciardi, David R. and Howell, Steve B. and Steffen, Jason H. and Borucki, William. J. and Koch, David G. and Winn, Joshua N. and Welsh, William F. and Uddin, Kamal and Tenenbaum, Peter and Still, M. and Seager, Sara and Quinn, Samuel N. and Mullally, F. and Miller, Neil and Marcy, Geoffrey W. and MacQueen, Phillip J. and Lucas, Phillip and Lissauer, Jack J. and Latham, David W. and Knutson, Heather and Kinemuchi, K. and Johnson, John A. and Jenkins, Jon M. and Isaacson, Howard and Howard, Andrew and Horch, Elliott and Holman, Matthew J. and Henze, Christopher E. and Haas, Michael R. and Gilliland, Ronald L. and Gautier, Thomas N., III and Ford, Eric B. and Fischer, Debra A. and Everett, Mark and Endl, Michael and Demory, Brice-Oliver and Deming, Drake and Charbonneau, David and Caldwell, Douglas and Buchhave, Lars and Brown, Timothy M. and Batalha, Natalie}, year={2011}, month=nov, pages={7} }

@ARTICLE{Anderson2012,
       author = {{Anderson}, Kassandra R. and {Adams}, Fred C.},
        title = "{Effects of Collisions with Rocky Planets on the Properties of Hot Jupiters}",
      journal = {\pasp},
         year = 2012,
        month = aug,
       volume = {124},
       number = {918},
        pages = {809},
          doi = {10.1086/667539},
archivePrefix = {arXiv},
       eprint = {1206.5857},
 primaryClass = {astro-ph.EP},
       adsurl = {https://ui.adsabs.harvard.edu/abs/2012PASP..124..809A}
}

@ARTICLE{JontofHutter2014,
       author = {{Jontof-Hutter}, Daniel and {Lissauer}, Jack J. and {Rowe}, Jason F. and {Fabrycky}, Daniel C.},
        title = "{Kepler-79's Low Density Planets}",
      journal = {\apj},
         year = 2014,
        month = apr,
       volume = {785},
       number = {1},
          eid = {15},
        pages = {15},
          doi = {10.1088/0004-637X/785/1/15},
archivePrefix = {arXiv},
       eprint = {1310.2642},
 primaryClass = {astro-ph.EP},
       adsurl = {https://ui.adsabs.harvard.edu/abs/2014ApJ...785...15J}
}

@ARTICLE{Santerne2019,
       author = {{Santerne}, A. and {Malavolta}, L. and {Kosiarek}, M.~R. and {Dai}, F. and {Dressing}, C.~D. and {Dumusque}, X. and {Hara}, N.~C. and {Lopez}, T.~A. and {Mortier}, A. and {Vanderburg}, A. and {Adibekyan}, V. and {Armstrong}, D.~J. and {Barrado}, D. and {Barros}, S.~C.~C. and {Bayliss}, D. and {Berardo}, D. and {Boisse}, I. and {Bonomo}, A.~S. and {Bouchy}, F. and {Brown}, D.~J.~A. and {Buchhave}, L.~A. and {Butler}, R.~P. and {Collier Cameron}, A. and {Cosentino}, R. and {Crane}, J.~D. and {Crossfield}, I.~J.~M. and {Damasso}, M. and {Deleuil}, M.~R. and {Delgado Mena}, E. and {Demangeon}, O. and {D{\'\i}az}, R.~F. and {Donati}, J. -F. and {Figueira}, P. and {Fulton}, B.~J. and {Ghedina}, A. and {Harutyunyan}, A. and {H{\'e}brard}, G. and {Hirsch}, L.~A. and {Hojjatpanah}, S. and {Howard}, A.~W. and {Isaacson}, H. and {Latham}, D.~W. and {Lillo-Box}, J. and {L{\'o}pez-Morales}, M. and {Lovis}, C. and {Martinez Fiorenzano}, A.~F. and {Molinari}, E. and {Mousis}, O. and {Moutou}, C. and {Nava}, C. and {Nielsen}, L.~D. and {Osborn}, H.~P. and {Petigura}, E.~A. and {Phillips}, D.~F. and {Pollacco}, D.~L. and {Poretti}, E. and {Rice}, K. and {Santos}, N.~C. and {S{\'e}gransan}, D. and {Shectman}, S.~A. and {Sinukoff}, E. and {Sousa}, S.~G. and {Sozzetti}, A. and {Teske}, J.~K. and {Udry}, S. and {Vigan}, A. and {Wang}, S.~X. and {Watson}, C.~A. and {Weiss}, L.~M. and {Wheatley}, P.~J. and {Winn}, J.~N.},
        title = "{An extremely low-density and temperate giant exoplanet}",
      journal = {arXiv e-prints},
         year = 2019,
        month = nov,
          eid = {arXiv:1911.07355},
        pages = {arXiv:1911.07355},
archivePrefix = {arXiv},
       eprint = {1911.07355},
 primaryClass = {astro-ph.EP},
       adsurl = {https://ui.adsabs.harvard.edu/abs/2019arXiv191107355S}
}

@ARTICLE{Akinsanmi2020,
       author = {{Akinsanmi}, B. and {Santos}, N.~C. and {Faria}, J.~P. and {Oshagh}, M. and {Barros}, S.~C.~C. and {Santerne}, A. and {Charnoz}, S.},
        title = "{Can planetary rings explain the extremely low density of HIP 41378 f?}",
      journal = {\aap},
         year = 2020,
        month = mar,
       volume = {635},
          eid = {L8},
        pages = {L8},
          doi = {10.1051/0004-6361/202037618},
archivePrefix = {arXiv},
       eprint = {2002.11422},
 primaryClass = {astro-ph.EP},
       adsurl = {https://ui.adsabs.harvard.edu/abs/2020A&A...635L...8A}
}

@InProceedings{ mckinney-proc-scipy-2010,
  author    = { Wes McKinney },
  title     = { Data Structures for Statistical Computing in Python },
  booktitle = { Proceedings of the 9th Python in Science Conference },
  pages     = { 51 - 56 },
  year      = { 2010 },
  editor    = { St\'efan van der Walt and Jarrod Millman }
}

@article{seaborn, title={seaborn: statistical data visualization}, volume={6}, rights={http://creativecommons.org/licenses/by/4.0/}, ISSN={2475-9066}, DOI={10.21105/joss.03021}, number={60}, journal={Journal of Open Source Software}, author={Waskom, Michael}, year={2021}, month=apr, pages={3021} }

@article{astroquery, title={astroquery: An Astronomical Web-querying Package in Python}, volume={157}, ISSN={0004-6256}, DOI={10.3847/1538-3881/aafc33}, note={ADS Bibcode: 2019AJ....157...98G}, journal={The Astronomical Journal}, publisher={IOP}, author={Ginsburg, Adam and Sipőcz, Brigitta M. and Brasseur, C. E. and Cowperthwaite, Philip S. and Craig, Matthew W. and Deil, Christoph and Guillochon, James and Guzman, Giannina and Liedtke, Simon and Lian Lim, Pey and Lockhart, Kelly E. and Mommert, Michael and Morris, Brett M. and Norman, Henrik and Parikh, Madhura and Persson, Magnus V. and Robitaille, Thomas P. and Segovia, Juan-Carlos and Singer, Leo P. and Tollerud, Erik J. and de Val-Borro, Miguel and Valtchanov, Ivan and Woillez, Julien and Astroquery Collaboration and a subset of astropy Collaboration}, year={2019}, month=mar, pages={98} }

@article{Astropy_v5, title={The Astropy Project: Sustaining and Growing a Community-oriented Open-source Project and the Latest Major Release (v5.0) of the Core Package}, volume={935}, ISSN={0004-637X}, DOI={10.3847/1538-4357/ac7c74}, note={ADS Bibcode: 2022ApJ...935..167A}, journal={The Astrophysical Journal}, publisher={IOP}, author={Astropy Collaboration and Price-Whelan, Adrian M. and Lim, Pey Lian and Earl, Nicholas and Starkman, Nathaniel and Bradley, Larry and Shupe, David L. and Patil, Aarya A. and Corrales, Lia and Brasseur, C. E. and Nöthe, Maximilian and Donath, Axel and Tollerud, Erik and Morris, Brett M. and Ginsburg, Adam and Vaher, Eero and Weaver, Benjamin A. and Tocknell, James and Jamieson, William and van Kerkwijk, Marten H. and Robitaille, Thomas P. and Merry, Bruce and Bachetti, Matteo and Günther, H. Moritz and Aldcroft, Thomas L. and Alvarado-Montes, Jaime A. and Archibald, Anne M. and Bódi, Attila and Bapat, Shreyas and Barentsen, Geert and Bazán, Juanjo and Biswas, Manish and Boquien, Médéric and Burke, D. J. and Cara, Daria and Cara, Mihai and Conroy, Kyle E. and Conseil, Simon and Craig, Matthew W. and Cross, Robert M. and Cruz, Kelle L. and D’Eugenio, Francesco and Dencheva, Nadia and Devillepoix, Hadrien A. R. and Dietrich, Jörg P. and Eigenbrot, Arthur Davis and Erben, Thomas and Ferreira, Leonardo and Foreman-Mackey, Daniel and Fox, Ryan and Freij, Nabil and Garg, Suyog and Geda, Robel and Glattly, Lauren and Gondhalekar, Yash and Gordon, Karl D. and Grant, David and Greenfield, Perry and Groener, Austen M. and Guest, Steve and Gurovich, Sebastian and Handberg, Rasmus and Hart, Akeem and Hatfield-Dodds, Zac and Homeier, Derek and Hosseinzadeh, Griffin and Jenness, Tim and Jones, Craig K. and Joseph, Prajwel and Kalmbach, J. Bryce and Karamehmetoglu, Emir and Kałuszyński, Mikołaj and Kelley, Michael S. P. and Kern, Nicholas and Kerzendorf, Wolfgang E. and Koch, Eric W. and Kulumani, Shankar and Lee, Antony and Ly, Chun and Ma, Zhiyuan and MacBride, Conor and Maljaars, Jakob M. and Muna, Demitri and Murphy, N. A. and Norman, Henrik and O’Steen, Richard and Oman, Kyle A. and Pacifici, Camilla and Pascual, Sergio and Pascual-Granado, J. and Patil, Rohit R. and Perren, Gabriel I. and Pickering, Timothy E. and Rastogi, Tanuj and Roulston, Benjamin R. and Ryan, Daniel F. and Rykoff, Eli S. and Sabater, Jose and Sakurikar, Parikshit and Salgado, Jesús and Sanghi, Aniket and Saunders, Nicholas and Savchenko, Volodymyr and Schwardt, Ludwig and Seifert-Eckert, Michael and Shih, Albert Y. and Jain, Anany Shrey and Shukla, Gyanendra and Sick, Jonathan and Simpson, Chris and Singanamalla, Sudheesh and Singer, Leo P. and Singhal, Jaladh and Sinha, Manodeep and Sipőcz, Brigitta M. and Spitler, Lee R. and Stansby, David and Streicher, Ole and Šumak, Jani and Swinbank, John D. and Taranu, Dan S. and Tewary, Nikita and Tremblay, Grant R. and de Val-Borro, Miguel and Van Kooten, Samuel J. and Vasović, Zlatan and Verma, Shresth and de Miranda Cardoso, José Vinícius and Williams, Peter K. G. and Wilson, Tom J. and Winkel, Benjamin and Wood-Vasey, W. M. and Xue, Rui and Yoachim, Peter and Zhang, Chen and Zonca, Andrea and Astropy Project Contributors}, year={2022}, month=aug, pages={167} }

@article{numpy, title={Array programming with NumPy}, volume={585}, ISSN={0028-0836, 1476-4687}, DOI={10.1038/s41586-020-2649-2}, number={7825}, journal={Nature}, author={Harris, Charles R. and Millman, K. Jarrod and Van Der Walt, Stéfan J. and Gommers, Ralf and Virtanen, Pauli and Cournapeau, David and Wieser, Eric and Taylor, Julian and Berg, Sebastian and Smith, Nathaniel J. and Kern, Robert and Picus, Matti and Hoyer, Stephan and Van Kerkwijk, Marten H. and Brett, Matthew and Haldane, Allan and Del Río, Jaime Fernández and Wiebe, Mark and Peterson, Pearu and Gérard-Marchant, Pierre and Sheppard, Kevin and Reddy, Tyler and Weckesser, Warren and Abbasi, Hameer and Gohlke, Christoph and Oliphant, Travis E.}, year={2020}, month=sept, pages={357–362}, language={en} }

@Article{PER-GRA:IPython,
  Author    = {P\'erez, Fernando and Granger, Brian E.},
  Title     = {{IP}ython: a System for Interactive Scientific Computing},
  Journal   = {Computing in Science and Engineering},
  Volume    = {9},
  Number    = {3},
  Pages     = {21--29},
  month     = may,
  year      = 2007,
  url       = "http://ipython.org",
  ISSN      = "1521-9615",
  doi       = {10.1109/MCSE.2007.53},
  publisher = {IEEE Computer Society},
}

@ARTICLE{Zeng2019,
       author = {{Zeng}, Li and {Jacobsen}, Stein B. and {Sasselov}, Dimitar D. and {Petaev}, Michail I. and {Vanderburg}, Andrew and {Lopez-Morales}, Mercedes and {Perez-Mercader}, Juan and {Mattsson}, Thomas R. and {Li}, Gongjie and {Heising}, Matthew Z. and {Bonomo}, Aldo S. and {Damasso}, Mario and {Berger}, Travis A. and {Cao}, Hao and {Levi}, Amit and {Wordsworth}, Robin D.},
        title = "{Growth model interpretation of planet size distribution}",
      journal = {Proceedings of the National Academy of Science},
         year = 2019,
        month = may,
       volume = {116},
       number = {20},
        pages = {9723-9728},
          doi = {10.1073/pnas.1812905116},
archivePrefix = {arXiv},
       eprint = {1906.04253},
 primaryClass = {astro-ph.EP},
       adsurl = {https://ui.adsabs.harvard.edu/abs/2019PNAS..116.9723Z}
}

@Misc{scipy,
  author =    {Eric Jones and Travis Oliphant and Pearu Peterson and others},
  title =     {{SciPy}: Open source scientific tools for {Python}},
  year =      {2001},
  url = "http://www.scipy.org/",
  note = {[Online; accessed <today>]}
}

@Article{Hunter:Matplotlib,
  Author    = {Hunter, J. D.},
  Title     = {Matplotlib: A 2D graphics environment},
  Journal   = {Computing In Science \& Engineering},
  Volume    = {9},
  Number    = {3},
  Pages     = {90--95},
  publisher = {IEEE COMPUTER SOC},
  doi       = {10.1109/MCSE.2007.55},
  year      = 2007
}

@conference{Kluyver:2016aa,
	Author = {Thomas Kluyver and Benjamin Ragan-Kelley and Fernando P{\'e}rez and Brian Granger and Matthias Bussonnier and Jonathan Frederic and Kyle Kelley and Jessica Hamrick and Jason Grout and Sylvain Corlay and Paul Ivanov and Dami{\'a}n Avila and Safia Abdalla and Carol Willing},
	Booktitle = {Positioning and Power in Academic Publishing: Players, Agents and Agendas},
	Editor = {F. Loizides and B. Schmidt},
	Organization = {IOS Press},
	Pages = {87 - 90},
	Title = {Jupyter Notebooks -- a publishing format for reproducible computational workflows},
	Year = {2016}}

@ARTICLE{Barnes2019,
       author = {{Barnes}, Rory and {Luger}, Rodrigo and {Deitrick}, Russell and {Driscoll}, Peter and {Quinn}, Thomas R. and {Fleming}, David P. and {Smotherman}, Hayden and {McDonald}, Diego V. and {Wilhelm}, Caitlyn and {Garcia}, Rodolfo and {Barth}, Patrick and {Guyer}, Benjamin and {Meadows}, Victoria S. and {Bitz}, Cecilia M. and {Gupta}, Pramod and {Domagal-Goldman}, Shawn D. and {Armstrong}, John},
        title = "{VPLanet: The Virtual Planet Simulator}",
      journal = {\pasp},
         year = 2020,
        month = feb,
       volume = {132},
       number = {1008},
          eid = {024502},
        pages = {024502},
          doi = {10.1088/1538-3873/ab3ce8},
archivePrefix = {arXiv},
       eprint = {1905.06367},
 primaryClass = {astro-ph.EP},
       adsurl = {https://ui.adsabs.harvard.edu/abs/2020PASP..132b4502B}
}

@ARTICLE{Hallatt2021,
       author = {{Hallatt}, Tim and {Lee}, Eve J.},
        title = "{Sculpting the Sub-Saturn Occurrence Rate via Atmospheric Mass Loss}",
      journal = {\apj},
         year = 2022,
        month = jan,
       volume = {924},
       number = {1},
          eid = {9},
        pages = {9},
          doi = {10.3847/1538-4357/ac32c9},
archivePrefix = {arXiv},
       eprint = {2105.12746},
 primaryClass = {astro-ph.EP},
       adsurl = {https://ui.adsabs.harvard.edu/abs/2022ApJ...924....9H}
}

@ARTICLE{Pu2017,
       author = {{Pu}, Bonan and {Valencia}, Diana},
        title = "{Ohmic Dissipation in Mini-Neptunes}",
      journal = {\apj},
         year = 2017,
        month = sep,
       volume = {846},
       number = {1},
          eid = {47},
        pages = {47},
          doi = {10.3847/1538-4357/aa826f},
archivePrefix = {arXiv},
       eprint = {1709.01642},
 primaryClass = {astro-ph.EP},
       adsurl = {https://ui.adsabs.harvard.edu/abs/2017ApJ...846...47P}
}

@ARTICLE{Adams2021,
       author = {{Adams}, Fred C. and {Meyer}, Michael R. and {Adams}, Arthur D.},
        title = "{A Theoretical Framework for the Mass Distribution of Gas Giant Planets Forming through the Core Accretion Paradigm}",
      journal = {\apj},
         year = 2021,
        month = mar,
       volume = {909},
       number = {1},
          eid = {1},
        pages = {1},
          doi = {10.3847/1538-4357/abdd2b},
archivePrefix = {arXiv},
       eprint = {2101.06714},
 primaryClass = {astro-ph.EP},
       adsurl = {https://ui.adsabs.harvard.edu/abs/2021ApJ...909....1A}
}

@ARTICLE{Howe2014,
       author = {{Howe}, Alex R. and {Burrows}, Adam and {Verne}, Wesley},
        title = "{Mass-radius Relations and Core-envelope Decompositions of Super-Earths and Sub-Neptunes}",
      journal = {\apj},
         year = 2014,
        month = jun,
       volume = {787},
       number = {2},
          eid = {173},
        pages = {173},
          doi = {10.1088/0004-637X/787/2/173},
archivePrefix = {arXiv},
       eprint = {1402.4818},
 primaryClass = {astro-ph.EP},
       adsurl = {https://ui.adsabs.harvard.edu/abs/2014ApJ...787..173H}
}

@ARTICLE{Pollack1996,
       author = {{Pollack}, James B. and {Hubickyj}, Olenka and {Bodenheimer}, Peter and {Lissauer}, Jack J. and {Podolak}, Morris and {Greenzweig}, Yuval},
        title = "{Formation of the Giant Planets by Concurrent Accretion of Solids and Gas}",
      journal = {\icarus},
         year = 1996,
        month = nov,
       volume = {124},
       number = {1},
        pages = {62-85},
          doi = {10.1006/icar.1996.0190},
       adsurl = {https://ui.adsabs.harvard.edu/abs/1996Icar..124...62P}
}

@article{Libby-Roberts-K51d-JWST, title={The James Webb Space Telescope NIRSpec-PRISM Transmission Spectrum of the Super-Puff, Kepler-51d}, url={https://ui.adsabs.harvard.edu/abs/2025arXiv250521358L}, DOI={10.48550/arXiv.2505.21358}, note={ADS Bibcode: 2025arXiv250521358L}, publisher={arXiv}, author={Libby-Roberts, Jessica E. and Bello-Arufe, Aaron and Berta-Thompson, Zachory K. and Cañas, Caleb I. and Chachan, Yayaati and Hu, Renyu and Kawashima, Yui and Murray, Catriona and Ohno, Kazumasa and Tokadjian, Armen and Mahadevan, Suvrath and Masuda, Kento and Hebb, Leslie and Morley, Caroline and Fu, Guangwei and Gao, Peter and Stevenson, Kevin B.}, year={2025}, month=may }

@ARTICLE{Ohno2021,
       author = {{Ohno}, Kazumasa and {Tanaka}, Yuki A.},
        title = "{Grain Growth in Escaping Atmospheres: Implications for the Radius Inflation of Super-Puffs}",
      journal = {\apj},
         year = 2021,
        month = oct,
       volume = {920},
       number = {2},
          eid = {124},
        pages = {124},
          doi = {10.3847/1538-4357/ac1516},
archivePrefix = {arXiv},
       eprint = {2107.07027},
 primaryClass = {astro-ph.EP},
       adsurl = {https://ui.adsabs.harvard.edu/abs/2021ApJ...920..124O}
}

@ARTICLE{Gao2020,
       author = {{Gao}, Peter and {Zhang}, Xi},
        title = "{Deflating Super-puffs: Impact of Photochemical Hazes on the Observed Mass-Radius Relationship of Low-mass Planets}",
      journal = {\apj},
         year = 2020,
        month = feb,
       volume = {890},
       number = {2},
          eid = {93},
        pages = {93},
          doi = {10.3847/1538-4357/ab6a9b},
archivePrefix = {arXiv},
       eprint = {2001.00055},
 primaryClass = {astro-ph.EP},
       adsurl = {https://ui.adsabs.harvard.edu/abs/2020ApJ...890...93G}
}

@article{Galarza_2024_engulfment_TOI1173, title={Detailed Abundances of the Planet-hosting TOI-1173 A/B System: Possible Evidence of Planet Engulfment in a Very Wide Binary}, volume={974}, ISSN={0004-637X}, DOI={10.3847/1538-4357/ad697f}, abstractNote={Over the last decade, studies of large samples of binary systems have identified chemical anomalies and shown that they might be attributed to planet formation or planet engulfment. However, both scenarios have primarily been tested in pairs without known exoplanets. In this work, we explore these scenarios in the newly detected planet-hosting wide binary TOI-1173 A/B (projected separation ∼11,400 au), using high-resolution MAROON-X and ARCES spectra. We determined photospheric stellar parameters both by fitting stellar models and via the spectroscopic equilibrium approach. Both analyses agree and suggest that they are cool main-sequence stars located in the thin disk. A line-by-line differential analysis between the components (B−A) displays an abundance pattern in the condensation temperature plane, where the planet-hosting star TOI-1173 A is enhanced in refractory elements such as iron by more than 0.05 dex. This suggests the engulfment of ∼18 M ⊕ of rocky material in star A. Our hypothesis is supported by the dynamics of the system (detailed in our companion paper), which suggest that the super-Neptune TOI-1173 A b might have been delivered to its current short period (∼7 days) through circularization and von Zeipel─Lidov─Kozai mechanisms, thereby triggering the engulfment of inner rocky exoplanets.}, note={ADS Bibcode: 2024ApJ...974..122Y}, journal={The Astrophysical Journal}, publisher={IOP}, author={Yana Galarza, Jhon and Reggiani, Henrique and Ferreira, Thiago and Lorenzo-Oliveira, Diego and Simon, Joshua D. and McWilliam, Andrew and Schlaufman, Kevin C. and Miquelarena, Paula and Flores Trivigno, Matias and Jaque Arancibia, Marcelo}, year={2024}, month=oct, pages={122} }

@article{David_2019_V1298TauDiscovery, title={Four Newborn Planets Transiting the Young Solar Analog V1298 Tau}, volume={885}, ISSN={0004-637X}, DOI={10.3847/2041-8213/ab4c99}, abstractNote={Exoplanets orbiting pre-main-sequence stars are laboratories for studying planet evolution processes, including atmospheric loss, orbital migration, and radiative cooling. V1298 Tau, a young solar analog with an age of 23 ± 4 Myr, is one such laboratory. The star is already known to host a Jupiter-sized planet on a 24 day orbit. Here, we report the discovery of three additional planets—all between the sizes of Neptune and Saturn—based on our analysis of K2 Campaign 4 photometry. Planets c and d have sizes of 5.6 and 6.4 {R}oplus , respectively, and with orbital periods of 8.25 and 12.40 days reside 0.25% outside of the nominal 3:2 mean-motion resonance. Planet e is 8.7 {R}oplus in size but only transited once in the K2 time series and thus has a period longer than 36 days, but likely shorter than 223 days. The V1298 Tau system may be a precursor to the compact multiplanet systems found to be common by the Kepler mission. However, the large planet sizes stand in sharp contrast to the vast majority of Kepler multiplanet systems, which have planets smaller than 3 {R}oplus . Simple dynamical arguments suggest total masses of <28 {M}oplus and <120 {M}oplus for the c-d and d-b planet pairs, respectively. The implied low masses suggest that the planets may still be radiatively cooling and contracting, and perhaps losing atmosphere. The V1298 Tau system offers rich prospects for further follow-up including atmospheric characterization by transmission or eclipse spectroscopy, dynamical characterization through transit-timing variations, and measurements of planet masses and obliquities by radial velocities.}, note={ADS Bibcode: 2019ApJ...885L..12D}, journal={The Astrophysical Journal}, publisher={IOP}, author={David, Trevor J. and Petigura, Erik A. and Luger, Rodrigo and Foreman-Mackey, Daniel and Livingston, John H. and Mamajek, Eric E. and Hillenbrand, Lynne A.}, year={2019}, month=nov, pages={L12} }

@ARTICLE{Lee2016,
       author = {{Lee}, Eve J. and {Chiang}, Eugene},
        title = "{Breeding Super-Earths and Birthing Super-puffs in Transitional Disks}",
      journal = {\apj},
         year = 2016,
        month = feb,
       volume = {817},
       number = {2},
          eid = {90},
        pages = {90},
          doi = {10.3847/0004-637X/817/2/90},
archivePrefix = {arXiv},
       eprint = {1510.08855},
 primaryClass = {astro-ph.EP},
       adsurl = {https://ui.adsabs.harvard.edu/abs/2016ApJ...817...90L}
}

@ARTICLE{Ohno2022,
       author = {{Ohno}, Kazumasa and {Fortney}, Jonathan J.},
        title = "{A Framework for Characterizing Transmission Spectra of Exoplanets with Circumplanetary Rings}",
      journal = {arXiv e-prints},
         year = 2022,
        month = jan,
          eid = {arXiv:2201.02794},
        pages = {arXiv:2201.02794},
archivePrefix = {arXiv},
       eprint = {2201.02794},
 primaryClass = {astro-ph.EP},
       adsurl = {https://ui.adsabs.harvard.edu/abs/2022arXiv220102794O}
}

@ARTICLE{Alam2022,
       author = {{Alam}, Munazza K. and {Kirk}, James and {Dressing}, Courtney D. and {L{\'o}pez-Morales}, Mercedes and {Ohno}, Kazumasa and {Gao}, Peter and {Akinsanmi}, Babatunde and {Santerne}, Alexandre and {Grouffal}, Salom{\'e} and {Adibekyan}, Vardan and {Barros}, Susana C.~C. and {Buchhave}, Lars A. and {Crossfield}, Ian J.~M. and {Dai}, Fei and {Deleuil}, Magali and {Giacalone}, Steven and {Lillo-Box}, Jorge and {Marley}, Mark and {Mayo}, Andrew W. and {Mortier}, Annelies and {Santos}, Nuno C. and {Sousa}, S{\'e}rgio G. and {Turtelboom}, Emma V. and {Wheatley}, Peter J. and {Vanderburg}, Andrew M.},
        title = "{The First Near-infrared Transmission Spectrum of HIP 41378 f, A Low-mass Temperate Jovian World in a Multiplanet System}",
      journal = {\apjl},
         year = 2022,
        month = mar,
       volume = {927},
       number = {1},
          eid = {L5},
        pages = {L5},
          doi = {10.3847/2041-8213/ac559d},
archivePrefix = {arXiv},
       eprint = {2201.02686},
 primaryClass = {astro-ph.EP},
       adsurl = {https://ui.adsabs.harvard.edu/abs/2022ApJ...927L...5A}
}

@ARTICLE{Wang2019_dustyoutflows,
       author = {{Wang}, Lile and {Dai}, Fei},
        title = "{Dusty Outflows in Planetary Atmospheres: Understanding {\textquotedblleft}Super-puffs{\textquotedblright} and Transmission Spectra of Sub-Neptunes}",
      journal = {\apjl},
         year = 2019,
        month = mar,
       volume = {873},
       number = {1},
          eid = {L1},
        pages = {L1},
          doi = {10.3847/2041-8213/ab0653},
archivePrefix = {arXiv},
       eprint = {1902.04188},
 primaryClass = {astro-ph.EP},
       adsurl = {https://ui.adsabs.harvard.edu/abs/2019ApJ...873L...1W}
}

@article{Yee_Vissapragada_2025_popcorn, title={“Popcorn Planets” are Not Actively Inflated by Eccentricity Tides}, url={https://ui.adsabs.harvard.edu/abs/2025arXiv251107746Y}, DOI={10.48550/arXiv.2511.07746}, note={ADS Bibcode: 2025arXiv251107746Y}, publisher={arXiv}, author={Yee, Samuel W. and Vissapragada, Shreyas}, year={2025}, month=nov }

@article{Barkaoui_2024_WASP193, title={An extended low-density atmosphere around the Jupiter-sized planet WASP-193 b}, volume={8}, ISSN={2397-3366}, DOI={10.1038/s41550-024-02259-y}, note={ADS Bibcode: 2024NatAs...8..909B}, journal={Nature Astronomy}, author={Barkaoui, Khalid and Pozuelos, Francisco J. and Hellier, Coel and Smalley, Barry and Nielsen, Louise D. and Niraula, Prajwal and Gillon, Michaël and de Wit, Julien and Müller, Simon and Dorn, Caroline and Helled, Ravit and Jehin, Emmanuel and Demory, Brice-Olivier and Van Grootel, Valerie and Soubkiou, Abderahmane and Ghachoui, Mourad and Anderson, David R. and Benkhaldoun, Zouhair and Bouchy, Francois and Burdanov, Artem and Delrez, Laetitia and Ducrot, Elsa and Garcia, Lionel and Jabiri, Abdelhadi and Lendl, Monika and Maxted, Pierre F. L. and Murray, Catriona A. and Pedersen, Peter Pihlmann and Queloz, Didier and Sebastian, Daniel and Turner, Oliver and Udry, Stephane and Timmermans, Mathilde and Triaud, Amaury H. M. J. and West, Richard G.}, year={2024}, month=july, pages={909–919} }

@article{Bayliss_2015_HATS8b, title={HATS-8b: A Low-density Transiting Super-Neptune}, volume={150}, ISSN={0004-6256}, DOI={10.1088/0004-6256/150/2/49}, note={ADS Bibcode: 2015AJ....150...49B}, journal={The Astronomical Journal}, publisher={IOP}, author={Bayliss, D. and Hartman, J. D. and Bakos, G. {\'A'}. and Penev, K. and Zhou, G. and Brahm, R. and Rabus, M. and Jordán, A. and Mancini, L. and de Val-Borro, M. and Bhatti, W. and Espinoza, N. and Csubry, Z. and Howard, A. W. and Fulton, B. J. and Buchhave, L. A. and Henning, T. and Schmidt, B. and Ciceri, S. and Noyes, R. W. and Isaacson, H. and Marcy, G. W. and Suc, V. and Lázár, J. and Papp, I. and Sári, P.}, year={2015}, month=aug, pages={49} }

@article{Beatty_2017, title={Determining Empirical Stellar Masses and Radii from Transits and Gaia Parallaxes as Illustrated by Spitzer Observations of KELT-11b}, volume={154}, ISSN={0004-6256}, DOI={10.3847/1538-3881/aa7511}, note={ADS Bibcode: 2017AJ....154...25B}, journal={The Astronomical Journal}, publisher={IOP}, author={Beatty, Thomas G. and Stevens, Daniel J. and Collins, Karen A. and Colón, Knicole D. and James, David J. and Kreidberg, Laura and Pepper, Joshua and Rodriguez, Joseph E. and Siverd, Robert J. and Stassun, Keivan G. and Kielkopf, John F.}, year={2017}, month=july, pages={25} }

@article{Bonomo_2023, title={Cold Jupiters and improved masses in 38 Kepler and K2 small planet systems from 3661 HARPS-N radial velocities. No excess of cold Jupiters in small planet systems}, volume={677}, ISSN={0004-6361}, DOI={10.1051/0004-6361/202346211}, note={ADS Bibcode: 2023A&A...677A..33B}, journal={Astronomy and Astrophysics}, publisher={EDP}, author={Bonomo, A. S. and Dumusque, X. and Massa, A. and Mortier, A. and Bongiolatti, R. and Malavolta, L. and Sozzetti, A. and Buchhave, L. A. and Damasso, M. and Haywood, R. D. and Morbidelli, A. and Latham, D. W. and Molinari, E. and Pepe, F. and Poretti, E. and Udry, S. and Affer, L. and Boschin, W. and Charbonneau, D. and Cosentino, R. and Cretignier, M. and Ghedina, A. and Lega, E. and López-Morales, M. and Margini, M. and Martínez Fiorenzano, A. F. and Mayor, M. and Micela, G. and Pedani, M. and Pinamonti, M. and Rice, K. and Sasselov, D. and Tronsgaard, R. and Vanderburg, A.}, year={2023}, month=sept, pages={A33} }

@article{Brahm_Hartman_2018_HATS, title={HATS-43b, HATS-44b, HATS-45b, and HATS-46b: Four Short-period Transiting Giant Planets in the Neptune-Jupiter Mass Range}, volume={155}, ISSN={0004-6256}, DOI={10.3847/1538-3881/aaa898}, note={ADS Bibcode: 2018AJ....155..112B}, journal={The Astronomical Journal}, publisher={IOP}, author={Brahm, R. and Hartman, J. D. and Jordán, A. and Bakos, G. {\'A'}. and Espinoza, N. and Rabus, M. and Bhatti, W. and Penev, K. and Sarkis, P. and Suc, V. and Csubry, Z. and Bayliss, D. and Bento, J. and Zhou, G. and Mancini, L. and Henning, T. and Ciceri, S. and de Val-Borro, M. and Shectman, S. and Crane, J. D. and Arriagada, P. and Butler, P. and Teske, J. and Thompson, I. and Osip, D. and Díaz, M. and Schmidt, B. and Lázár, J. and Papp, I. and Sári, P.}, year={2018}, month=mar, pages={112} }

@article{Hadden_Lithwick_2017, title={Kepler Planet Masses and Eccentricities from TTV Analysis}, volume={154}, ISSN={0004-6256}, DOI={10.3847/1538-3881/aa71ef}, note={ADS Bibcode: 2017AJ....154....5H}, journal={The Astronomical Journal}, publisher={IOP}, author={Hadden, Sam and Lithwick, Yoram}, year={2017}, month=july, pages={5} }

@article{Hartman_2019_HATS, title={HATS-60b-HATS-69b: 10 Transiting Planets from HATSouth}, volume={157}, ISSN={0004-6256}, DOI={10.3847/1538-3881/aaf8b6}, note={ADS Bibcode: 2019AJ....157...55H}, journal={The Astronomical Journal}, publisher={IOP}, author={Hartman, J. D. and Bakos, G. {\'A'}. and Bayliss, D. and Bento, J. and Bhatti, W. and Brahm, R. and Csubry, Z. and Espinoza, N. and Henning, Th. and Jordán, A. and Mancini, L. and Penev, K. and Rabus, M. and Sarkis, P. and Suc, V. and de Val-Borro, M. and Zhou, G. and Addison, B. and Arriagada, P. and Butler, R. P. and Crane, J. and Durkan, S. and Shectman, S. and Tan, T. G. and Thompson, I. and Tinney, C. G. and Wright, D. J. and Lázár, J. and Papp, I. and Sári, P.}, year={2019}, month=feb, pages={55} }

@article{Hellier_2017, title={WASP-South transiting exoplanets: WASP-130b, WASP-131b, WASP-132b, WASP-139b, WASP-140b, WASP-141b and WASP-142b}, volume={465}, ISSN={0035-8711}, DOI={10.1093/mnras/stw3005}, note={ADS Bibcode: 2017MNRAS.465.3693H}, journal={Monthly Notices of the Royal Astronomical Society}, publisher={OUP}, author={Hellier, C. and Anderson, D. R. and Collier Cameron, A. and Delrez, L. and Gillon, M. and Jehin, E. and Lendl, M. and Maxted, P. F. L. and Neveu-VanMalle, M. and Pepe, F. and Pollacco, D. and Queloz, D. and Ségransan, D. and Smalley, B. and Southworth, J. and Triaud, A. H. M. J. and Udry, S. and Wagg, T. and West, R. G.}, year={2017}, month=mar, pages={3693–3707} }

@article{Jontof-Hutter_2014_Kepler79, title={Kepler-79’s Low Density Planets}, volume={785}, ISSN={0004-637X}, DOI={10.1088/0004-637X/785/1/15}, note={ADS Bibcode: 2014ApJ...785...15J}, journal={The Astrophysical Journal}, publisher={IOP}, author={Jontof-Hutter, Daniel and Lissauer, Jack J. and Rowe, Jason F. and Fabrycky, Daniel C.}, year={2014}, month=apr, pages={15} }

@article{Karalis_2025, title={Separating Super-puffs versus Hot Jupiters among Young Puffy Planets}, volume={978}, ISSN={0004-637X, 1538-4357}, DOI={10.3847/1538-4357/ad946c}, number={1}, journal={The Astrophysical Journal}, author={Karalis, Amalia and Lee, Eve J. and Thorngren, Daniel P.}, year={2025}, month=jan, pages={46} }

@article{Liang_2021_Kepler90, title={Kepler-90: Giant Transit-timing Variations Reveal a Super-puff}, volume={161}, ISSN={0004-6256, 1538-3881}, DOI={10.3847/1538-3881/abe6a7}, number={4}, journal={The Astronomical Journal}, author={Liang, Yan and Robnik, Jakob and Seljak, Uroš}, year={2021}, month=apr, pages={202} }

@article{Mantovan_Malavolta_2024, title={The GAPS programme at TNG. XLIX. TOI-5398, the youngest compact multi-planet system composed of an inner sub-Neptune and an outer warm Saturn}, volume={682}, ISSN={0004-6361}, DOI={10.1051/0004-6361/202347472}, note={ADS Bibcode: 2024A&A...682A.129M}, journal={Astronomy and Astrophysics}, publisher={EDP}, author={Mantovan, G. and Malavolta, L. and Desidera, S. and Zingales, T. and Borsato, L. and Piotto, G. and Maggio, A. and Locci, D. and Polychroni, D. and Turrini, D. and Baratella, M. and Biazzo, K. and Nardiello, D. and Stassun, K. and Nascimbeni, V. and Benatti, S. and John, A. Anna and Watkins, C. and Bieryla, A. and Lissauer, J. J. and Twicken, J. D. and Lanza, A. F. and Winn, J. N. and Messina, S. and Montalto, M. and Sozzetti, A. and Boffin, H. and Cheryasov, D. and Strakhov, I. and Murgas, F. and D’Arpa, M. and Barkaoui, K. and Benni, P. and Bignamini, A. and Bonomo, A. S. and Borsa, F. and Cabona, L. and Cameron, A. C. and Claudi, R. and Cochran, W. and Collins, K. A. and Damasso, M. and Dong, J. and Endl, M. and Fukui, A. and Fűrész, G. and Gandolfi, D. and Ghedina, A. and Jenkins, J. and Kabáth, P. and Latham, D. W. and Lorenzi, V. and Luque, R. and Maldonado, J. and McLeod, K. and Molinaro, M. and Narita, N. and Nowak, G. and Orell-Miquel, J. and Pallé, E. and Parviainen, H. and Pedani, M. and Quinn, S. N. and Relles, H. and Rowden, P. and Scandariato, G. and Schwarz, R. and Seager, S. and Shporer, A. and Vanderburg, A. and Wilson, T. G.}, year={2024}, month=feb, pages={A129} }

@ARTICLE{Masuda2014,
       author = {{Masuda}, Kento},
        title = "{Very Low Density Planets around Kepler-51 Revealed with Transit Timing Variations and an Anomaly Similar to a Planet-Planet Eclipse Event}",
      journal = {\apj},
         year = 2014,
        month = mar,
       volume = {783},
       number = {1},
          eid = {53},
        pages = {53},
          doi = {10.1088/0004-637X/783/1/53},
archivePrefix = {arXiv},
       eprint = {1401.2885},
 primaryClass = {astro-ph.EP},
       adsurl = {https://ui.adsabs.harvard.edu/abs/2014ApJ...783...53M}
}

@article{McKee_Montet_2023, title={Transit Depth Variations Reveal TOI-216 b to be a Super-puff}, volume={165}, ISSN={0004-6256, 1538-3881}, DOI={10.3847/1538-3881/accd66}, number={6}, journal={The Astronomical Journal}, author={McKee, Brendan J. and Montet, Benjamin T.}, year={2023}, month=june, pages={236} }

@article{Mills_2016, title={A resonant chain of four transiting, sub-Neptune planets}, volume={533}, ISSN={0028-0836}, DOI={10.1038/nature17445}, note={ADS Bibcode: 2016Natur.533..509M}, journal={Nature}, author={Mills, Sean M. and Fabrycky, Daniel C. and Migaszewski, Cezary and Ford, Eric B. and Petigura, Erik and Isaacson, Howard}, year={2016}, month=may, pages={509–512} }

@article{Močnik_2017, title={Starspots on WASP-107 and pulsations of WASP-118}, volume={469}, ISSN={0035-8711}, DOI={10.1093/mnras/stx972}, note={ADS Bibcode: 2017MNRAS.469.1622M}, journal={Monthly Notices of the Royal Astronomical Society}, publisher={OUP}, author={Močnik, T. and Hellier, C. and Anderson, D. R. and Clark, B. J. M. and Southworth, J.}, year={2017}, month=aug, pages={1622–1629} }

@article{Ofir_2014, title={An independent planet search in the Kepler dataset. II. An extremely low-density super-Earth mass planet around Kepler-87}, volume={561}, ISSN={0004-6361}, DOI={10.1051/0004-6361/201220935}, note={ADS Bibcode: 2014A&A...561A.103O}, journal={Astronomy and Astrophysics}, publisher={EDP}, author={Ofir, Aviv and Dreizler, Stefan and Zechmeister, Mathias and Husser, Tim-Oliver}, year={2014}, month=jan, pages={A103} }

@article{Ofir_2025, title={Planetary Mass Determinations from a Simplified Photodynamical Model—Application to the Complete Kepler Dataset}, volume={169}, ISSN={0004-6256}, DOI={10.3847/1538-3881/ad91a7}, note={ADS Bibcode: 2025AJ....169...90O}, journal={The Astronomical Journal}, publisher={IOP}, author={Ofir, Aviv and Yoffe, Gideon and Aharonson, Oded}, year={2025}, month=feb, pages={90} }

@article{Petigura_2018_K2-24PARAMS, title={Dynamics and Formation of the Near-resonant K2-24 System: Insights from Transit-timing Variations and Radial Velocities}, volume={156}, ISSN={0004-6256}, DOI={10.3847/1538-3881/aaceac}, note={ADS Bibcode: 2018AJ....156...89P}, journal={The Astronomical Journal}, publisher={IOP}, author={Petigura, Erik A. and Benneke, Björn and Batygin, Konstantin and Fulton, Benjamin J. and Werner, Michael and Krick, Jessica E. and Gorjian, Varoujan and Sinukoff, Evan and Deck, Katherine M. and Mills, Sean M. and Deming, Drake}, year={2018}, month=sept, pages={89} }

@article{Polanski_2024, title={The TESS-Keck Survey. XX. 15 New TESS Planets and a Uniform RV Analysis of All Survey Targets}, volume={272}, ISSN={0067-0049}, DOI={10.3847/1538-4365/ad4484}, note={ADS Bibcode: 2024ApJS..272...32P}, journal={The Astrophysical Journal Supplement Series}, publisher={IOP}, author={Polanski, Alex S. and Lubin, Jack and Beard, Corey and Akana Murphy, Joseph M. and Rubenzahl, Ryan and Hill, Michelle L. and Crossfield, Ian J. M. and Chontos, Ashley and Robertson, Paul and Isaacson, Howard and Kane, Stephen R. and Ciardi, David R. and Batalha, Natalie M. and Dressing, Courtney and Fulton, Benjamin and Howard, Andrew W. and Huber, Daniel and Petigura, Erik A. and Weiss, Lauren M. and Angelo, Isabel and Behmard, Aida and Blunt, Sarah and Brinkman, Casey L. and Dai, Fei and Dalba, Paul A. and Fetherolf, Tara and Giacalone, Steven and Hirsch, Lea A. and Holcomb, Rae and Kosiarek, Molly R. and Mayo, Andrew W. and MacDougall, Mason G. and Močnik, Teo and Pidhorodetska, Daria and Rice, Malena and Rosenthal, Lee J. and Scarsdale, Nicholas and Turtelboom, Emma V. and Tyler, Dakotah and Van Zandt, Judah and Yee, Samuel W. and Coria, David R. and Dulz, Shannon D. and Hartman, Joel D. and Householder, Aaron and Lange, Sarah and Langford, Andrew and Louden, Emma M. and Siegel, Jared C. and Gilbert, Emily A. and Gonzales, Erica J. and Schlieder, Joshua E. and Boyle, Andrew W. and Christiansen, Jessie L. and Clark, Catherine A. and Fernandes, Rachel B. and Lund, Michael B. and Savel, Arjun B. and Gill, Holden and Beichman, Charles and Matson, Rachel and Matthews, Elisabeth C. and Furlan, E. and Howell, Steve B. and Scott, Nicholas J. and Everett, Mark E. and Livingston, John H. and Ershova, Irina O. and Cheryasov, Dmitry V. and Safonov, Boris and Lillo-Box, Jorge and Barrado, David and Morales-Calderón, María}, year={2024}, month=june, pages={32} }

@article{Sanchis-Ojeda_2012, title={Alignment of the stellar spin with the orbits of a three-planet system}, volume={487}, ISSN={0028-0836}, DOI={10.1038/nature11301}, note={ADS Bibcode: 2012Natur.487..449S}, journal={Nature}, author={Sanchis-Ojeda, Roberto and Fabrycky, Daniel C. and Winn, Joshua N. and Barclay, Thomas and Clarke, Bruce D. and Ford, Eric B. and Fortney, Jonathan J. and Geary, John C. and Holman, Matthew J. and Howard, Andrew W. and Jenkins, Jon M. and Koch, David and Lissauer, Jack J. and Marcy, Geoffrey W. and Mullally, Fergal and Ragozzine, Darin and Seader, Shawn E. and Still, Martin and Thompson, Susan E.}, year={2012}, month=july, pages={449–453} }

@article{Santerne_2016_SOPHIE_PARAMS, title={SOPHIE velocimetry of Kepler transit candidates. XVII. The physical properties of giant exoplanets within 400 days of period}, volume={587}, ISSN={0004-6361}, DOI={10.1051/0004-6361/201527329}, note={ADS Bibcode: 2016A&A...587A..64S}, journal={Astronomy and Astrophysics}, publisher={EDP}, author={Santerne, A. and Moutou, C. and Tsantaki, M. and Bouchy, F. and Hébrard, G. and Adibekyan, V. and Almenara, J. -M. and Amard, L. and Barros, S. C. C. and Boisse, I. and Bonomo, A. S. and Bruno, G. and Courcol, B. and Deleuil, M. and Demangeon, O. and Díaz, R. F. and Guillot, T. and Havel, M. and Montagnier, G. and Rajpurohit, A. S. and Rey, J. and Santos, N. C.}, year={2016}, month=mar, pages={A64} }

@article{Wang_Liu_2024, title={Photo-dynamical Analysis of Circumbinary Multi-planet System TOI-1338: A Fully Coplanar Configuration with a Puffy Planet}, volume={168}, ISSN={0004-6256}, DOI={10.3847/1538-3881/ad4a60}, note={ADS Bibcode: 2024AJ....168...31W}, journal={The Astronomical Journal}, publisher={IOP}, author={Wang, Mu-Tian and Liu, Hui-Gen}, year={2024}, month=july, pages={31} }
\bibliographystyle{aasjournalv7}



\end{document}